\documentclass[twoside, onecolumn,11pt]{article}
\usepackage{extsizes}
\usepackage[super,sort&compress,comma]{natbib} 
\usepackage[version=3]{mhchem}
\usepackage[left=1.5cm, right=1.5cm, top=1.785cm, bottom=2.0cm]{geometry}
\usepackage{balance}
\usepackage{times,mathptmx}
\usepackage{sectsty}
\usepackage{graphicx} 
\usepackage{lastpage}
\usepackage[format=plain,justification=justified,singlelinecheck=true,font={stretch=1.125,small,sf},labelfont=bf,labelsep=space]{caption}
\usepackage{float}
\usepackage{fancyhdr}
\usepackage{fnpos}
\usepackage[english]{babel}
\addto{\captionsenglish}{%
  
}
\usepackage{array}
\usepackage{droidsans}
\usepackage{charter}
\usepackage[T1]{fontenc}
\usepackage[usenames,dvipsnames]{xcolor}
\usepackage{setspace}
\usepackage[compact]{titlesec}
\usepackage{hyperref}
\usepackage{amsfonts}
\usepackage{amsmath,amssymb}

\usepackage{epstopdf}

\definecolor{cream}{RGB}{222,217,201}

\newcommand{\blue}[1]{{\color{black} #1}}

\usepackage{titlesec}
\titlelabel{\thetitle.\quad}

\usepackage{etoolbox}

\makeatletter
\let\oldcite\cite
\pretocmd{\listoffigures}{\def\cite{\ignorespaces\@gobble}}{}{}
\apptocmd{\listoffigures}{\let\cite\oldcite}{}{}
\pretocmd{\listoftables}{\def\cite{\ignorespaces\@gobble}}{}{}
\apptocmd{\listoftables}{\let\cite\oldcite}{}{}

\makeatother

\usepackage{mathtools}

\begin{document}


\makeFNbottom
\makeatletter
\renewcommand\LARGE{\@setfontsize\LARGE{15pt}{17}}
\renewcommand\Large{\@setfontsize\Large{12pt}{14}}
\renewcommand\large{\@setfontsize\large{10pt}{12}}
\renewcommand\footnotesize{\@setfontsize\footnotesize{7pt}{10}}
\makeatother

\renewcommand{\thefootnote}{\fnsymbol{footnote}}
\renewcommand\footnoterule{\vspace*{1pt}%
\color{cream}\hrule width 3.5in height 0.4pt \color{black}\vspace*{5pt}} 
\setcounter{secnumdepth}{5}

\makeatletter 
\renewcommand\@biblabel[1]{#1}            
\renewcommand\@makefntext[1]%
{\noindent\makebox[0pt][r]{\@thefnmark\,}#1}
\makeatother 
\renewcommand{\figurename}{\small{Figure}}
\sectionfont{\sffamily\Large}
\subsectionfont{\normalsize}
\subsubsectionfont{\bf}
\setstretch{1.125} 
\setlength{\skip\footins}{0.8cm}
\setlength{\footnotesep}{0.25cm}
\setlength{\jot}{10pt}
\titlespacing*{\section}{0pt}{4pt}{4pt}
\titlespacing*{\subsection}{0pt}{15pt}{1pt}

\makeatletter 
\newlength{\figrulesep} 
\setlength{\figrulesep}{0.5\textfloatsep} 

\newcommand{\topfigrule}{\vspace*{-1pt}%
\noindent{\color{cream}\rule[-\figrulesep]{\columnwidth}{1.5pt}} }

\newcommand{\botfigrule}{\vspace*{-2pt}%
\noindent{\color{cream}\rule[\figrulesep]{\columnwidth}{1.5pt}} }

\newcommand{\dblfigrule}{\vspace*{-1pt}%
\noindent{\color{cream}\rule[-\figrulesep]{\columnwidth}{1.5pt}} }

\makeatother

\begin{center}
\noindent\LARGE\textbf{Honeycomb Layered Frameworks with Metallophilic Bilayers}\\
\noindent\LARGE\centering{}
\end{center}

\noindent\large{Godwill Mbiti Kanyolo,\textit{$^{a, b *}$} Titus Masese,\textit{$^{a, c *}$} Yoshinobu Miyazaki,\textit{$^d$} Shintaro Tachibana,\textit{$^e$}, Chengchao Zhong,\textit{$^e$} Yuki Orikasa,\textit{$^e$} Tomohiro Saito\textit{$^d$} }\\

\noindent{\textit{$^{a}$Research Institute of Electrochemical Energy, National Institute of Advanced Industrial Science and Technology (AIST), 1-8-31 Midorigaoka, Ikeda, Osaka 563-8577, Japan.} Email: titus.masese@aist.go.jp; gm.kanyolo@aist.go.jp
\\
\textit{$^{b}$Department of Engineering Science, The University of Electro-Communications, 1-5-1 Chofugaoka, Chofu, Tokyo 182-8585, Japan. } Email: gmkanyolo@mail.uec.jp
\\
\textit{$^{c}$AIST-Kyoto University Chemical Energy Materials Open Innovation Laboratory (ChEM-OIL), Sakyo-ku, Kyoto 606-8501, Japan.}\\
\textit{$^{d}$Tsukuba Laboratory, Sumika Chemical Analysis Service (SCAS), Ltd.,Tsukuba, Ibaraki 300-3266, Japan.}\\
\textit{$^{e}$Graduate School of Life Sciences, Ritsumeikan University, 1-1-1 Noji-higashi, Kusatsu, Shiga 525-8577, Japan.}\\

\noindent\normalsize{
\textbf{Honeycomb layered frameworks have garnered traction in a wide range of disciplines owing not only to their unique honeycomb configuration, but also to the plenitude of physicochemical and topological properties such as fast ionic conduction, diverse coordination chemistry and structural defects amongst others typically exploited for energy storage applications. In turn, honeycomb layered frameworks manifesting metallophilic bilayer arrangements of cations sandwiched between the transition metal cation slabs have recently garnered attention due to the presence of anomalous fractional valency state of the cations always accompanied by metallophilic interactions constituting the cationic bonds within the bilayered structure. The concepts needed to characterise the aforementioned peculiarities and other phenomena such as conductor-semiconductor-insulator phase transition and magnetoresistance in these materials cut across multi-disciplines ranging from materials science and solid-state chemistry to condensed matter physics, suggesting applications that fall beyond energy storage. This \textit {Review} highlights the exciting advancements in the science of honeycomb layered frameworks with metallophilic bilayers. First, the latest tactics and techniques including but not limited to X-ray absorption spectroscopy (XAS) and high-resolution transmission electron microscopy (HRTEM) particularly necessary for characterising recent honeycomb layered frameworks with metallophilic bilayers are described, with emphasis on silver-based oxides. Second, new strategies and concepts related to topochemically- or temperature-induced cationic-deficient phases expanding the compositional space of honeycomb layered frameworks focused on cationic bilayer architectures are also accentuated.  Third, the latest condensed matter theoretic advances towards a full, atomistic description of the bilayered structure in such frameworks are detailed, especially related to critical phenomena at the cusp of the monolayer-bilayer phase transition. This entails, in part, describing honeycomb layered frameworks as optimised lattices within the congruent sphere packing problem, equivalent to a particular two-dimensional (2D) conformal field theory. Within this picture, the monolayer-bilayer phase transition represents the bifurcation of the honeycomb lattice into its bipartite constituents, related to a 2D-to-3D crossover. Altogether, it is hoped that this \textit {Review} will give the reader a panoramic view of the honeycomb layered frameworks with important applications within the emerging field of quantum matter, potentially redefining their frontier. Thus, the scope of this \textit {Review} is expected to be worthwhile for recent graduates and emerging experts alike not only in the materials science and chemistry community but also in other diverse fields of interest.}}

\begin{figure*}[!t] 
 \centering
 \includegraphics[width=1.0\columnwidth]{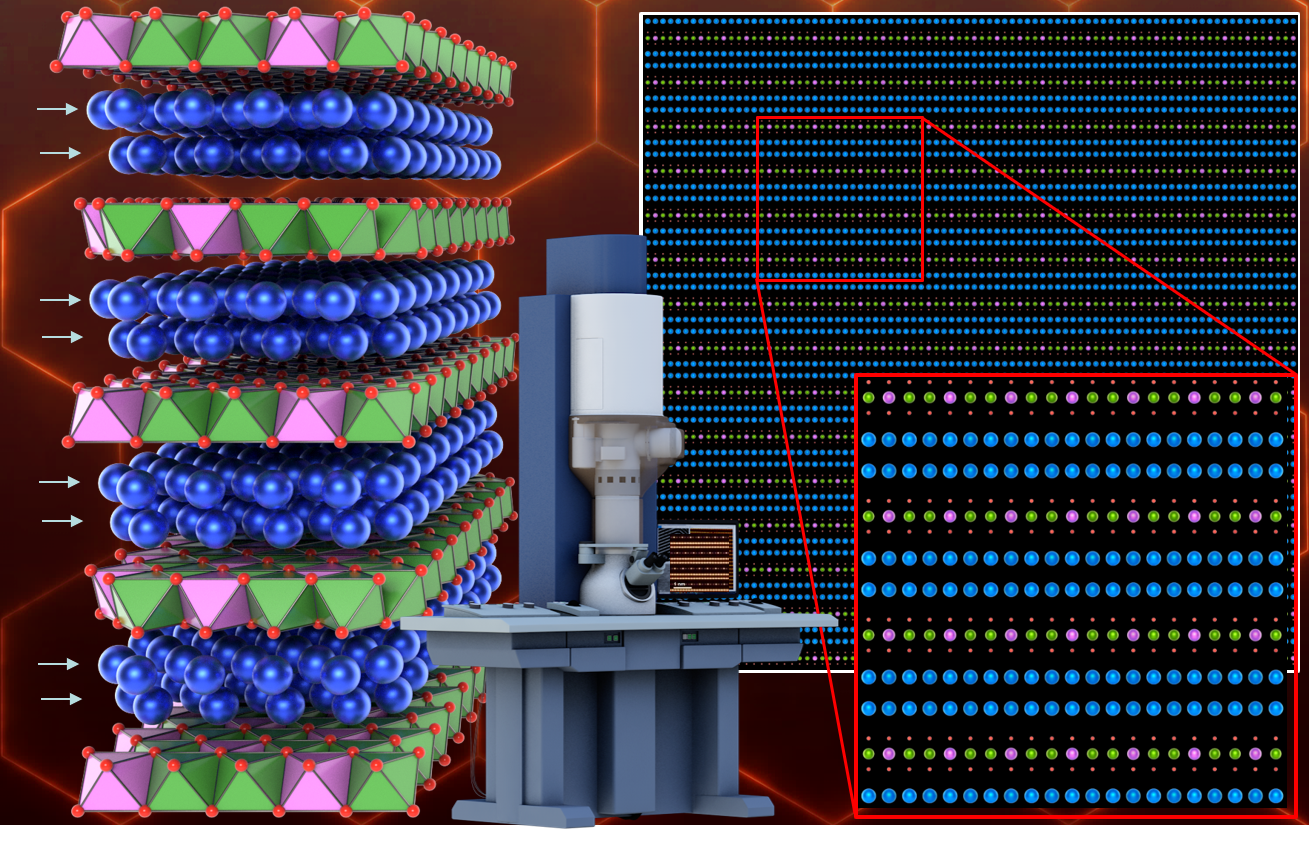}
 \label{Figure_rendition}
\end{figure*}

\renewcommand*\rmdefault{bch}\normalfont\upshape
\rmfamily
\section*{}
\vspace{-1cm}

\newpage

\tableofcontents

\newpage

\newpage

\newpage

\section{\label{Section: Introduction} Introduction}

The groundbreaking achievement of isolating graphene, a two-dimensional (2D) material, from the three-dimensional (3D) graphite introduced an immensely powerful tool for investigating and manipulating a range of intriguing phenomena.\cite{Novoselov2004, Kitagawa2018} One such phenomenon is the quantum Hall effect, intricately linked to the confined spatial dimensions of the 2D graphene structure. These remarkable 2D phenomena also manifest prominently in layered materials such as honeycomb layered frameworks, which hold immense research importance in terms of their characterisation, functionality, and subsequent potential for commercialisation. The fundamental comprehension of the inherent 2D condensed matter phenomena exhibited by these honeycomb layered frameworks is pivotal to unlocking this potential, thus revolutionising the realm of multifunctional materials.

Specifically, honeycomb layered frameworks are characterised by their distinctive composition, typically comprising alternating layers of transition metal slabs encapsulating monolayer lattices consisting of alkali metal atoms (\textit {e.g.}, K, Na, Li, \textit {etc}.), alkaline-earth metal atoms (Ba, Ca, Mg, \textit {etc}.), or coinage metal atoms (Cu, Ag, Au, \textit {etc}.). Each individual slab or lattice within these frameworks consists of 2D arrays of atoms meticulously arranged in a honeycomb or hexagonal pattern.\cite{kanyolo2021honeycomb} Consequently, these materials exhibit a rich array of properties that are inherently linked to the study of densely-packed (optimised) lattices. Remarkably, these optimised lattices venture beyond the conventional boundaries of energy storage investigations, encompassing broader scientific domains where their implications unfold in captivating ways. Notably, the honeycomb lattice is not only a non-Bravais bipartite lattice, but is also the unique lattice which optimises the area of a 2D enclosure whilst minimising its perimeter (honeycomb conjecture\cite{hales2001honeycomb}) -- a fact exploited by evolution which naturally selected worker bees that minimised their labour for maximal honey storage. The honeycomb conjecture has biomimetically been employed to rationalise the honeycomb diffusion pathways of cations in honeycomb layered materials\cite{kanyolo2022cationic, masese2021mixed}, provided the honeycomb area and perimeter can be associated with the thermodynamic entropy and free energy respectively in the diffusion process (entropy proportional to area is also an important feature in black hole thermodynamics\cite{kanyolo2022local, kanyolo2022advances2}), rendering honeycomb layered materials as possible analogue materials for `quantum gravity' research. Indeed, applying mathematical techniques employed in the congruent sphere packing problem in their characterisation catapults research in honeycomb layered frameworks into diverse fields of study such as pure mathematics, quantum computing (particularly code optimisation and error correction), and more recently, quantum gravity research where such notions are ubiquitous.\cite {masese2023honeycomb, kanyolo2021honeycomb, kanyolo2022advances, kanyolo2020idealised, kanyolo2022cationic, kanyolo2021partition, kanyolo2022local, masese2021topological, kitaev2006anyons, kanyolo2021reproducing, tada2022implications, kanyolo2022advances2} Thus, honeycomb layered frameworks hold promise in a myriad of research disciplines as showcased in \textbf{Figure \ref{Figure_1}}. 

\begin{figure*}[!t] 
 \centering
 \includegraphics[width=\columnwidth]{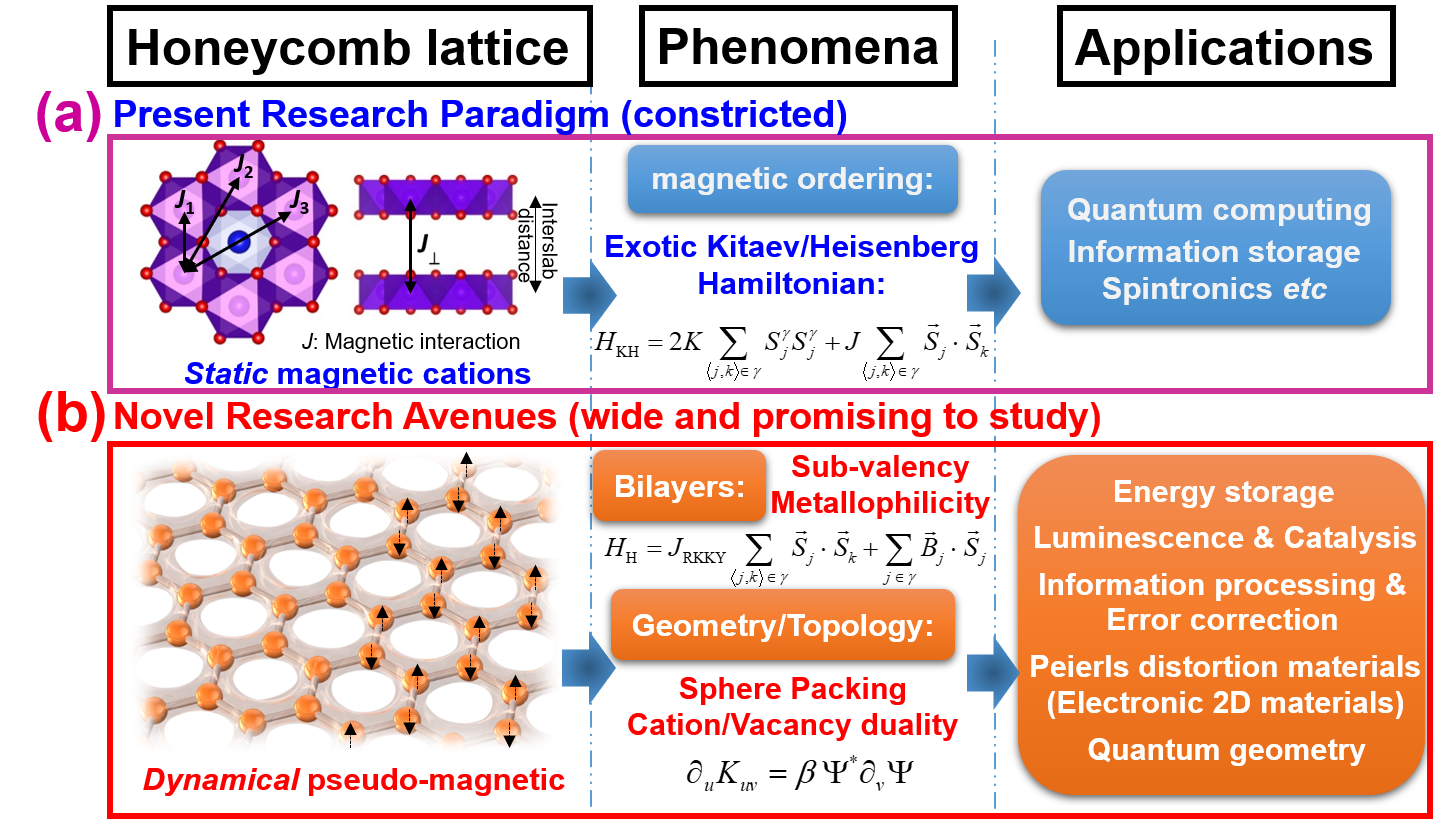}
 \caption{Non-exhaustive illustrations of the diverse applications of honeycomb layered materials, encompassing: (a) Ongoing research endeavours focused on layered materials featuring static magnetic cations arranged within a honeycomb lattice; and (b) Emerging frontiers that explore layered materials exhibiting a honeycomb configuration of mobile cations.\cite {masese2023honeycomb, kanyolo2021honeycomb, kanyolo2022advances, kanyolo2020idealised, kanyolo2022cationic, masese2021topological, kanyolo2021partition, tada2022implications, kitaev2006anyons, kanyolo2021reproducing, kanyolo2022local, kanyolo2022advances2, kanyolo2023pseudo}}
 \label{Figure_1}
\end{figure*}

In particular, the present fundamental research paradigm focuses on the phenomena found within the slabs of static \textit{albeit} potentially magnetic cations. These include exotic phenomena such as magnetic ordering described by Kitaev and Heisenberg interaction terms,\cite {kitaev2006anyons} or even competing anti-symmetric exchange interaction terms such as the Dzyaloshinskii-Moriya interaction coupled to magnetic degrees of freedom.\cite{dzyaloshinsky1958} Such interaction terms are expected to display a plenitude of magnetic phenomena such as magnetoresistance, topological models such as Skyrmions and Kitaev-Heinsenberg physics such as (anti-)ferromagnetism and quantum spin liquid, which have applications in spintronics devices used for information storage and quantum computing.\cite{fert2017, fert2013, Jackeli2010} The plethora of success in studying the condensed matter physics of the static cations within the slab has relegated the dynamical aspects of the cations sandwiched between them to the realm of applied physics and materials science, whereby applications are considered primarily in energy storage as cathode or solid-state electrolytes.\cite{kanyolo2021honeycomb} 
Despite this, the fact that the interslab cations are dynamical implies a wider and more promising applications space. Specifically, under sufficient activation energies, the interslab cations become dynamical introducing complex crystalline structures such as cationic vacancies, stress, strain and other topological features.\cite{kanyolo2021honeycomb, kanyolo2022advances, kanyolo2020idealised, kanyolo2022cationic, masese2021topological, kanyolo2021partition, kanyolo2022advances2} Thus, quantum degrees of freedom on such undulating geometries mimics quantum electrodynamics on a dynamical background. Moreover, it has been suggested that the vacancy creation (de-intercalation) and annihilation (intercalation) processes in the spacing between the slabs is characterised by 2D Liouville conformal field/emergent gravity theory that acts as the dynamical background.\cite {kanyolo2022cationic, kanyolo2020idealised, kanyolo2021partition, kanyolo2022advances2} This theory can be shown to be dual to the congruent sphere packing problem.

Moreover, the symmetries of the honeycomb unit cell can be mapped to an important symmetry group in number theory known as the modular symmetry group.\cite{kanyolo2022cationic} A material exhibiting modular symmetry remains undeformed and unperturbed under discrete rotations and rescaling, maintaining its geometric and structural integrity under successive intercalation and de-intercalation processes.\cite{kanyolo2022cationic, kanyolo2021partition, kanyolo2022advances2} Thus, honeycomb or hexagonal lattices offer the best structurally stable alkali or alkaline-earth metal-based cathode materials for energy storage applications. On the other hand, pseudo-spin models have been employed in the physics of graphene to describe the effects of contortion and strain on the honeycomb lattice.\cite{allen2010honeycomb, mecklenburg2011spin, georgi2017tuning, kvashnin2014phase} Thus, realising pseudo-spin models in honeycomb layered materials can break modular symmetry hence introducing new physics and chemistry on the honeycomb lattice. Such novelties include unconventional bonding of pairs of metal cations within each unit cell in honeycomb layered materials, analogous to the dimerisation of carbon atoms in polyacetylene which leads to a conductor-semiconductor-insulator phase transition via Peierls distortion.\cite{peierls1955quantum, peierls1979surprises} Thus, honeycomb layered materials with pseudo-spin honeycomb lattices of cations as the only structurally unstable counterexamples, as is indicative of experimental literature of unconventional bonding known as metallophilicity found, \textit {inter alia}, in silver-based layered materials.\cite{masese2023honeycomb, schmidbaur2015argentophilic, schreyer2002synthesis, kanyolo2022advances, kanyolo2022advances2}

\begin{figure*}[!b]
 \centering
 \includegraphics[width=0.9\columnwidth]{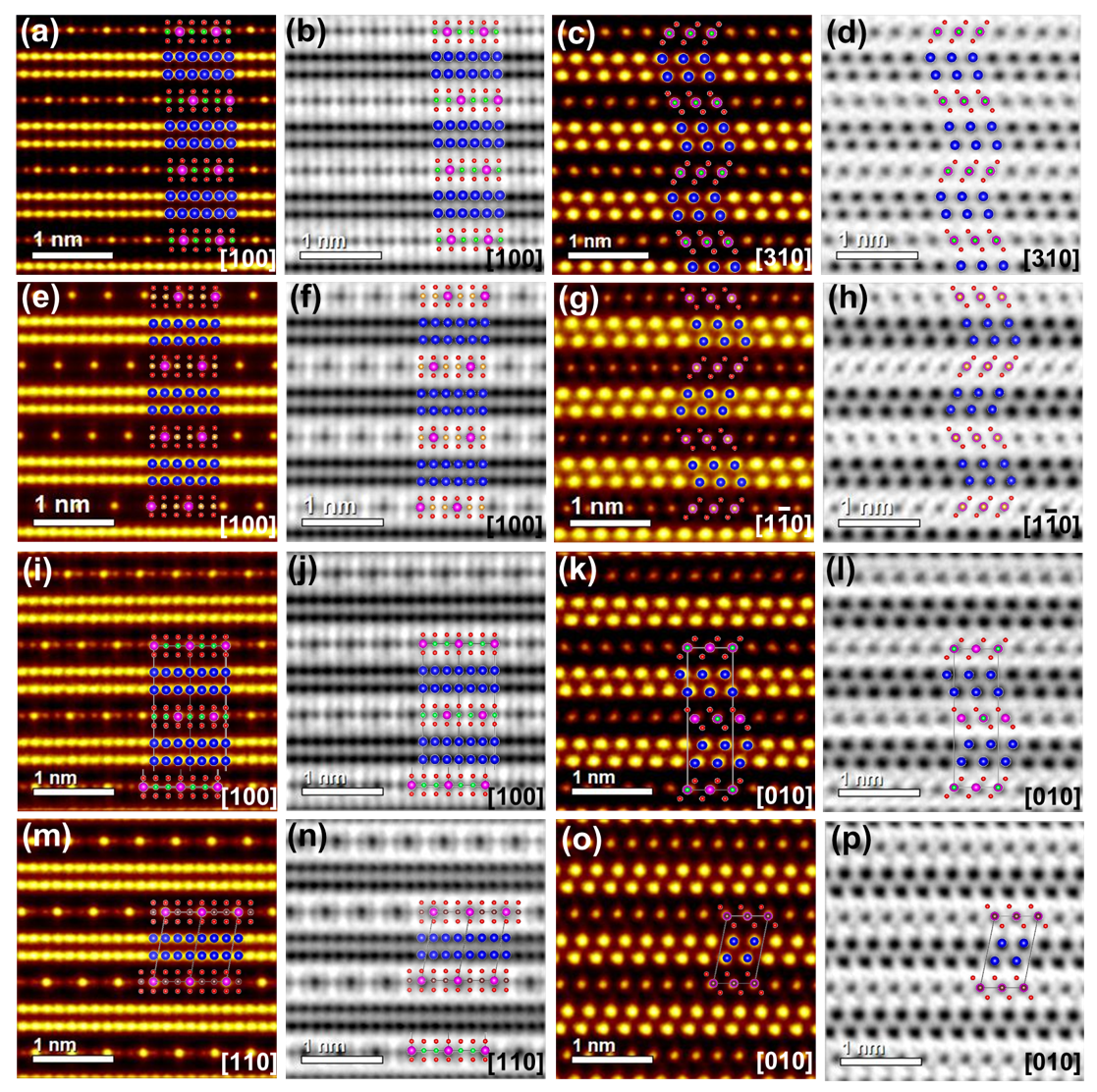}
 \caption{ High-resolution transmission electron microscopy (HRTEM) images of honeycomb layered ${\rm Ag_6}M_2\rm TeO_6$, ($M = \rm Ni, Co, Mg,Cu$ \textit{etc}), comprising metallophilic Ag atom bilayers. (a) High-angle annular dark-field scanning transmission electron microscopy (HAADF-STEM) and (b) Annular bright-field (ABF)-STEM images of $\rm Ag_6Ni_2TeO_6$ taken along the [100] zone axis showing the arrangement of Ag atom bilayers sandwiched between slabs of Ni and Te atoms coordinated with oxygen. (c) HAADF-STEM and (d) ABF-STEM images of $\rm Ag_6Ni_2TeO_6$ taken along the [310] zone axis. (e) HAADF-STEM and (f) ABF-STEM images of $\rm Ag_6Mg_2TeO_6$ taken along the [100] zone axis. (g) HAADF-STEM and (h) ABF-STEM images of $\rm Ag_6Mg_2TeO_6$ taken along the [1-10] zone axis. (i) HAADF-STEM and (j) ABF-STEM images of $\rm Ag_6Co_2TeO_6$ taken along the [100] zone axis. (k) HAADF-STEM and (l) ABF-STEM images of $\rm Ag_6Co_2TeO_6$ taken along the [010] zone axis. (m) HAADF-STEM and (n) ABF-STEM images of $\rm Ag_6Cu_2TeO_6$ taken along the [110] zone axis. (o) HAADF-STEM and (p) ABF-STEM images of $\rm Ag_6Cu_2TeO_6$ taken along the [010] zone axis. An atomistic model of the average structures of $\rm Ag_6{\it M}_2TeO_6$ ($\rm {\it M} = Co, Cu, Ni, Zn, Mg$) acquired based on STEM analyses have been embedded on the STEM images as a guide. Ag atoms are shown in blue. Oxygen atoms are shown in red.  Transition metal {\it M} atoms  atoms  are shown in green  whilst Te atoms are shown in pink. Reproduced with permission.\cite {masese2023honeycomb}Copyright 2023 Wiley-VCH.}
 \label{Figure_2}
\end{figure*}

\subsection{Honeycomb Layered Frameworks with Metallophilic Bilayers 
}

Silver-based materials such as the suboxide $\rm Ag_4^{1/2+}O^{2-}$ were reported by W\"{o}hler and later Faraday as early as 1839 to the disbelief of the scientists of the day.\cite{bailey1887xliii} Although valence theory had to wait a century or more for the formulation of quantum theory of the atom equipped to tackle valency theory, the anomalous nature of such silver compounds had already began to be appreciated in the 1920s especially due to extensive experimental studies with the subfluoride $\rm Ag_2^{1/2+}F^{1-}$.\cite{terrey1928ccclxxii, kawamura1974, ott1928, freed1940magnetic, williams1989neutron, ido1988, andres1966superconductivity, fujita1974, ikezawa1985, wang1991, tong2011, argay1966redetermination} Thus, almost a century later, the issue had resurfaced by the discovery of the subfluoride, $\rm Ag_2^{1/2+} F^{1-}$ (1928), and subsequently a series of suboxides such as $\rm Ag_6^{2/3+} O_2^{2-}$ (1981), $\rm Ag_2^{1/2+}{\it M}^{3+}O_2^{2-}$ with $M = \rm Fe, Ni, Cr, Mn$, \textit {etc} (2002 $\sim$ 2020) and $\rm Ag_{16}^{1/2+}B_4^{3+}O_{10}^{2-}$ (2020), finally culminating in the discovery of $\rm Ag_6^{1/2+}{\it M}_2^{2+}Te^{5+}O_6^{2-}$ with $M = \rm Co, Ni, Cu, Mg, Zn$, \textit{etc} recently reported (2023).\cite{yoshida2020static, schreyer2002synthesis, matsuda2012partially, ji2010orbital, kovalevskiy2020uncommon, lobato2021comment, yin2021reply, masese2023honeycomb}}
\textbf{Figure \ref{Figure_2}} shows the recently reported crystal structure of honeycomb layered oxide ${\rm Ag_6}M_2\rm TeO_6$, comprising metallophilic Ag atom bilayers coordinated with oxygen atoms of the adjacent metal slabs.\cite{masese2023honeycomb} Unlike Sill\'en-Aurivillius layered structures that manifest cation bilayers with anions (such as O, Cl, {\it etc}.) separating the bilayers,\cite{ozaki2021, tian2022} this class of honeycomb layered frameworks have metallophilic bilayers devoid of any anions separating the metallophilic cations. The transition metal slabs sandwiching the metallophilic bilayers are composed of transition metal atoms arranged in a hexagonal (triangular) or honeycomb lattice.\cite {kanyolo2021honeycomb} In general, this class of honeycomb layered frameworks embody a bilayer arrangement of metallophilic cations such as Ag, interposed between layers of mainly transition metals (suchlike Zn, Cu, Ni, Co, Fe, Mn, {\it etc}.) aligned in a honeycomb or hexagonal lattice.\cite{yoshida2008unique, yoshida2006spin, schreyer2002synthesis, sorgel2007ag3ni2o4, ji2010orbital, yoshida2020partially, yoshida2020static, taniguchi2020butterfly, yoshida2011novel, masese2023honeycomb} 

It was recently conjectured that, within this class of silver-based materials with metallophilic bilayers, the 2D emergent gravity appears to be replaced by the familiar 3D Newtonian gravity via a 2D-to-3D crossover due to Peierls distortion.\cite{masese2023honeycomb, kanyolo2022advances, kanyolo2022advances2} For honeycomb layered frameworks, this crossover manifests as a stable 3D bilayered structure between the slabs, owing to a pseudo-spin model on the 2D honeycomb lattice. The Heisenberg interaction term corresponds to the conduction-electron-mediated Ruderman-Kittel-Kasuya-Yosida (RKKY) exchange interactions due to hybridisation of $5s^2$ and $4d_{z2}^1$ silver orbitals whereas the pseudo-magnetic field takes the form of stress, strain or vacancy number density in the 2D honeycomb lattice.\cite{masese2023honeycomb, kanyolo2022advances, kanyolo2022advances2} The finite pseudo-magnetic field can then be related to the peculiar metallophilic interactions between two silver cations known as argentophilicity,\cite {schmidbaur2015argentophilic} responsible for the stability of the metallophilic bilayers, provided novel electromagnetic symmetries such as the special unitary group of degree $2$ ($\rm SU(2)$) is emergent.\cite{masese2023honeycomb, kanyolo2022advances, kanyolo2022advances2} These electromagnetic symmetry, alongside the conventional $\rm U(1)$ symmetry of electromagnetism describe the metallophilic interactions as a mass term between degenerate valence silver states. Consequently, the interesting emergent physics and chemistry of silver in metallophilic bilayers confers new and exciting research avenues, suchlike luminescence, catalysis, information processing, error correction, Peierls distortion materials and quantum geometry analogues, which fall beyond energy storage applications.\cite{kanyolo2021honeycomb, kanyolo2022advances, kanyolo2020idealised, kanyolo2022cationic, masese2021topological, kanyolo2021partition, kanyolo2022advances2}

\subsection{Objective and Scope of the Review}

Currently, honeycomb layered frameworks have niche applications not only in energy storage, but also stand as exemplary pedagogical models showcasing the countless capabilities of nanomaterials across a broad spectrum of disciplines, {\it inter alia}, condensed matter physics, solid-state electrochemistry and materials science.\cite{masese2023honeycomb, kanyolo2021honeycomb, kanyolo2021honeycomb, kanyolo2022advances, kanyolo2020idealised, kanyolo2022cationic, masese2021topological, kanyolo2021partition, kanyolo2022advances2} Besides their structural advantages, honeycomb frameworks have opened new paradigms of theories and computational techniques quintessential in the field of condensed matter physics and quantum material science, that are bound to catapult the discovery of novel forms of materials with highly controlled and unique catalytic, magnetic, topological or optical properties. For instance, honeycomb layered frameworks manifesting metallophilic bilayers have recently been understood to exhibit fascinating critical phenomena and phase transitions that can alter the crystalline symmetries leading to the possibility of engineering desirable conduction-insulator properties by varying thermodynamic or structural properties such as temperature or strain respectively. Therefore, such honeycomb layered frameworks harbour a great potential to not only address issues in energy storage pertinent to our environment, but also breed new telecommunications technologies. Unfortunately, the dearth of a concise and critical {\it Review} delving into honeycomb layered frameworks with metallophilic bilayers not only highlighting the quest towards the exploration of their material functionalities but also introducing their more fundamental atomistic mechanisms appears to stifle their adaptation across the relevant research fields and subsequent applications, and hence serves as the impetus for the timeliness, relevance and scope of our proposed Review topic.

This {\it Review} therefore aims to comprehensively discuss the present state of advancements made in the preparation, theory, characterisations and relevant applications of honeycomb layered frameworks exhibiting metallophilic bilayers through proper understanding of their fundamental chemistry, physicochemical and topological properties. We shall make distinctions between metallophilic materials (such as $\rm Ag_{16}B_4O_{10}$ \cite {kovalevskiy2020uncommon} and $\rm [{\it X}-{\it M}-PH_3]_3$ ($X$ = $\rm I, Br, Cl$; $M$ = $\rm Au, Ag, Cu$) \cite {assadollahzadeh2008comparison} which only exhibit metallophilicity without a clear-cut bilayered structure) and bilayered materials (such as the Sill\'en-Aurivillius $\rm BiOIO_3$ and $\rm BaBi_5Ti_3O_{14}Cl$ \cite{ozaki2021, tian2022}) with conventional valence bonds constituting the bilayers that do not exhibit metallophilicity. Thus, this \textit{Review} solely concerns the present state of experimental, computational and theoretical research of the intersection of the two paradigms, which we refer to as \textit{metallophilic bilayered materials} in the title.
Particularly, the nature of silver anomalous valency states $1/2+, 2/3+$ \textit{etc} (herein referred to as sub-valency/sub-valent states) in metallophilic bilayers has remained a matter of great mystery to the scientific community since their discovery. Unconventional silver-silver interactions (metallophilicity) with characteristically shorter bonds than elemental silver, localised paired valence electrons responsible for semi-conductivity and low temperature superconductivity amongst other peculiarities indicate the physics of silver in metallophilic bilayered materials is poorly understood. Whilst the present state of experimental research mainly concerns Ag-based metallophilic bilayered materials, we seek to review the entire body of work concerning \textit{metallophilic bilayered} materials, with the review of non-Ag based materials not yet found experimentally augmented by including the present state of computations and theory towards their characterisation and experimental realisation. Consequently, we critique the challenges and knowledge gaps that have been buried in the ever-expanding literature space and accentuate new scopes for future research development of honeycomb layered frameworks manifesting such bilayered structures.
Due to the intriguing concepts to be elucidated within this {\it Review}, we anticipate that this work will not only cater to a broader community of experimentalists and theoreticians engaged in various domains of Condensed Matter research, including solid-state Ionics, solid-state physics, solid-state chemistry, and materials science, but also to those exploring the realm of Electromagnetism, encompassing photonics and electromagnetic dynamics. Additionally, the insights shared in this work will prove highly relevant to researchers at all career stages, from established experts to early career investigators, as it serves as a comprehensive survey of the extensive body of research findings and applications accumulated over the years. Moreover, this {\it Review} boldly charts the frontier of knowledge, pointing towards promising avenues of prospective research within the aforementioned fields.

To ensure readers remain well-informed about the remarkable advancements in the materials development of layered frameworks that exhibit metallophilic bilayers, we present a comprehensive overview of the significant milestones attained in the creation of novel material compositions, accompanied by a thorough exposition of their synthesis techniques in \textbf{Section \ref{Section: Synthesis_Structure_crystal}}. Furthermore, we provide a comprehensive outline of the characterisation techniques employed to elucidate the diverse functionalities inherent in this class of materials (\textbf{Sections \ref{Section: Properties_Functionalities} and \ref{Section: Characterisation_Techniques}}). In \textbf{Section \ref{Section: Theoretical_Characterisations}}, we delve into the exploration of emerging phenomena exclusive to layered frameworks that showcase metallophilic bilayers, specifically emphasising their invaluable role as experimental platforms for investigating ideas within the realms of topological field theories and 2D emergent quantum gravity. Finally, in \textbf{Section \ref{Section: Prospects}}, we conclude by directing attention towards potential avenues for future research endeavours of honeycomb layered frameworks with metallophilic bilayers, encompassing both experimental and theoretical pursuits, with the ultimate aim of realising new frontiers in the vast expanse of material compositional space.

\newpage

\section{\label{Section: Synthesis_Structure_crystal} Syntheses Techniques, Structural Features and Crystal Chemistry}

\subsection{Milestones in Development of Layered Frameworks with Metallophilic Bilayers}
\begin{figure*}[!b]
 \centering
 \includegraphics[width=0.95\columnwidth]{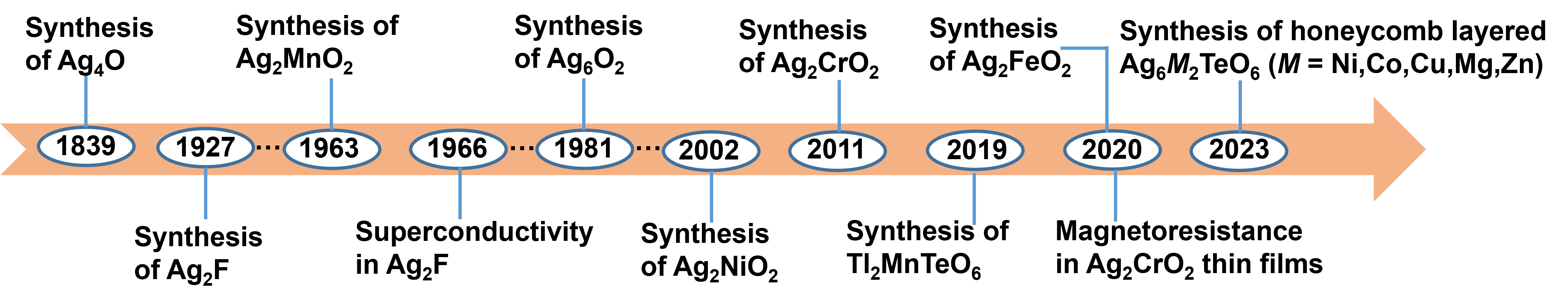}
 \caption{Timeline for the development of representative layered materials manifesting bilayer arrangement of cations.\cite{rienacker1963, ott1928, yoshida2020partially, taniguchi2020butterfly, bailey1887xliii} Note that although the existence of the suboxide of silver with a chemical composition of $\rm Ag_4O$ remains disputed,\cite{bailey1887xliii} it has been included in the timeline for historical purposes.}
 \label{Figure_3}
\end{figure*}

\begin{figure*}[!t]
 \centering
 \includegraphics[width=0.95\columnwidth]{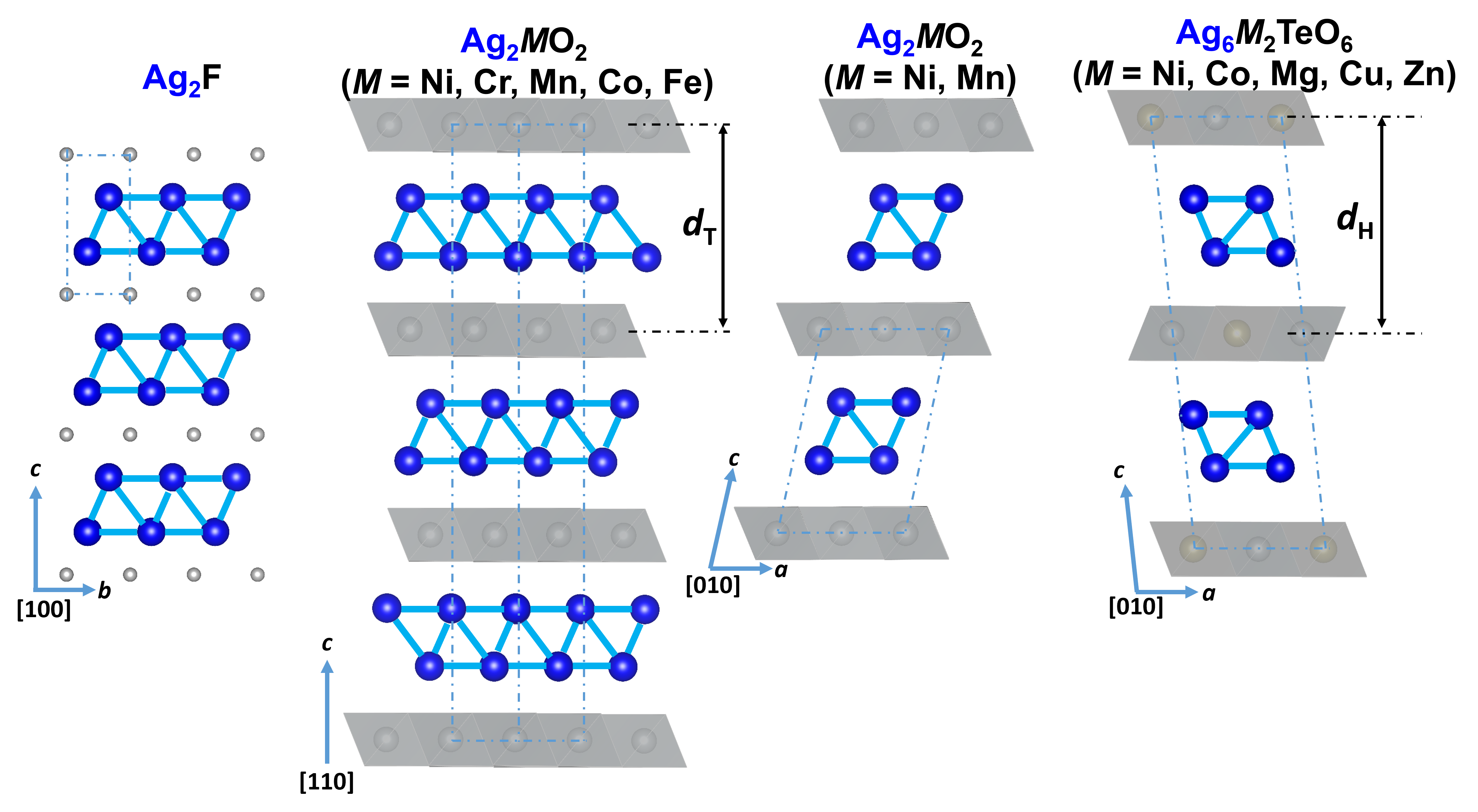}
 \caption{Crystal structure of layered materials exhibiting a bilayer arrangement of $\rm Ag$ cations. The crystallographic unit cell is indicated by the dashed line. The interslab distance for $\rm Ag_2{\it M}O_2$ with a triangular (hexagonal) lattice of transition metal $M$ atoms is denoted as $d_{\rm T}$ and follows the sequence: $\rm Ag_2NiO_2$ (8.03 \AA) < $\rm Ag_2FeO_2$ (8.34 \AA) < $\rm Ag_2CoO_2$ (8.60 \AA) < $\rm Ag_2MnO_2$ (8.64 \AA) < $\rm Ag_2FeO_2$ (8.66 \AA).\cite {yoshida2020partially, taniguchi2020butterfly, yoshida2020static, schreyer2002synthesis} As for $\rm Ag_6{\it M}_2TeO_6$ with a honeycomb lattice of transition metal $M$ atoms, the interslab distance (denoted as $d_H$) is in the sequence: $\rm Ag_6Ni_2TeO_6$ (8.58 \AA) < $\rm Ag_6Mg_2TeO_6$ (8.7 \AA) < $\rm Ag_6Cu_2TeO_6$ (8.73 \AA) < $\rm Ag_6Co_2TeO_6$ (8.77 \AA). \cite {masese2023honeycomb} }
 \label{Figure_4}
\end{figure*}

A comprehensive chronology of the achieved milestones in the ongoing development of layered frameworks with metallophilic bilayers is presented in \textbf {Figure \ref{Figure_3}}. Therein, the first experimental report dates back to 1839 with a report on a suboxide of silver with a composition of $\rm Ag_4^{1/2+}O$ \textit{albeit} disputed in 1887.\cite{bailey1887xliii} 
Subsequently, the discovery of the subfluoride $\rm Ag_2^{1/2+}F$ in 1927\cite {ott1928} designed via the electrochemical reduction of an aqueous $\rm AgF$ solution hallmarked the unequivocal existence of silver subvalency. The structure of subvalent $\rm Ag_2F$ exhibits a layered arrangement, as depicted in \textbf {Figure \ref{Figure_4}}. It can be visualised as comprising hexagonally packed layers of $\rm F$ and $\rm Ag$ atoms that are vertically stacked/aligned in the sequence of $\rm -F-Ag-Ag-F-Ag-Ag-F-$ along the $c$-axis. Within the $\rm Ag-Ag$ bilayer, the interatomic distance between the silver atoms is approximately 2.84 \AA, which is shorter than the closest interatomic distance observed in bulk $\rm Ag$ metal. The Hall effect measurements conducted on $\rm Ag_2F$ indicate that the material harbours one conduction electron per molecule of $\rm Ag_2F$, aligning with the stoichiometric conjecture: $\rm Ag^{0}$ + $\rm Ag^{1+}F^{1-}$. 
However, the precise equivalence of the silver ions/atoms within the $\rm Ag_2F$ crystal necessitates equitable consideration for the alternative formulation: $\rm AgF^{1-}$ + $\rm Ag^{1+}$. This ambiguity in assigning definitive valency states to the silver ions is conjectured to contribute to the electrical conductivity observed in $\rm Ag_2F$.\cite{andres1966superconductivity} Moreover, experimental investigations have established that $\rm Ag_2F$ demonstrates not only metallic conductivity but also superconductivity with a transition temperature of 66 mK.\cite {andres1966superconductivity} This is rather peculiar, since excellent conductors tend to be poor Bardeen-Cooper-Schrieffer (BCS) superconductors due to weak electron-phonon coupling.\cite{bardeen1957theory} Moreover, $\rm Ag_2F$ is prone to decomposition into metallic $\rm Ag$ and $\rm AgF$ under exposure to ultraviolet radiation, moisture, or in high temperatures exceeding 70$^{\circ}$ C.\cite {kawamura1974} Consequently, such intriguing physicochemical properties inherent in the physics and chemistry of $\rm Ag_2F$ have stimulated extensive research endeavours.\cite {freed1940magnetic, williams1989neutron, ido1988, fujita1974, ikezawa1985, wang1991, argay1966redetermination} 

In 1963,\cite {rienacker1963} the scientific community witnessed the initial discovery of $\rm Ag_2MnO_2$, wherein Ag atoms are arranged in bilayers and positioned between magnetically active $\rm Mn$ transition metal layers, forming a triangular (hexagonal) lattice. $\rm Ag_2MnO_2$ was synthesised through a solid-state reaction at high temperatures, involving the reaction\cite {rienacker1963}:  
\begin{align}
    2\,{\rm Ag} + 1/3\,{\rm Mn_3O_4} + 1/3\,{\rm O_2} \rightarrow {\rm Ag_2MnO_2}.
\end{align}
Subsequently, in 1981,\cite {beesk1981x} a study unveiled $\rm Ag_6O_2$ 
(equivalently, $\rm Ag_3O$), which exhibited a similar Ag bilayer structure, as confirmed by diffraction analyses. $\rm Ag_3O$ was initially proposed to adopt a composition of $\rm Ag_2O$. However, subsequent investigations based on diffraction and elemental analyses proposed instead a subvalent composition of $\rm Ag_3^{2/3+}O$. The compositional space was further expanded in 2002 by the synthesis of $\rm Ag_2NiO_2$, reported to also possess $\rm Ag$ bilayers, with the sub-valency of $\rm Ag$ in $\rm Ag_2^{1/2+}Ni^{3+}O_2^{2-}$ confirmed via spectroscopic techniques.\cite{schreyer2002synthesis} $\rm Ag_2NiO_2$ was synthesised via a high-temperature solid-state reaction at high oxygen pressures, involving the reaction\cite {schreyer2002synthesis}:  
\begin{align}
    \,{\rm Ag_2O} + \,{\rm NiO} \rightarrow {\rm Ag_2NiO_2}.
\end{align}

Unlike $\rm Ag_2F$, $\rm Ag_2NiO_2$ is insensitive to air and moisture.  
Subsequently, numerous stable layered materials of the form $\rm Ag_2{\it M}O_2$ (where $M$ represents elements such as Co, Cr, Fe, \textit {etc}.) with $\rm Ag$ bilayer arrangements have been documented,\cite {yoshida2020partially, sugiyama2009static, yoshida2020static} with much of the research focusing on their magnetic and electronic properties.\cite {yoshida2011novel} Notably, magnetoresistance has been observed in thin films of $\rm Ag_2CrO_2$.\cite {taniguchi2020butterfly} In 2019, the concept of topochemical ion exchange routes was utilised to design layered $\rm Tl_2MnTeO_6$, which hallmarked the discovery of the first non-silver-based material exhibiting metallophilic bilayers (of thallium atoms).\cite {nalbandyan2019Tl2MnTeO6} Finally, in 2023, a similar topochemical ion exchange approach was employed to develop a swath of honeycomb layered frameworks with a global composition of $\rm Ag_2{\it M}_2TeO_6$ ($M$ = Co, Cu, Mg, Zn, Ni) that featured domains of $\rm Ag_6^{\nu+}{\it M}_2TeO_6$ with metallophilic bilayers of silver atoms as ascertained by scanning transmission electron microscopy energy-dispersive X-ray spectroscopy (STEM-EDX), \textit{albeit} an unascertained Ag subvalency experimentally constrained within the range, $+1/3 \leq \nu \leq 2/3+$.\cite {masese2023honeycomb} 

\subsection{Synthesis Techniques for (Honeycomb) Layered Frameworks with Metallophilic Bilayers}

Solid-state calcination methods offer a straightforward and highly scalable approach. Consequently, these techniques have found extensive application in synthesising diverse layered frameworks that display metallophilic bilayers. In high-temperature solid-state reactions, the process typically involves the meticulous blending of powdered starting reagents, subsequent compaction into pellets, and ultimately subjecting them to controlled furnace heating under specific atmospheric conditions. The reaction temperature is contingent upon both the desired product and the intrinsic properties of the starting reagents. During the initial stages of the reaction, the diffusion of the starting reagents ensues gradually, relying on the ionic mobility amongst them to facilitate the formation of the desired product.

When embarking on the synthesis of desired layered compounds, careful consideration is given to the structural characteristics, volatility, and reactivity of the starting materials. For example, starting reagents like $\rm MnO_2$, $\rm NiO$, $\rm Fe_2O_3$, $\rm Cr_2O_3$, $\rm Co_3O_4$ and others employed in the preparation of layered oxides such as $\rm Ag_2{\it M}O_2$ ($M$ = $\rm Mn, Ni, Fe, Cr, Co,$ \textit {etc}.)\cite {schreyer2002synthesis, yoshida2020partially, sugiyama2009static, yoshida2020static, taniguchi2020butterfly, rienacker1963} exhibit high inertness and necessitate elevated temperatures and pressures to initiate the desired reactions. Conversely, reagents such as $\rm Ag_2O$ are volatile at higher temperatures and may undergo loss during the reaction process. Additionally, the reactivity between the materials and the reaction vessel within the furnace must be taken into account, prompting the utilisation of gold or platinum trays as standard reaction containers. The choice of firing atmosphere is another crucial factor to consider, with oxygen being the predominant environment for synthesising most layered materials of this classification. For instance, the synthesis of $\rm Ag_2CoO_2$ can be accomplished through a high-pressure and high-temperature solid-state reaction. In this process, precise proportions of $\rm Ag$, $\rm Ag_2O$, and $\rm Co_3O_4$ reagents are combined and encapsulated within a Pt cell. Subsequently, the reaction assembly is subjected to a temperature of 900$^{\circ}$ C for a duration of 1 hour, employing a pressure of 6 GPa. This specific reaction condition facilitates the formation of $\rm Ag_2CoO_2$.\cite {yoshida2020static}

\begin{table*}
\caption{Synthetic conditions of layered frameworks manifesting mobile cation bilayers. \cite{masese2023honeycomb, schreyer2002synthesis, susloparova2022, nalbandyan2019Tl2MnTeO6, yoshida2020static, taniguchi2020butterfly, yoshida2011novel, matsuda2012partially, yoshida2008unique, yoshida2006spin, sorgel2007ag3ni2o4, tong2011, argay1966redetermination, wang1991, ikezawa1985, fujita1974, andres1966superconductivity, ido1988, freed1940magnetic, yoshida2020partially, sugiyama2009static, ji2010orbital, liu2007, kawamura1974, Yoshida2007impurity, Yoshida2007impurity}}\label{Table_1}
\begin{center}
\scalebox{0.55}{
\begin{tabular}{llll} 
\hline
\textbf{Compound} & \textbf{Synthesis technique} & \textbf{Precursors} & \textbf{Firing condition }\\ 
 & & & \textbf{(Temperature, atmosphere)}\\ 
\hline\hline
$\rm Ag_2F$\cite{wang1991, ikezawa1985, fujita1974, andres1966superconductivity, ido1988, williams1989neutron, freed1940magnetic, kawamura1974} (pleochroic powders) & Electrolysis & aqueous $\rm AgF$ solution & 50-80$^\circ$C\\
$\rm Ag_2F$\cite{tong2011} & Mechanochemical route & $\rm Ag$, $\rm AgF$ & high-energy milling (room temperature); 45 min in He\\
$\rm Ag_2NiO_2$\cite{yoshida2006spin, schreyer2002synthesis} (lustrous black powder) & Solid-state reaction & $\rm Ag_2O$, $\rm NiO$ & 550$^\circ$C; in $\rm O_2$ (65-70 MPa)\\
$\rm Ag_2NiO_2$\cite{yoshida2006spin, schreyer2002synthesis} & Solid-state reaction & $\rm Ag_2O$, $\rm Ni(OH)_2$ & 550$^\circ$C; 24h in $\rm O_2$ (65 MPa)\\
$\rm Ag_2NiO_2$\cite{liu2007} & Electrochemical synthesis & $\rm AgNiO_2$ & 270-330$^\circ$C\\
$\rm Ag_3Ni_2O_4$\cite{sorgel2007ag3ni2o4} & Solid-state reaction & $\rm Ag_2O$, $\rm AgNiO_2$ & 550$^\circ$C; 70h in $\rm O_2$ (130 MPa)\\
$\rm Ag_2MnO_2$\cite{yoshida2008unique, sugiyama2009static, ji2010orbital}(brown powder with metallic lustre) & Solid-state reaction & $\rm Ag$, $\rm MnO_2$ & 650-750$^\circ$C; 24h in $\rm O_2$\\
$\rm Ag_2MnO_2$\cite{rienacker1963} & Solid-state reaction & $\rm Ag$, $\rm Mn_3O_4$ & 600$^\circ$C; 23h in $\rm O_2$\\
$\rm Ag_2CoO_2$\cite{yoshida2020static} & Solid-state reaction & $\rm Ag$, $\rm Ag_2O$, $\rm Co_3O_4$ & 900$^\circ$C; 1h (6 GPa)\\
$\rm Ag_2FeO_2$\cite{yoshida2020partially} & Solid-state reaction & $\rm Ag$, $\rm Ag_2O$, $\rm Fe_2O_3$ & 900$^\circ$C; 1h (6 GPa)\\
$\rm Ag_2Ni_{1-{\it x}}Fe_{\it x}O_2$ ($0\leq x \leq0.3$)\cite{Yoshida2007impurity} & Solid-state reaction & $\rm Ag_2O$, $\rm Fe_2O_3$, $\rm NiO$ & 550$^\circ$C; 24h in $\rm O_2$ pressure (65 MPa)\\
$\rm Ag_2CrO_2$\cite{taniguchi2020butterfly, yoshida2011novel, matsuda2012partially}(dark brown powder with metallic lustre) & Solid-state reaction & $\rm Ag$, $\rm Ag_2O$, $\rm Cr_2O_3$ & 1200$^\circ$C; 1h (6 GPa)\\
$\rm Ag_6{\it M}_2TeO_6$ ($\rm {\it M} = Co, Cu, Mg, Ni$ and $\rm Zn$)\cite{masese2023honeycomb} & Topochemical ion-exchange & $\rm AgNO_3$, $\rm Na_2{\it M}_2TeO_6$ ($\rm {\it M} = Co, Cu, Mg, Ni$ and $\rm Zn$) & 250$^\circ$C; 5 days in air\\
$\rm Tl_2MnTeO_6$\cite{susloparova2022, nalbandyan2019Tl2MnTeO6} & Topochemical ion-exchange & $\rm TlNO_3$, $\rm Na_2MnTeO_6$ & 300$^\circ$C; 2h in air\\
\hline
\end{tabular}}
\end{center}
\end{table*}

\begin{figure*}[!t]
 \centering
 \includegraphics[width=\columnwidth]{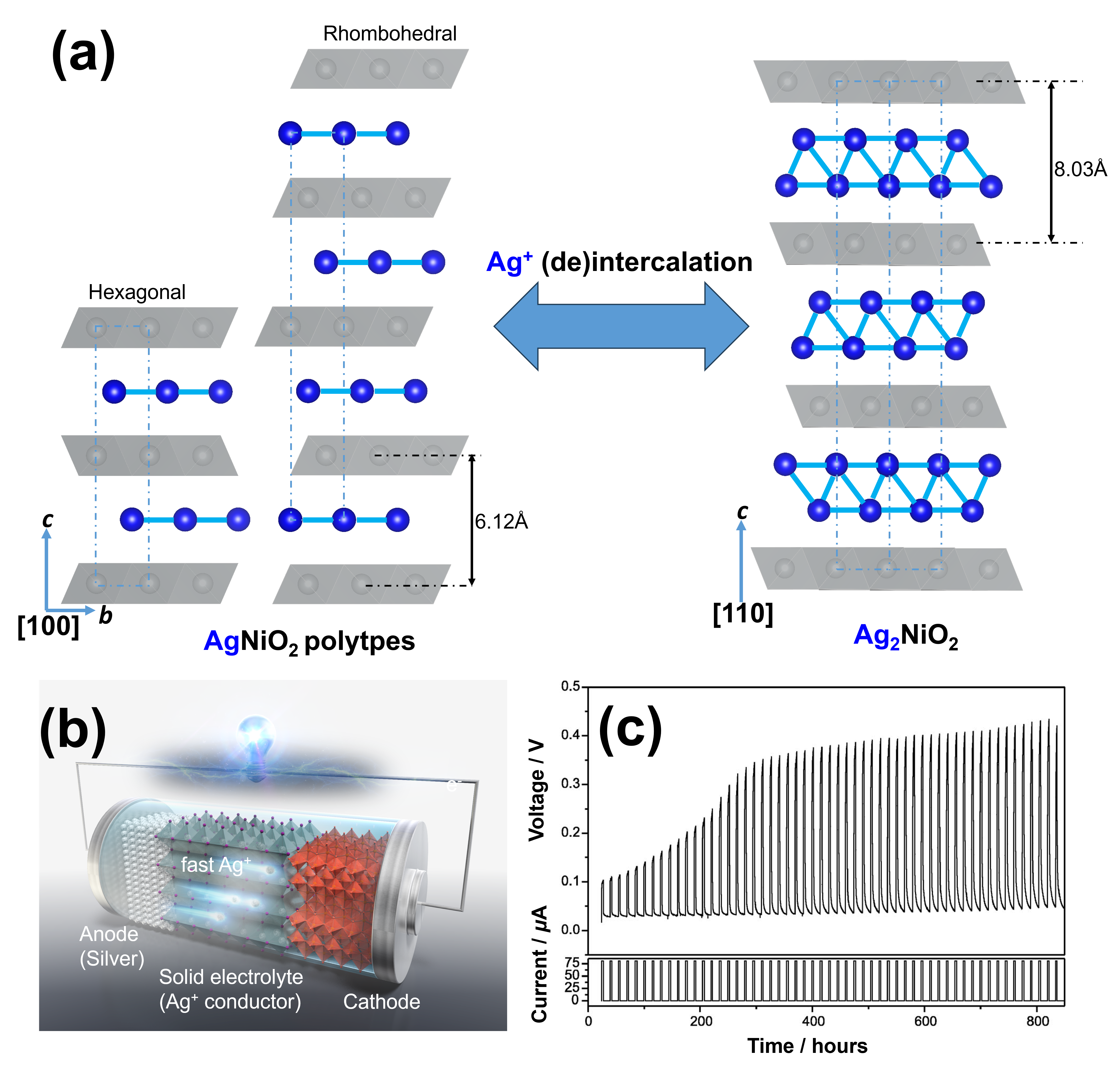}
 \caption{ (a) Crystal structure of $\rm AgNiO_2$ polytypes and $\rm Ag_2NiO_2$ formed via topochemical Ag-ion intercalation and de-intercalation in a solid-state $\rm Ag$-ion battery setup.\cite {liu2007} $\rm Ag$ atoms are highlighted in blue. The crystallographic unit cell is indicated by the dashed line. (b) A schematic illustration of the all-solid-state $\rm Ag$-ion battery set-up (comprising $\rm Ag$ metal as anode (negative electrode), $\rm AgI$ as solid-electrolyte and $\rm Ag_2NiO_2$ as the cathode (positive electrode) utilised to perform topochemical $\rm Ag$-ion (de)intercalation. (c) Corresponding voltage response curves of the all-solid-state $\rm Ag$-ion battery setup.}
 \label{Figure_5}
\end{figure*}

The design of materials (such as $\rm Ag_2{\it M}O_2$ (where $M$ denotes $\rm Fe, Ni, Mn, Co, Cr, {\it etc.}$)) featuring Ag atom bilayers primarily entails the implementation of rigorous synthetic conditions, including the application of giga-Pascal scale pressures and the solid-state reaction of precursors under elevated oxygen pressures and temperatures, as provided in \textbf {Table \ref{Table_1}}.\cite{masese2023honeycomb, schreyer2002synthesis, susloparova2022, nalbandyan2019Tl2MnTeO6, yoshida2020static, taniguchi2020butterfly, yoshida2011novel, matsuda2012partially, yoshida2008unique, yoshida2006spin, sorgel2007ag3ni2o4, tong2011, argay1966redetermination, wang1991, ikezawa1985, fujita1974, andres1966superconductivity, ido1988, freed1940magnetic, yoshida2020partially, sugiyama2009static, ji2010orbital, liu2007} However, it is noteworthy that layered frameworks characterised by metallophilic bilayers, such as those containing Ag and Tl, can still be obtained through alternative means. Specifically, low-temperature metathetic (topochemical ion-exchange) synthetic routes utilising molten salts offer a viable approach to access these intriguing layered frameworks. For instance, $\rm Tl_2MnTeO_6$ exhibiting Tl atom bilayers can be prepared via topochemical ion-exchange of $\rm Na_2MnTeO_6$ using $\rm TlNO_3$ as the molten salt at 300$^{\circ}$ C via the following reaction\cite {nalbandyan2019Tl2MnTeO6}:
\begin{align}
    \,{\rm Na_2MnTeO_6} + 2\,{\rm TlNO_3}  \rightarrow {\rm Tl_2MnTeO_6} + 2\,{\rm NaNO_3},
\end{align}
In addition, electrochemical Ag-ion intercalation of layered delafossite $\rm AgNiO_2$  (as schematically shown in \textbf {Figure \ref{Figure_5}}) has also been shown to yield bilayered $\rm Ag_2NiO_2$ based on the following reaction\cite {liu2007}:
\begin{align}\label{AgNiO2_Ag_eq}
    \,{\rm Ag} + \,{\rm AgNiO_2}  \rightarrow {\rm Ag_2NiO_2},
\end{align}
Moreover, mechanochemical reaction of $\rm AgF$ and $\rm Ag$ via high-energy ball milling at room temperature under He environment has successfully been shown to yield $\rm Ag_2F$ based on the following reaction\cite {tong2011}:

\begin{align}\label{AgF_Ag_eq}
    \,{\rm Ag} + \,{\rm AgF}  \rightarrow {\rm Ag_2F},
\end{align}
Electrolysis has also been utilised as a route to synthesise $\rm Ag_2F$ via the electrochemical reduction of aqueous $\rm AgF$ solution.\cite {wang1991, ikezawa1985, fujita1974, andres1966superconductivity, ido1988, williams1989neutron, freed1940magnetic, kawamura1974} Although $\rm Ag_2F$ is stable in aromatic hydrocarbons such as benzene, it reacts with water and undergoes partial decomposition, resulting in the formation of $\rm Ag$ and $\rm AgF$ in the presence of moisture. Consequently, the optimal preservation of $\rm Ag_2F$ requires storage under meticulously controlled atmospheric conditions.\cite {kawamura1974} Furthermore, $\rm Ag_2F$ is photosensitive and thus requires storage in a dark environment.\cite {andres1966superconductivity}

Honeycomb layered oxides encompassing the global composition of ${\rm Ag_2}M_2{\rm TeO_6}$ (where $M = \rm Mg, Cu, Co, Ni$) displaying bilayer ${\rm Ag_6}M_2{\rm TeO_6}$ domains (shown in \textbf {Figure \ref{Figure_2}}) can be synthesised via the topochemical $\rm Na^+ / Ag^+$ or $\rm K^+ / Ag^+$ ion-exchange of ${\rm Na_2}M_2{\rm TeO_6}$ or ${\rm K_2}M_2{\rm TeO_6}$ precursors alongside a molten flux of $\rm AgNO_3$ at 250 $^{\circ}$C in air, based on the following reaction\cite {masese2023honeycomb}:

\begin{align}
    2\,{\rm AgNO_3} + \,{\rm Na_2{\it M}_2TeO_6} ({\it M}= \rm Cu, Co, Mg, Ni, Zn) \rightarrow 2\,{\rm NaNO_3} + {\rm Ag_2{\it M}_2TeO_6}, 
\end{align}
To facilitate a complete ion-exchange reaction, an excess amount of $\rm AgNO_3$ is typically employed. The resulting product undergoes thorough washing with distilled water, in order to dissolve the residual nitrates, namely the byproduct $\rm NaNO_3$ and the remaining $\rm AgNO_3$. The solution is vigorously agitated using a magnetic mixer and subsequently subjected to filtration and drying.

\subsection{Stacking Disorders and Defects}

Layered frameworks comprise structural arrangements in which coinage, alkaline-earth, or alkali metal atoms are situated between parallel slabs. These slabs consist of transition metals or $s$-block metals, interconnected by oxygen atoms. In this arrangement, oxygen atoms from the slabs coordinate with coinage, alkaline-earth, or alkali metal atoms, thereby forming interlayer bonds. It is important to note that the strength of these interlayer bonds is considerably weaker compared to the covalent bonds present within the slabs. The specific nature and strength of interlayer coordination primarily depend on the Shannon-Prewitt radii of the coinage, alkaline-earth, or alkali metal atoms.\cite {shannon1971chemistryb, shannon1971chemistrya} These radii play a crucial role in determining the interlayer distance within the resulting configuration or heterostructures, characterised by various layer stacking sequences and possible disorders.\cite {kanyolo2021honeycomb, kanyolo2022advances}

\begin{figure*}[!t]
 \centering
 \includegraphics[width=\columnwidth]{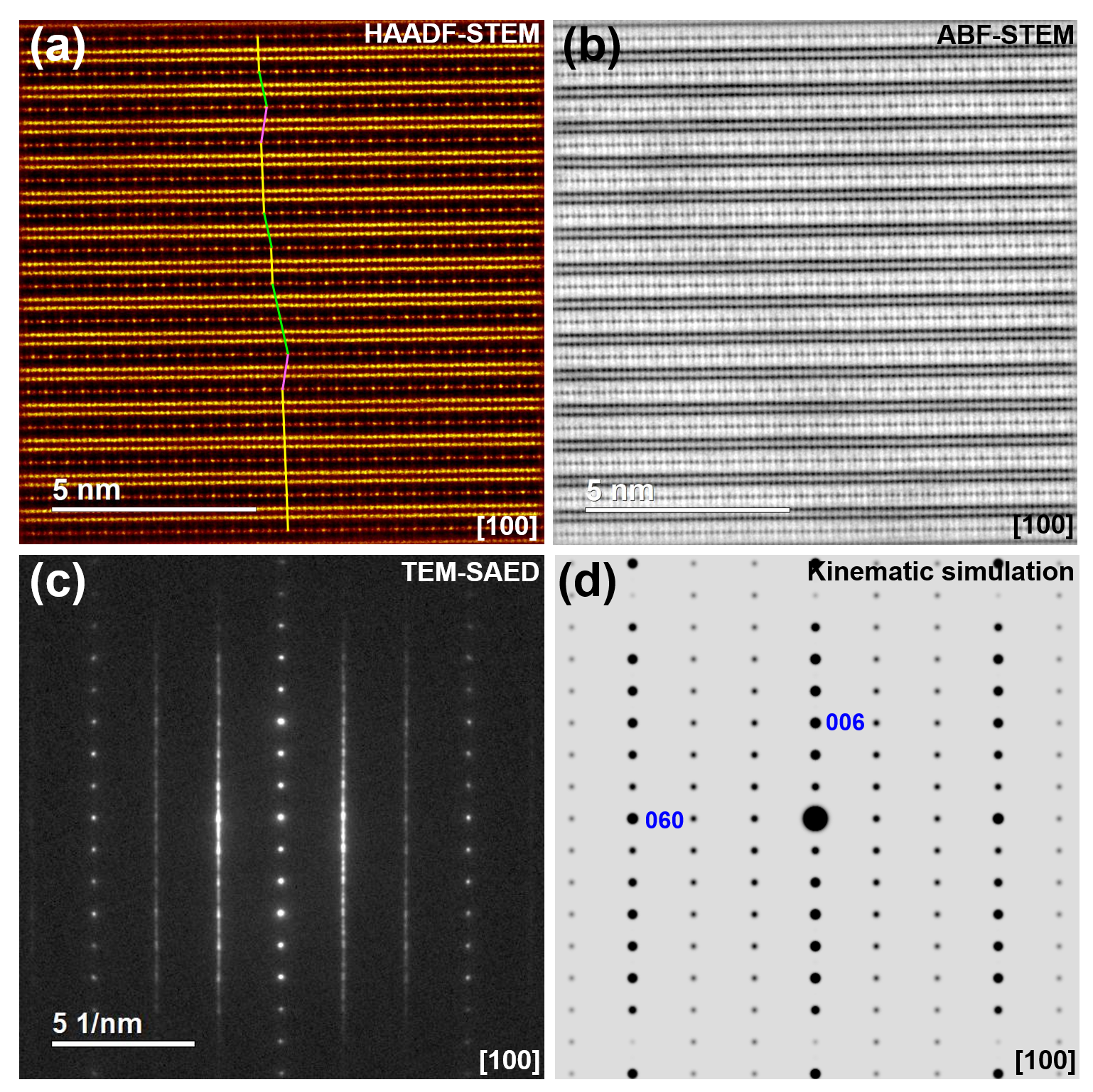}
 \caption{Stacking disorders in layered materials manifesting mobile cation bilayers. (a) HAADF-STEM image of $\rm Ag_6Ni_2TeO_6$ showing the aperiodic ordering sequence of $\rm Te$ and $\rm Ni$ atoms in successive slabs taken along [100] zone axis. Green and pink lines denote the right and left shift of $\rm Te$/$\rm Ni$ atoms in the adjacent slabs, respectively. (b) Corresponding ABF-STEM image. (c) Selected area electron diffraction (SAED) patterns taken along the [100] zone axis revealing streaks and spot shifts that are suggestive of the existence of aperiodicity. (d) Corresponding kinematic simulations based on the atomistic structural model shown in \textbf{ Figure \ref{Figure_2}}. Note that a disordered/faulted model has not been taken into consideration. Figures reproduced with permission.\cite {masese2023honeycomb} Copyright 2023 Wiley-VCH.}
 \label{Figure_6}
\end{figure*}

Aberration-corrected scanning transmission electron microscopy (STEM) has been employed to reveal local atomic structural disorders intrinsic in layered frameworks manifesting metallophilic bilayers, characterised by aperiodic stacking and incoherency in the bilayer arrangement of $\rm Ag$ atoms.\cite {masese2023honeycomb} 
High-angle annular dark-field (HAADF) scanning transmission electron microscopy (STEM) imaging captures electrons that have been scattered to high angles beyond the bright field range, utilising an annular detector. In this ADF-STEM mode, the electron signal is primarily governed by elastic Rutherford scattering.\cite {carter2016transmission} As the Coulombic interaction between incident electrons and atomic cores intensifies with an increase in the effective nuclear charge, the intensity of the ADF signal becomes intricately related to the atomic number ($Z$) of the constituent elements within the specimen (sample). Generally, the ADF signal intensity ($I$) can be expressed as being proportional to $\rm {\it Z}^{\alpha}$, where the parameter $\rm \alpha $ is contingent upon the collection angle, microscope settings, and the specific properties of the sample, typically ranging between 1.2 and 1.8.\cite {pennycook2006, pennycook1988, pennycook2006materials, wang2011quantitative, nellist2000principles} Consequently, HAADF-STEM images are readily interpretable, as individual atomic columns in a crystalline structure of a sample manifest themselves as luminous dots amidst a dark background, with the brightness of each dot scaling proportionally with the average atomic number ($Z$) present within the respective atomic column. \textbf {Figure \ref{Figure_6}a} shows a high-resolution HAADF-STEM image of $\rm Ag_6Ni_2TeO_6$ as viewed along the [100] zone axis. The inversion in imaging contrast between $\rm Ag$ ($Z$ = 47) and $\rm Te$ ($Z$ = 52) arises from the absence of dense atomic columns proximate to $\rm Te$, in stark contrast to $\rm Ag$, where brighter contrasts ensue as a consequence of the overlapping presence of adjacent $\rm Ag$ columns.

One limitation inherent to HAADF-STEM imaging pertains to its weak sensitivity to low atomic number ($Z$) elements, resulting in substantial fluctuations in ADF signal intensity for materials with different chemical compositions. Notably, light atoms, such as oxygen, exhibit minimal or no discernible visibility within the context of HAADF-STEM imaging, as exemplified in \textbf {Figure \ref{Figure_6}a}. This inherent challenge is further exacerbated when attempting to discern the precise locations and quantify the presence of light elements (such as oxygen), especially when juxtaposed with heavier counterparts (such as $\rm Ni$ or $\rm Te$). For instance, in the HAADF-STEM image presented in \textbf {Figure \ref{Figure_6}a}, oxygen atoms are not visible. In contrast, annular bright-field (ABF)-STEM imaging capitalises on either a physical or virtual annular detector to capture electrons predominantly scattered via coherent mechanisms at the outer periphery of the bright-field disk.\cite {findlay2009robust, findlay2017annular} The collection angles are significantly reduced to attain sufficient signal from light elements, such as oxygen. By virtue of the intricate interplay between electron intensity and electron channeling effects in this region of the bright field disk, ABF-STEM imaging imparts the unique capability of simultaneously visualising both lightweight and heavyweight elements. Distinct from HAADF-STEM imaging, the contrast within ABF-STEM imaging exhibits a weak dependency on atomic number ($I$ $\rm  \propto $ $\rm {\it Z}^{1/3}$), rendering this imaging technique particularly suitable for the visualisation of lightweight elements. \textbf {Figure \ref{Figure_6}b} shows a high-resolution ABF-STEM image of $\rm Ag_6Ni_2TeO_6$ as viewed along the [100] zone axis. From the [100] zone axis, the alignment of $\rm Ni$ atoms marked by darker amber spots ($Z$ = 28) and $\rm Te$ atoms represented by yellow spots ($Z$ = 52), manifest a $\rm -Te-Ni-Ni-Te-$ sequence (see \textbf {Figure \ref{Figure_2}}) as should be expected for a honeycomb slab structure. $\rm Ag$ atoms ($Z$ = 47) manifest bright yellow spots in a bilayer arrangement sandwiched between the $\rm Te$ and $\rm Ni$ slabs. The kinematically simulated electron diffraction pattern (\textbf {Figure \ref{Figure_6}d}) is in good accord with the obtained SAED patterns (\textbf {Figure \ref{Figure_6}c}). 

However, the emergence of `streak-like' array of spots in the SAED pattern (as can be seen in \textbf {Figure \ref{Figure_6}c}) instead of discrete spots confirms the stacking disorder (faults) of atoms across the slab (along the $c$-axis). 
HAADF- and ABF-STEM images taken along the [310] zone axis (\textbf {Figures \ref{Figure_2}c and \ref{Figure_2}d}) further show structural disorders in the arrangement of $\rm Ag$ atom bilayers. Whilst the orientation in the oxygen atoms across the slab follow a periodic sequence, shifts in the bilayer alignment of $\rm Ag$ atoms along the \textit {ab} plane (perpendicular to the $c$-axis) is observed. There is no coherency in the orientation of $\rm Ag$ atom bilayers along the $c$-axis, as the orientation between adjacent $\rm Ag$ bilayer planes is observed to frequently invert with no periodicity across the slabs (\textbf {Figures \ref{Figure_6}a and \ref{Figure_6}b}). The atomic-resolution STEM images highlight layered frameworks manifesting metallophic bilayers such as $\rm Ag_6Ni_2TeO_6$ to engender a multitude of structural defects / disorders, that are beyond the reach of diffraction measurements. Stacking faults may be attributed to the occurrence resulting from the comparatively weak $\rm Ag-O$ interactions existing between individual $\rm Ag$ bilayers and the adjacent neighbouring $\rm Te$/$\rm Ni$ slabs. Consequently, the stacking arrangement of these layers tends to be imperfect, as also observed in related layered materials.\cite {masese2023honeycomb, masese2021topological, masese2021unveiling}

\newpage

\section{\label{Section: Properties_Functionalities} Properties and Functionalities of Honeycomb Layered Frameworks with Metallophilic Bilayers}

Geometric frustration plays a pivotal role in engendering unconventional phenomena and novel phases within spin systems residing on diverse triangular lattices.\cite {nakatsuji2005} When spin-carrying atoms are arranged in triangular arrays, with antiferromagnetic interactions governing their neighbouring spins, their endeavours to achieve ordered arrangements are thwarted, leading to the emergence of exotic states instead of the conventional long-range order. Moreover, in alignment with recent developments, quantum fluctuations effectively suppress long range order of the ground state, particularly in low spin systems such as $S = 1/2$. Theoretical predictions indicate that quantum spins residing on a triangular lattice fail to adopt a fixed configuration, even at the lowest temperatures, giving rise to a distinctive state referred to as a `spin liquid'.\cite {moessner2006, broholm2020, zhou2017, chamorro2020} The spin liquid is characterised by a large (often infinite) number of spin degeneracy. At finite temperature, other considerations predict a partial disorder can exist in some systems where the ground state degeneracy is reduced by thermal fluctuations thus selecting high entropy quantum states, in what has been dubbed `order by disorder'.\cite{nakatsuji2005spin, capriotti1999long, rodriguezfrustrated} Enthralled by the potential for uncovering intriguing physical phenomena, experimentalists have diligently sought out actual materials that encapsulate such intriguing behaviour.\cite {ye2007, nakatsuji2005, lin2021}

The layered frameworks of $\rm Ag_2{\it M}O_2$ compounds, where $M$ denotes elements such as $\rm Ni, Cr, Co, Fe,$ amongst others\cite {yoshida2020partially, schreyer2002synthesis, yoshida2020static, taniguchi2020butterfly, yoshida2011novel, matsuda2012partially, yoshida2008unique, yoshida2006spin}, possess metallophilic bilayers, rendering them distinctive material systems. These systems are particularly intriguing due to the anticipated interplay between localised quantum spins residing on a frustrated lattice and the conduction electrons. Such interplay holds the promise of yielding fascinating phenomena and behaviours within these compounds. An illustrative example is $\rm Ag_2NiO_2$,\cite {schreyer2002synthesis} which exhibits an arrangement of $\rm Ni$ transition metal slabs alternating with staggered silver bilayers, with each layer forming a hexagonal lattice. The $\rm Ni$ atoms within this compound are octahedrally coordinated with oxygen atoms, resulting in a spin-$1/2$ triangular lattice characterised by two-fold spin and two-fold orbital degeneracy, as confirmed by magnetic susceptibility measurements.\cite {schreyer2002synthesis} Furthermore, the observed metallic conductivity in $\rm Ag_2NiO_2$ (based on resistivity measurements) has been attributed to the subvalent nature of silver,\cite {schreyer2002synthesis} leading to the partially filled $\rm Ag$ 5$s$ orbital band, akin to the behaviour observed in $\rm Ag_2F$. Consequently, $\rm Ag_2NiO_2$ represents a unique system where an intriguing interplay between frustrated spins/orbitals and itinerant electrons is anticipated.
In this section, we delve into the diverse magnetic and electrical properties exhibited by layered frameworks such as $\rm Ag_2NiO_2$. Additionally, we explore their potential applications in various fields, including electronics.

\subsection{Electromagnetic Behaviour and Exotic Phenomena}

\begin{figure*}[!t]
 \centering
 \includegraphics[width=0.95\columnwidth]{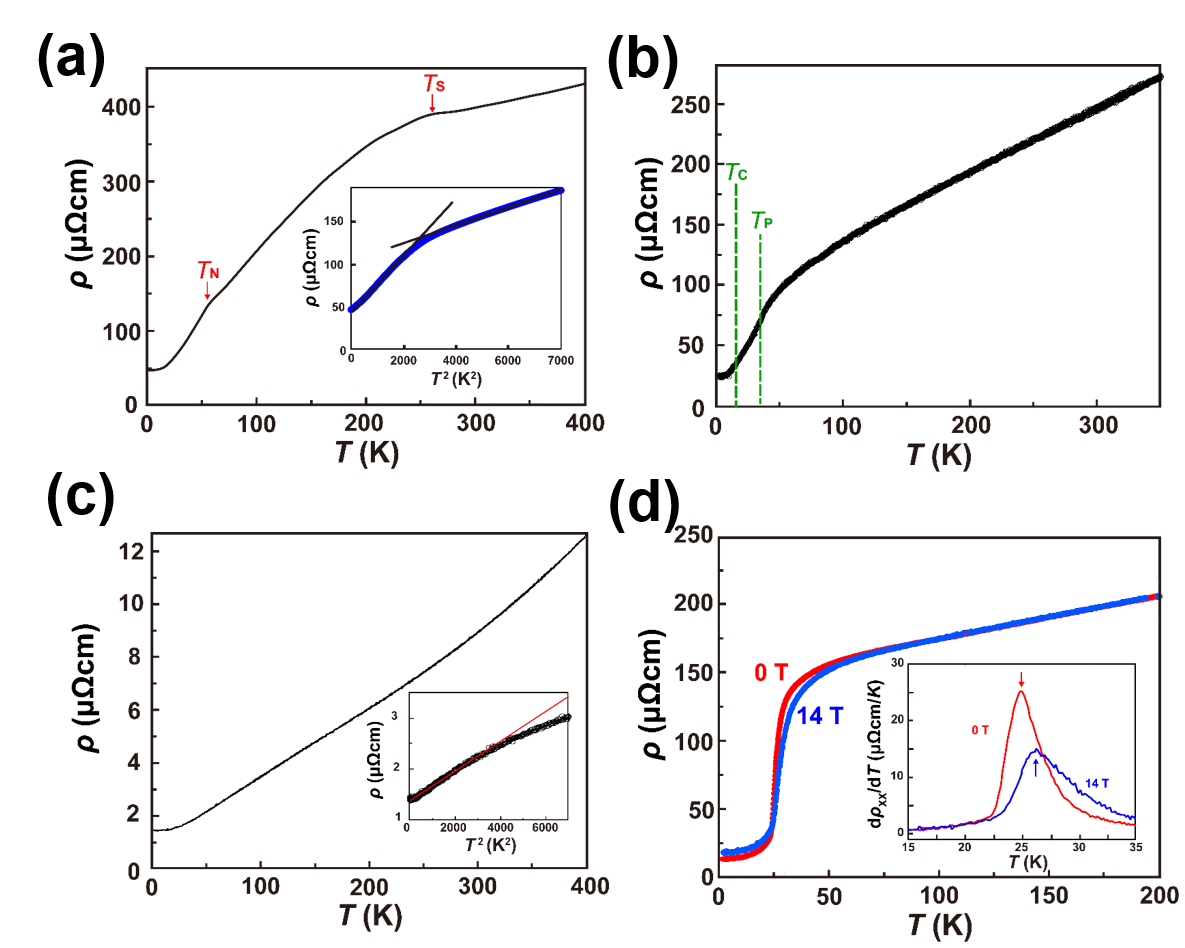}
 \caption{(a) Resistivity of  $\rm Ag_2NiO_2$  measured on a compressed pellet. $T_{\rm s}$ and $T_{\rm N}$ are the temperatures of structural and magnetic transitions, respectively. Inset shows a plot below about 90 K. The lines are guides to the eye. (b) Electrical resistivity of $\rm Ag_2FeO_2$. (c) Resistivity of $\rm Ag_2MnO_2$ measured on heating and cooling. The lower inset shows a plot at low temperatures with a linear fit presented by a solid line (d) Temperature dependence of the electrical resistivity ($\rho_{xx}$) on polycrystalline samples of $\rm Ag_2CrO_2$ in magnetic fields of 0 T and 14 T. The inset is the temperature derivative of the $\rho_{xx}$. Figures reproduced with permission.\cite {yoshida2020partially, yoshida2008unique, yoshida2006spin, schreyer2002synthesis, kida2015transport} Copyright 2002 Wiley-VCH, Copyright 2006 American Physical Society, Copyright 2008 Japan Physical Society, Copyright 2020 American Physical Society, Copyright 2015 Elsevier.}
\label{Figure_7}
\end{figure*}

$\rm Ag_2{\it M}O_2$ ($M$ = $\rm Ni, Cr, Fe, Co,$ \textit {etc}) exhibit intriguing electric transport properties of significant interest.\cite {yoshida2008unique, yoshida2020partially, yoshida2006spin} $\rm Ag_2NiO_2$ demonstrates good metallic conductivity, boasting a specific resistance of $2.18 \times 10^{-4}$ $\Omega$ cm at $298$ K.\cite {schreyer2002synthesis} Furthermore, it manifests a positive temperature coefficient of $70.78$ $\mu\Omega$ cm K$^{-1}$,\cite {schreyer2002synthesis} placing it on par with the electric transport characteristics observed in $\rm Ag_3O$ and $\rm Ag_2F$. Notably, $\rm Ag_2NiO_2$ exhibits persistent metallic conductivity without undergoing a superconducting transition even at ultra-low temperatures down to $0.4$ K, as evidenced in \textbf {Figure \ref {Figure_7}a}.\cite {yoshida2006spin} A discernible minor perturbation emerges at $260$ K, followed by a sudden decline below 56 K. The N\'eel temperature ($T_{\rm N}$) of $\rm Ag_2NiO_2$ stands at 56 K, corresponding to the onset of antiferromagnetic ordering of $\rm Ni$ spins, as confirmed by magnetic susceptibility measurements.  
\blue{Note that, the temperature dependence of the resistivity in the low temperature regime in the absence of superconductivity is generally given by\cite{bass1990temperature},
\begin{align}\label{rho_eq}
    \rho (T) = \rho_0 + aT^2 + bT^n + c\ln \left (\frac{\mu}{T}\right ),
\end{align}
where $\rho_0$ is the residual resistivity\cite{liboff1996electron}, and $n$, $a$, $b$, $c$ and $\mu$ are constants associated with properties of electrons, phonons and magnetic impurities. The term $aT^2$ can be derived from the properties of fermi liquids\cite{van2011common}, whereas $bT^n$ arises from electron-(acoustic) phonon coupling at low temperatures (under the Bloch-Gr\"{o}neisen model, simple alkali-metal based compounds yields $n = 5$, whereas, $sd$ inter-band scattering in transition metals yields $n = 3$ ).\cite{liboff1996electron, mott1936electrical} Meanwhile, the natural logarithm term arises from the Kondo effect, corresponding to electron scattering with magnetic impurities.\cite{kondo1964resistance} Proceeding, the resistivity against temperature ($\rho$ $-$ $T$) plots in the inset of \textbf {Figure \ref {Figure_7}a} portray a $T^{2}$ dependence at the vicinity of the N\'{e}el temperature ($T_{\rm N}$), where the $a \propto m_{\rm e}^2$ coefficient in eq. (\ref{rho_eq}) exhibits a three-fold increase below $T_{\rm N}$ ($a_{\rm below} = 3a_{\rm above} \simeq  0.033 \mu\Omega$ cm/K$^2$), interpreted as the result of pronounced electron correlations which renormalise the effective electron mass $m_{\rm e} \rightarrow m_{\rm e}^*$ in heavy fermion systems.\cite{stewart1984heavy} 

Heavy fermion systems were first reported in $f$ electron (magnetic, non-magnetic and superconducting) materials, characterised by large specific heat $\gamma(0)$ value and magnetic susceptibility $\chi(T)$ at low temperatures.\cite{stewart1984heavy} Heavy fermions were also reported in the $3d$ electron material $\rm LiV_2O_4$,\cite{kondo1997liv, kondo1999synthesis, johnston1999specific, urano2000liv} which lends credence to the heavy fermion interpretation for the peculiar $T^2$ dependence in the Ni $3d$ electron system, $\rm Ag_2NiO_2$. Particularly, the heavy fermion superconductors tend to have a Wilson ratio $R_{\rm W} < 1$, contrary to ordinary metals ($R_{\rm W} \simeq 1$) and highly magnetic materials ($R_{\rm W} > 1$), where,  
\begin{align}
    R_{\rm W} = \frac{\pi^2k_{\rm B}^2\chi(0)}{\mu_{\rm eff}^2\gamma(0)},
\end{align}
with $k_{\rm B}$ the Boltzmann constant, $\chi(T) = C'/(T - \Theta_{\rm CW})$ the Curie-Weiss law, $\Theta_{\rm CW}$ the Curie-Weiss temperature, $C'$ the curie constant, $\mu_{\rm eff} = g\sqrt{S(S + 1)}\,\mu_{\rm B}$ the magnitude of the atomic dipole moment, $S$ the spin of the magnetic atom, $g$ the Land\'{e} $g-$factor, $\mu_{\rm B}$ the Bohr magneton, $\gamma(T) = C(T)/T - \beta T^2 \sim nN(E_{\rm F})m_{\rm e}^*/m_{\rm e}$ the electronic specific heat (Sommerfeld) coefficient, $C(T)$ the specific heat, $\beta \propto n/\Theta_{\rm D}$, $\Theta_{\rm D}$ the Debye coefficient, $N(E_{\rm F})$ the band-structure electronic density of states at the Fermi energy $E_{\rm F}$ and $n$ the number of atoms in a chemical formula unit.\cite{stewart1984heavy} Due to the linear dependence of $\gamma$ on the effective heavy fermion mass and transition metals, the Kadowaki-Woods ratio $R_{\rm KW} = a/\gamma^2(0)$ is typically a material-independent constant for similar conditions such as dimensionality, carrier density, unit cell volume and multi-band structure.\cite{jacko2009unified}  
Specific heat measurements of $\rm Ag_2NiO_2$ corroborate the occurrence of two anomalies at $T_{\rm N} = 56$ K and $T_{\rm s} = 260 K$, substantiating the existence of second order phase transitions at the N\'{e}el and structural temperatures respectively.\cite{yoshida2006spin} 
Meanwhile, the metallic conductivity observed in $\rm Ag_2NiO_2$ without a superconducting transition down to 2 K, has been attributed to the partially-filled $\rm Ag$ 5$s$ bands, analogous to the behaviour displayed by $\rm Ag_2F$.\cite {andres1966superconductivity} Although $\rm Ag_2NiO_2$ has a higher specific heat value $\gamma(0) = 18.8$ mJ/K$^2$ mol compared to $\rm Ag_2F$ ($\gamma(0) = 0.62$ mJ/K$^2$ mol),\cite{yoshida2006spin} significant magnetic ordering due to spin $1/2$ $\rm Ni^{3+}$ cations octahedrally coordinated with oxygen atoms in the slabs ($t_{2g}^6e_g^1$) can be thought to lead to a large Wilson ratio $R_{\rm W} > 1$, potentially explaining the absence of superconductivity (or possibly a small transition temperature of milli-Kelvin order similar to $\rm Ag_2F$ (66 mK), since low temperature experiments have only been conducted down to $0.4$ K.\cite {yoshida2006spin}).}

In a similar vein, $\rm Ag_2FeO_2$ shows metallic characteristics persisting down to 2 K, as clearly depicted in \textbf {Figure \ref {Figure_7}b}.\cite {yoshida2020partially} However, a gradual decrease in conductivity is discernible around 40 K, coinciding with a magnetic anomaly denoted as $T_{\rm p}$. This decline in conductivity has been attributed to the hindrance of magnetic scattering of the itinerant electrons of $\rm Ag$ 5$s$ orbitals caused by the establishment of an ordered state of $\rm Fe$ spins below $T_{\rm p}$. Meanwhile, $\rm Ag_2MnO_2$ demonstrates metallic nature akin to $\rm Ag_2NiO_2$. This assertion was substantiated by the temperature-dependent resistivity data of $\rm Ag_2MnO_2$, as illustrated in \textbf {Figure \ref {Figure_7}c}, where the resistivity values were measured during both cooling and subsequent heating processes.\cite {yoshida2008unique} Remarkably, $\rm Ag_2MnO_2$ exhibits good metallic behaviour, characterised by consistently low resistivity values, even when the measurements were performed on a compressed pellet. Notably, no discernible anomalies arising from superconductivity or phase transitions are observed down to 2 K. This behaviour stands in stark contrast to the distinctive features observed in $\rm Ag_2NiO_2$, where two anomalies, namely at 56 K ($T_{\rm N}$) and 260 K ($T_{\rm s}$), manifest prominently. At low temperatures ($T$), the resistivity ($\rho$) of a metal primarily influenced by the electron-phonon scattering process is anticipated to exhibit a proportionality to $T^n$ ($n = 3, 5$). Nevertheless, like in the case of $\rm Ag_2NiO_2$, the resistivity behaviour of $\rm Ag_2MnO_2$ deviates from this expected trend and instead demonstrates a proportionality to $T^2$, as depicted in the inset of \textbf {Figure \ref {Figure_7}c}. Such a characteristic is commonly associated with heavy fermion system featuring substantial electron correlations.\cite{stewart1984heavy} Namely, the $T^2$ dependence in the resistivity of $\rm Ag_2MnO_2$ suggests the presence of electron-electron interactions that play a dominant role in the scattering of charge carriers, over electron-phonon scattering which leads to a renormalised electron mass and enhances the $a$ coefficient in eq. (\ref{rho_eq}).\cite {yoshida2008unique} 
As remarked earlier, this heavy fermion behaviour is often observed in some $f$ and $d$ electron systems where electron-electron interactions are significant and can influence the transport properties of the material.\cite{kondo1997liv, kondo1999synthesis, johnston1999specific, urano2000liv, stewart1984heavy} 
In the case of $\rm Ag_2MnO_2$, results in neutron scattering experiments suggest the electrons in Mn $3d$ and O $2p$ states are spin-polarised, leading to the heavy fermion effect.\cite{ji2010orbital}

\textbf {Figure \ref {Figure_7}d} shows the temperature dependence of the electrical resistivity exhibited by $\rm Ag_2CrO_2$ under magnetic fields $0$ T and 14 T.\cite {kida2015transport} Similar to $\rm Ag_2MnO_2$ and $\rm Ag_2NiO_2$, $\rm Ag_2CrO_2$ showcases metallic behaviour characterised by a positive temperature gradient, which has been attributed to the presence of itinerant electrons within the partially-filled $\rm Ag$ 5$s$ bands.\cite {taniguchi2020butterfly, kida2015transport, yoshida2011novel, matsuda2012partially} However, the most striking aspect of $\rm Ag_2CrO_2$ is the abrupt decrease in resistivity observed at its magnetic transition temperature ($T_{\rm N}$ = 24 K). Although a similar decrease in resistivity has been observed in $\rm Ag_2NiO_2$, the decrease in $\rm Ag_2CrO_2$ is orders of magnitude steeper.
This sudden and substantial decline in resistivity is believed to originate from a significant $s$-$d$ orbitals interaction, specifically the Ruderman-Kittel-Kasuya-Yosida (RKKY) interaction, between the itinerant $\rm Ag$ 5$s$ electrons and the localised $\rm Cr$ 3$d$ spins residing on the triangular lattice.\cite {taniguchi2020butterfly, kida2015transport, yoshida2011novel, matsuda2012partially} The RKKY interaction is postulated to play a crucial role in alleviating magnetic frustration and facilitating 3D long-range ordering at the N\'eel temperature ($T_{\rm N}$). Consequently, the itinerant Ag 5$s$ electrons experience considerable scattering due to the fluctuations of paramagnetic spins localised at the $\rm Cr$ sites, resulting in spin-disorder resistivity above $T_{\rm N}$. Furthermore, the onset temperature at which the resistivity drop occurs in $\rm Ag_2CrO_2$ exhibits a slight increase with applied magnetic fields, as evidenced by the inset in \textbf {Figure \ref {Figure_7}d}. This phenomenon has been attributed to the enhanced stability of the long-range ordered magnetic state under the influence of the magnetic field.\cite {kida2015transport} Overall, $\rm Ag_2CrO_2$ showcases intriguing electrical resistivity characteristics, with the abrupt resistivity decrease at $T_{\rm N}$ highlighting the pronounced influence of the RKKY interaction and the interplay between itinerant electrons and localised magnetic spins.

The temperature dependence of the resistivity in $\rm Ag_2CoO_2$ has also been investigated, revealing metallic conductivity.\cite {yoshida2020static} However, unlike $\rm Ag_2NiO_2$ and $\rm Ag_2CrO_2$ which exhibit a substantial resistivity decrease at the magnetic transition temperature due to a relatively strong coupling between itinerant electrons and localised spins, no notable anomaly was observed at the magnetic transition in $\rm Ag_2CoO_2$.\cite {yoshida2020static} Despite the presence of magnetic interactions, the resistivity behaviour in $\rm Ag_2CoO_2$ remains relatively unaffected, suggesting a different underlying mechanism governing its electrical conductivity. 

\subsection{Tunability of Electromagnetic Responses}

$\rm Ag_2NiO_2$ possesses the characteristics of an antiferromagnetic metal, exhibiting a Neel temperature ($T_{\rm N}$) of 56 K.\cite {nozaki2008neutron, schreyer2002synthesis, yoshida2006spin} Notably, the coexistence of metallic conductivity and static antiferromagnetic order below $T_{\rm N}$ down to 0.4 K is an uncommon occurrence within the Ni-based triangular-lattice layered frameworks. Typically, the presence of static antiferromagnetic order tends to preclude the itinerancy of $d$ electrons associated with the Ni ions, thereby hindering the manifestation of metallic behaviour. For instance, $\rm NaNiO_2$ and $\rm LiNiO_2$ are recognised as antiferromagnetic insulators,\cite {kitaoka1998, chatterji2005, darie2005, lewis2005, baker2005} whilst $\rm AgNiO_2$ exhibits an antiferromagnetic semi-metallic behaviour owing to a charge ordering phenomenon involving the Ni ions.\cite {schreyer2002synthesis}

It is noteworthy that the $\rm Ni^{3+}$ ions within these compounds adopt a low-spin state characterised by a $S=1/2$ ($t^6_{2g}e^1_{g}$) configuration. This arises from a pronounced crystalline electric field splitting effect imposed upon the $d$ orbitals residing in the $\rm NiO_6$ octahedra. Consequently, layered nickelates are perceived as fascinating systems that encapsulate the intriguing physics inherent to an ideal 2D triangular lattice. In the case of $\rm Ag_2NiO_2$, a structural transition from a rhombohedral (high-temperature) phase to a (monoclinic) low-temperature phase occurs at 260 K ($T_{\rm s}$).\cite {wedig2006studies, yoshida2006spin, schreyer2002synthesis} This transition is understood within the context of Jahn-Teller distortion and subsequent orbital ordering\cite{pavarini2016Quantum} of the $e^1_{g}$ orbitals (cooperative Jahn-Teller transition similar to the case of $\rm NaNiO_2$\cite{lewis2005ordering, darie2005magnetic}), where the orbital ordering leads to a gap between $3d_{z^2 - r^2} (3 d_{z^2})$ and $3d_{x^2 - y^2}$ giving rise to two dominant anti-ferromagnetic interactions and one ferromagnetic interaction in the Ni triangle.\cite {yoshida2006spin} However, this conclusion has been questioned on the basis of muon-spin-rotation and relaxation ($\mu^+$SR) spectroscopy, and first-principles calculations.\cite{sugiyama2006Incommensurate, johannes2007formation}
In particular, it was reported that, since the (weak) transverse $\mu^+$SR measurements particularly sensitive to crystal field shifts showed no obvious anomalies at $T_{\rm s}$ unlike the susceptibility curves, this structural transition cannot be associated with a cooperative Jahn-Teller transition, with the high likelihood that the crystal structure remains rhombohedral down to 5 K.\cite{sugiyama2006Incommensurate} Moreover, magnetic behaviour in nickelates and cobaltates is typically characterised by the emergence of A-type anti-ferromagnetism (\textit {i.e.}, manifestation of ferromagnetic ordering within the $\rm Ni$/$\rm Co$ metal slabs but antiferromagnetic ordering between the two adjacent $\rm Ni$/$\rm Co$ metal slabs.) at low temperatures. 
Thus, although $\rm Ag_2NiO_2$ possesses a N\'eel temperature of $\rm 55-56$ K,\cite {nozaki2008neutron, schreyer2002synthesis, yoshida2006spin} the exact nature of the anti-ferromagnetic magnetic ordering remains unknown, giving rise to speculations that it may deviate from the A-type pattern. Moreover, an additional transition at $T_{\rm m} = 22$ K was detected by both $\mu^+$SR and susceptibility measurements, throwing the phenomenological understanding of the phase transition at $T_{\rm N}$ to further disarray.\cite{sugiyama2006Incommensurate} 
Presently, this has led to suggestions that the magnetic order in $\rm Ag_2NiO_2$ is antiferromagnetic within the planes and potentially incommensurate in nature.\cite {sugiyama2006Incommensurate} Meanwhile, in a bid to also explain the unusual Ag subvalency of $1/2+$, the first-principle calculations reported strong bonding-antibonding splitting of the Ag bilayer, whose lower $sp$ band structure falls below O $p$ band structure, leading to a ligand hole ($\underline{L}$) supplanted in the $\rm Ni^{2+}$ ($d^8\underline{L}$) ground state, but nonetheless measured as a $d^7$ ground state by high covalency spectroscopic measurements. The heavy fermion behaviour is accounted for by strongly fluctuating spins due to competing anti-ferromagnetic interplane superexchange and metallic FM double exchange. A high-symmetry rhombohedral structure ground state obtained using appropriate pseudopotential, disfavoured the Jahn-Teller distorted ground state.\cite{johannes2007formation} Further comprehensive details regarding this phenomenon beyond the provided references are yet to be elucidated.
These peculiar magnetic and structural properties observed in $\rm Ag_2NiO_2$ distinguish it from most similar materials, highlighting the intricate and intriguing nature of its magnetic behaviour and the complexities associated with the interplay between structural distortions and magnetic phenomena within the material.

The magnetic behaviour of $\rm Ag_2FeO_2$ exhibits a series of successive changes through distinct phase transitions.\cite {yoshida2020partially} At $T_{\rm p}$=36 K, a second-order phase transition occurs, followed by a crossover at $T_{\rm c}$=20 K. Below $T_{\rm p}$, a partially disordered state manifests, where approximately two-thirds of the spins become ordered whilst the remaining one-third undergoes fluctuation. Remarkably, this partially disordered state was noted to persist at least down to 5 K. A notable increase in the spin correlation length was observed starting at $T_{\rm p}$, indicating the development of magnetic correlations within the system. However, the spin correlation length was noted to remain relatively short below $T_{\rm c}$. This intriguing magnetic behaviour observed in $\rm Ag_2FeO_2$ is closely associated with the presence of strong frustration within the classic antiferromagnetic triangular lattice. The interplay between frustration and the underlying lattice structure gives rise to this exotic magnetism, adding to the complex and captivating magnetic nature of $\rm Ag_2FeO_2$.\cite {yoshida2020partially} $\rm Ag_2MnO_2$ also exhibits a sequence of magnetic transitions, progressing from a Curie-Weiss paramagnetic phase to a short-range magnetic ordered phase at $T_{\rm N}$ = 80 K, and subsequently to a spin-glass phase at 22 K.

\begin{figure*}[!t]
 \centering
 \includegraphics[width=0.95\columnwidth]{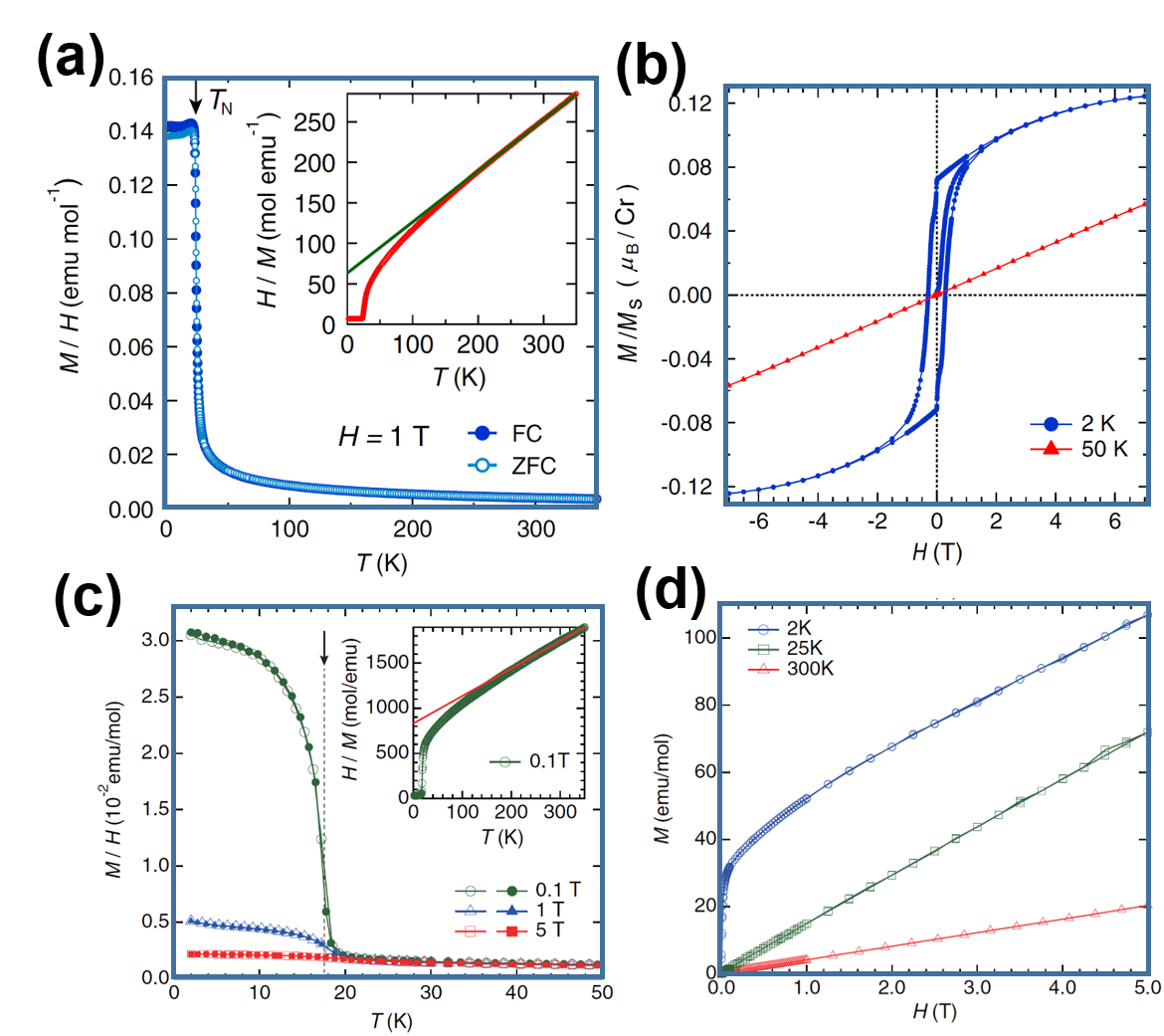}
 \caption{(a) Temperature dependence of magnetic susceptibility of $\rm Ag_2CrO_2$ along with reference compounds. Inverse susceptibility data is shown in inset. Curie-Weiss fitting is shown in solid lines. (b) Magnetisation curves of $\rm Ag_2CrO_2$ taken at 50 K and 2 K. (c) Temperature dependency of the magnetic susceptibility of $\rm Ag_2CoO_2$ under various magnetic fields. Field-cooled (FC) and zero-field-cooled (ZFC) processes are shown in solid circles and open circles, respectively. (d) Magnetisation curves of $\rm Ag_2CoO_2$ at various temperatures.
Figures reproduced with permission.\cite {yoshida2020static, yoshida2011novel} Copyright 2011 Japan Physical Society and Copyright 2020 American Physical Society.}
 \label{Figure_8}
\end{figure*}

$\rm Ag_2CrO_2$ exhibits a prominent antiferromagnetic long-range order at its N\'eel temperature ($T_{\rm N}$) of 24 K, concomitant with the presence of weak ferromagnetic moments.\cite {yoshida2011novel} \textbf {Figure \ref {Figure_8}a} illustrates the temperature-dependent magnetic susceptibility curves of $\rm Ag_2CrO_2$. The inset portrays the relationship between the inverse of susceptibility and temperature. Field-cooled (FC) process encompasses subjecting a sample to a controlled cooling procedure whilst exposed to an applied magnetic field, whereas the zero-field-cooled (ZFC) process entails the subsequent heating of sample after the application of a magnetic field at the base temperature. It is worth noting that the susceptibility data conforms to the Curie-Weiss law within the high-temperature regime. By conducting a linear regression analysis on the dataset spanning 200 to 350 K, the estimated value of the effective magnetic moment closely aligns with the expected spin-only value for a high-spin state of $\rm Cr^{3+}$ ($t^3_{2g}$, $S = 3/2$). This finding strongly suggests the localisation of spins on the triangular lattices. Additionally, the magnetic susceptibility exhibits an increase at $T_{\rm N}$ = 24 K, accompanied by a slight temperature hysteresis observed between the zero-field-cooling (ZFC) and field-cooling (FC) data below $T_{\rm N}$. This behaviour has been reported as compelling evidence for the magnetic ground state of $\rm Ag_2CrO_2$ being characterised by an antiferromagnetic long-range order and the presence of weak ferromagnetic moments. These properties have been understood as parasitic weak ferromagnetic behaviour.\cite {yoshida2011novel}

\textbf {Figure \ref {Figure_8}b} presents the magnetisation curves of $\rm Ag_2CrO_2$ at two distinct temperatures, namely 50 K (>$T_{\rm N}$) and 2 K (<$T_{\rm N}$). The data set obtained at 50 K demonstrates a linear and reversible magnetisation behaviour, indicative of a paramagnetic state. In contrast, the data set acquired at 2 K exhibits a pronounced hysteresis loop characterised by spontaneous magnetisation. Furthermore, the magnetisation observed in the high-field range falls notably below the anticipated total magnetic moment of $\rm Cr^{3+}$ ($S=3/2$) spins. This discrepancy has been interpreted as compelling evidence for the existence of a canted antiferromagnetic long-range order as the magnetic ground state inconclusively linked to the presence of finite Dzyaloshinskii-Moriya (DM) interactions, at least below the critical temperature ($T_{\rm N}$). 

\textbf {Figure \ref {Figure_8}c} depicts the temperature-dependent magnetic susceptibility of $\rm Ag_2CoO_2$, measured under various magnetic field strengths.\cite {yoshida2020static} In the high-temperature regime, the susceptibility conforms to the Curie-Weiss law. However, below approximately 20 K, a sudden increase in susceptibility is observed, followed by a tendency to saturate as the temperature decreases. This behaviour strongly indicates the occurrence of magnetic ordering in $\rm Ag_2CoO_2$ below this transition temperature. Through the normalisation of the experimental saturation value with the theoretically anticipated value, it has been proposed that the magnetic ordering in $\rm Ag_2CoO_2$ corresponds to a canted-antiferromagnetic transition.\cite {yoshida2020static} Additionally, \textbf {Figure \ref {Figure_8}d} presents magnetisation curves recorded at several temperatures. Above 25 K, these curves demonstrate a linear field dependence consistent with the paramagnetic state of $\rm Ag_2CoO_2$. Conversely, at 2 K, a pronounced enhancement of magnetisation is observed under low magnetic field conditions, attributed to the canted spin arrangement in $\rm Ag_2CoO_2$.

\subsection{Giant Magneto-Resistance and Potential Applications in Spintronics}

Spintronic devices integrating antiferromagnets possess considerable promise as viable contenders for future applications. Significantly, a myriad of intriguing physical phenomena have been extensively documented in relation to antiferromagnetism-based devices. The utilisation of layered frameworks featuring metallophilic bilayers has been successfully exemplified through a noteworthy instance, namely, the triangular-lattice $\rm Ag_2CrO_2$ antiferromagnet. This specific device configuration, obtained by means of the mechanical exfoliation technique, remarkably demonstrates a distinctive butterfly-shaped magnetoresistance (MR) effect within a crystal of micrometer-scale dimensions.\cite {taniguchi2020butterfly}

\begin{figure*}[!t]
 \centering
 \includegraphics[width=0.9\columnwidth]{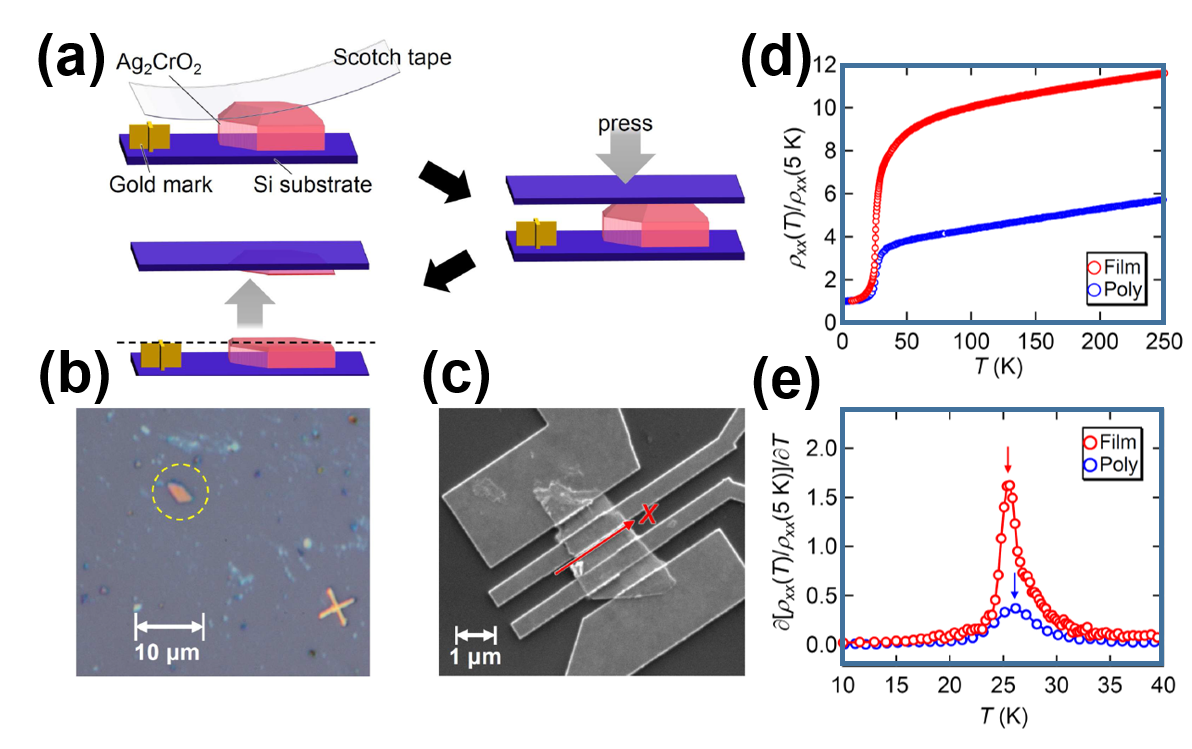}
 \caption{Electric transport properties of $\rm Ag_2CrO_2$ thin films.\cite {taniguchi2018fabrication} (a) Illustration of the procedure to fabricate $\rm Ag_2CrO_2$ thin films. (b) Optical microscope image of the attained $\rm Ag_2CrO_2$ thin films. The film selected to assess the transport properties is shown in dashed yellow circles. (c) Scanning electron micrograph of the $\rm Ag_2CrO_2$ thin film device. (d) Comparison of the normalised resistivity ($\rho_{xx}$ (T) / $\rho_{xx}$ (5 K)) between the polycrystalline sample and the thin film (e) $\partial$ [$\rho_{xx}$ (T) / $\rho_{xx}$ (5 K)] / $\partial$T versus T curves for both the polycrystalline sample and the thin film. The arrows show the peak positions of the derivatives and the lines are guided for eyes. Copyright 2018 American Institute of Physics.
}
 \label{Figure_9}
\end{figure*}

The comprehensive fabrication procedure employed for generating thin films of $\rm Ag_2CrO_2$, accompanied by the subsequent characterisation of their transport properties, is presented in \textbf {Figure \ref {Figure_9}}.\cite {taniguchi2018fabrication} The initial step involved the pulverisation of $\rm Ag_2CrO_2$ samples onto a glass plate, resulting in the generation of small fragments. Subsequently, utilising a scotch tape, minute grains of $\rm Ag_2CrO_2$ were meticulously collected and affixed onto a silicon substrate that featured multiple gold marks, each with a thickness of several hundred nanometers. Upon carefully peeling off the scotch tape from the substrate (\textbf {Figure \ref {Figure_9}a}), another silicon substrate, devoid of any gold marks, was prepared and firmly pressed onto the silicon substrate containing the $\rm Ag_2CrO_2$ flakes, in addition to the 100 nm thick gold marks (\textbf {Figures \ref {Figure_9}b and \ref {Figure_9}c}). The thicknesses of the flakes were subsequently verified through the utilisation of atomic force microscopy. Meanwhile, as mentioned earlier concerning \textbf{Figure \ref{Figure_7}d}, RKKY interactions thought to arise from the Ag $5s$ and the partially disordered $S = 3/2$ Cr $3d$ localised \textit{albeit} thermally-fluctuating spins play a significant role in the electron-spin scattering, reproducing the significant resistivity drop below $T_{\rm N}$, whereby magnetic ordering favours low resistivity whereas the thermally-fluctuating spins above $T_{\rm N}$ interfere with electron itinerancy leading to a considerably elevated resistivity. In \textbf{Figure \ref {Figure_9}d}, the resistivity disparity between polycrystalline samples and thin films of $\rm Ag_2CrO_2$ is presented as a function of temperature. Notably, the normalised resistivity of the thin film exhibits a significantly greater and more pronounced magnitude at the N\'eel temperature ($T_{\rm N}$) in comparison to the polycrystalline counterpart. Accompanying plots in \textbf{Figure \ref {Figure_9}e} depict the derivatives of the normalised resistivity for both the polycrystalline sample and the thin film. It is noteworthy that the peaks, indicated by arrows, manifest at nearly identical temperatures (approximately 25 K), a proximity that closely aligns with $T_{\rm N}$ ascertained from heat capacity measurements ($T_{\rm N}$ = 24 K). Moreover, the thin film exhibits a peak width that is more asymmetric and narrower in comparison to the polycrystalline sample, signifying a superior crystalline nature in the exfoliated $\rm Ag_2CrO_2$ thin film relative to its polycrystalline counterparts. It has been conjectured that the thin film devices attained through the mechanical exfoliation technique contain significantly fewer grain boundaries in comparison to the polycrystalline bulk samples.\cite {taniguchi2018fabrication} This difference has been suggested to arise from a substantially higher quality of $\rm Ag_2CrO_2$ in the thin films.\cite {taniguchi2018fabrication} The demonstrated fabrication of $\rm Ag_2CrO_2$ thin films exhibiting enhanced performance holds the promise of ushering in novel avenues for the application of layered antiferromagnetic materials in electronic devices. Butterfly-shaped magnetoresistance (MR) has been observed in micrometer-sized $\rm Ag_2CrO_2$ flakes prepared through the mechanical exfoliation technique,\cite {taniguchi2020butterfly} akin to the process depicted in \textbf{Figure \ref {Figure_9}a}.

\begin{figure*}[!b]
 \centering
 \includegraphics[width=0.9\columnwidth]{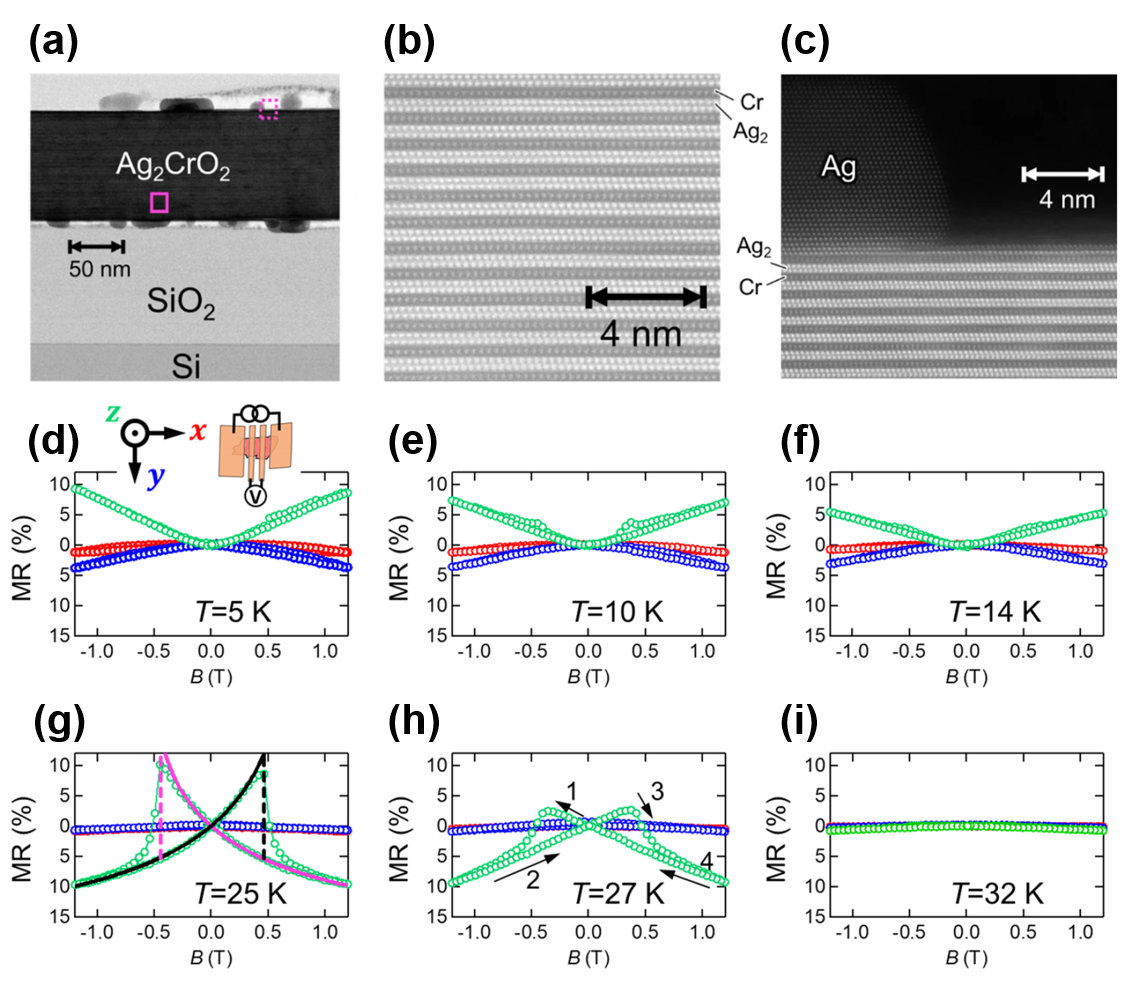}
 \caption{(a)  Bright-field STEM images of $\rm Ag_2CrO_2$ pasted onto a thermally oxidised silicon ($\rm Si$/$\rm SiO_2$) substrate. (b) High-angle annular dark-field (HAADF) STEM image of the area shown with pink solid square in (a). The dark and bright spheres correspond to $\rm Cr$ and $\rm Ag$ atoms, respectively. (c) HAADF-STEM image of the area highlighted in dashed pink line in (a). The disparity in imaging contrast of $\rm Ag$ atoms between the region labeled as `$\rm Ag$' and the regions labeled as `$\rm Ag_2CrO_2$' can be attributed to variations in sample thickness. (d-i) Magnetoresistance (MR) curves taken at various temperatures. The green, blue and red curves show the magnetoresistance values when the magnetic field ($B$) is applied along the $z$-, $y$- and $x$-axes, respectively. Figures reproduced with permission.\cite {taniguchi2020butterfly} Copyright 2020 Springer Nature.
}
 \label{Figure_10}
\end{figure*}

Notably, the $\rm Ag_2CrO_2$ flakes employed for transport measurements were confirmed to approximate a single crystal, as depicted in \textbf{Figures \ref {Figure_10}a, \ref {Figure_10}b and \ref {Figure_10}c}. The Cr metal slabs and $\rm Ag$ bilayers are observed to be stacked alternately in a perpendicular fashion relative to the $\rm Si$/$\rm SiO_2$ substrate. Whilst clusters of $\rm Ag$ can be discerned at the top and bottom surfaces of the $\rm Ag_2CrO_2$ flakes, no significant clusters are evident within the $\rm Ag_2CrO_2$ flakes themselves, leading to the conclusion that the flakes closely resemble a single crystal. Magnetoresistance (MR) measurements were conducted on the fabricated $\rm Ag_2CrO_2$ device employing three distinct magnetic field ($B$) orientations, as illustrated schematically in \textbf{Figure \ref {Figure_10}d}.\cite {taniguchi2020butterfly} Specifically, measurements were performed along the $z$-axis (out-of-plane perpendicular to the current direction), $y$-axis (in-plane perpendicular to the current direction), and $x$-axis (in-plane along the current direction). \textbf {Figures \ref {Figure_10}d, \ref {Figure_10}e, \ref {Figure_10}f, \ref {Figure_10}g, \ref {Figure_10}h and \ref {Figure_10}i} display the MR values obtained for each of the three orientations, measured at various temperatures.

At high magnetic fields, a distinct positive magnetoresistance (MR) value is observed when the magnetic field ($B$) is applied along the $z$-direction at $T$ = 5 K ($\ll T_{\rm N}$). This can be attributed to ordinary magnetoresistance in metals, typically arising from the Lorenz force term, whose resistivity scales with applied magnetic field. At this temperature, scattering of the quarter-filled Ag $5s$ electrons by the localized spins of the Cr $d$ electrons has been significantly suppressed by anti-ferromagnetic ordering. As the temperature increases, the positive slope gradually flattens. As the temperature is raised, the positive slope gradually becomes flatter. This is due to a slight increase of electron scattering favoured by the slight thermal agitation of the magnetic spins. At temperatures in the vicinity of the N\'{e}el temperature, ordinary magneto-resistivity fails to out-compete the thermally agitated spin fluctuations. This leads to the negative MR values are observed at 25 K ($\sim T_{\rm N}$).\cite{taniguchi2020butterfly} 
Another notable characteristic is the butterfly-shaped magnetoresistance observed at $B$ $\sim$ $\pm$ 0.5 T, as illustrated in \textbf{Figure \ref {Figure_10}g}. The amplitude of the butterfly-shaped magnetoresistance is relatively small when $T$ $<<$ $T_{\rm N}$. However, as the temperature approaches $T_{\rm N}$, it becomes more pronounced and reaches a maximum at 25 K ($\sim$ $T_{\rm N}$). The maximum value exceeds 10\% at $B$ = 0.5 T, which is remarkably large for conventional ferromagnetic materials. Furthermore, the amplitude of the magnetoresistance shows a sudden decline as the temperature continues to rise, eventually reaching zero above $T$ = 32 K.

Similar magnetoresistance (MR) effects are frequently observed not only in current-in-plane (CIP) giant magnetoresistance (GMR) devices but also in antiferromagnetic and ferromagnetic materials. However, the butterfly-shaped MR observed in $\rm Ag_2CrO_2$ appears to be fundamentally distinct from those phenomena.\cite {taniguchi2020butterfly} In commonly-used CIP-GMR devices, the magnetisation is in-plane and aligned parallel to the current direction. In contrast, $\rm Ag_2CrO_2$ exhibits a perpendicular magnetisation to the current direction and the basal plane, setting it apart from typical MR behaviours. In essence, the butterfly-shaped magnetoresistance (MR) is exclusively observed when the magnetic field is applied along the $c$-axis, suggesting a robust uniaxial anisotropy in $\rm Ag_2CrO_2$ of the form\cite{taniguchi2020butterfly},
\begin{align}\label{anisotropy_eq}
    H = - \Delta \sum_i S_i^zS_i^z - B\sum_iS_i^z,
\end{align}
associated with a residual unexplained small \textit{albeit} finite spontaneous magnetisation below $T_{\rm N}$, where $S_i^z$ is the residual spin strictly in the $z$ direction, $\Delta$ is the anisotropy and $B$ is the applied magnetic field. This distinctive MR pattern exhibits a peak value of 15\% in proximity to the transition temperature, indicating the significance of spin fluctuations. To elucidate this phenomenon, a theoretical model rooted in a two-dimensional (2D) magnetic system with Ising anisotropy has been proposed.\cite {taniguchi2020butterfly} The magnetically frustrated system's interaction with conducting electrons undoubtedly promises a wealth of intriguing physics. Particularly noteworthy is the substantial magnetoresistance observed at low switching fields in the frustrated $\rm Ag_2CrO_2$ spin system, which holds significant promise for its potential application in advanced spintronic devices.

\newpage

\section{\label{Section: Characterisation_Techniques} Characterisation Techniques of Honeycomb Layered Frameworks with Metallophilic Bilayers}

Honeycomb layered frameworks exhibiting metallophilic bilayers represent a rapidly expanding class of materials, and the quest for discovering and optimising additional materials can be expedited through advancements in characterisation techniques that unveil the atomic-level origins of material functionality and performance. Within this section, we present a comprehensive overview of the primary analytical techniques rooted in diffraction and spectroscopy that have been employed to elucidate the diverse electronic and structural aspects inherent to this captivating class of honeycomb layered frameworks.

\subsection{Structural Analysis Methods}

X-ray diffraction (XRD) and neutron scattering have been the \textit {de facto} methods of choice to ascertain the atomic structural framework of layered frameworks manifesting metallophilic bilayers. The latter has intensively been pursued, for instance, by investigating the spin-wave excitations and magnetic orders of layered frameworks (such as the triangular lattice antiferromagnet $\rm Ag_2{\it M}O_2$ ($M$ =$\rm Ni, Cr, Fe, Co, \textit {etc}$) and $\rm Ag_2F$) using inelastic neutron scattering.\cite {yoshida2020partially, williams1989neutron, susloparova2022, yoshida2020static, matsuda2012partially, schreyer2002synthesis, nozaki2008neutron, nozaki2013magnetic}

Neutron diffraction experiments conducted on $\rm Ag_2CrO_2$ revealed the material to possess a partially disordered state with a five-fold magnetic unit cell (a long-range magnetic structure comprising five sub-lattices) below the N\'eel temperature ($T_{\rm N}$ = 24 K).\cite {matsuda2012partially} The long-range magnetic structure is accompanied by a trigonal-to-monoclinic structural distortion, indicating that spin-lattice coupling stabilises the N\'eel order. Inelastic neutron scattering experiments performed above $T_{\rm N}$ revealed diffuse scattering originating from short-range magnetic correlations. Furthermore, the magnetic diffuse scattering helped demonstrate that $\rm Ag_2CrO_2$ prefers a four-sub-lattice structure over the five-sub-lattices structure above $T_{\rm N}$.

Neutron diffraction studies performed on $\rm Ag_2NiO_2$ have clarified its magnetic structure as a modulated A-type antiferromagnet, wherein $\rm Ni$ spins align antiparallel along the $c$-axis through the $\rm Ag$ bilayer planes. Although the magnetic $\rm Ni$ metal slabs are believed to be well spatially separated by the $\rm Ag$ bilayer planes, the interaction between adjacent $\rm Ni$ metal slabs is thought to play a pivotal role in forming the antiferromagnetic order across the slabs (\textit {i.e.}, along the $c$-axis), indicating the presence of a magnetic interaction path via the $\rm Ag$ bilayer planes. The neutron diffraction studies have clarified that the $\rm Ag$ bilayer planes are responsible for the metallic conductivity and are essential for the formation of long-range antiferromagnetic order on the $\rm Ni$ metal slabs through an RKKY-type 3$d$-5$s$ interaction.

Moreover, neutron diffraction studies combined with XRD have been performed on $\rm Ag_2NiO_2$ to elucidate the antiferromagnetic structure at the N\'eel temperature ($T_{\rm N}$ = 56 K) and to understand the mechanism of the structural phase transition occurring at $T_{\rm s}$ $\sim$ 270 K. The formation of long-range antiferromagnetic order was confirmed through neutron diffraction measurements, which supported the results obtained from muon-spin spectroscopy measurements. This confirms the appearance of static incommensurate antiferromagnetic order below $T_{\rm N}$ in the $\rm Ni$ metal slabs, suggesting that the half-filled 2D triangular lattice of the $\rm Ni$ slabs constitutes an antiferromagnetic-coupled frustrated system. Weak intensities of certain magnetic peaks were observed in $\rm Ag_2NiO_2$, indicating a 2D nature of the antiferromagnetic order, although the specific spin structure remains unknown.

In another neutron diffraction study performed on $\rm Ag_2CrO_2$ and $\rm Ag_2MnO_2$, it was noted that the triangular lattices of $\rm Cr^{3+}$ and $\rm Mn^{3+}$ ions are distorted at low temperatures in both compounds, resulting in isosceles triangular lattices. Jahn-Teller distortion of the $\rm {\it M}O_6$ octahedra was rationalised to be responsible for the macroscopic structural distortion in $\rm Ag_2{\it M}O_2$, although the distortion is very small for $\rm Ag_2NiO_2$ compared to those for $\rm Ag_2CrO_2$ and $\rm Ag_2MnO_2$. Such structural distortions tend to enhance the anisotropic magnetic interaction in the $\rm {\it M}O_6$ slabs; hence, believed to be the reason behind the absence of long-range antiferromagnetic order in $\rm Ag_2CrO_2$ and $\rm Ag_2MnO_2$.

\subsection{Spectroscopic Investigations}
Spectroscopic techniques such as X-ray absorption spectroscopy (XAS) and muon spin rotation and relaxation spectroscopy ($\mu$SR) have enriched our understanding about the local electronic and magnetic structures of layered frameworks manifesting metallophilic bilayers.\cite {wadati2005valence, schreyer2002synthesis}

\begin{figure*}[!b]
\centering
\includegraphics[width=0.9\columnwidth]{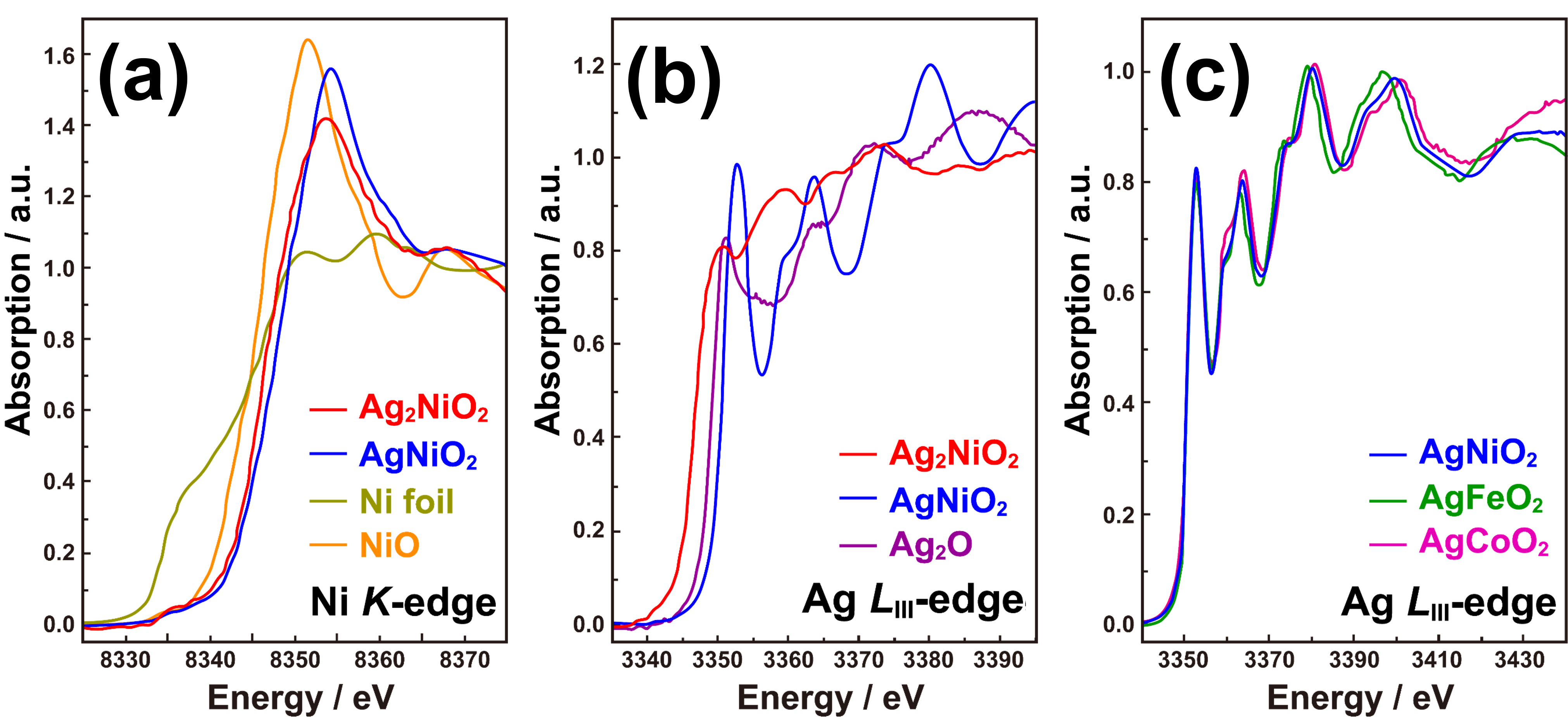}
\caption{X-ray absorption spectroscopic measurements of $\rm Ag_2NiO_2$. (a)  $\rm Ni$ $K$-edge XANES spectra of $\rm Ag_2NiO_2$ along with reference compounds. (b) $\rm Ag$ $\rm {\it L}_{III}$-edge XANES spectra of $\rm Ag_2NiO_2$ along with reference compounds. (c) $\rm Ag$ $\rm {\it L}_{III}$-edge XANES spectra of $\rm Ag_2{\it M}O_2$ ($M$ = $\rm Ni, Fe, Co$) reference compounds. Figures reproduced with permission.\cite {schreyer2002synthesis, wadati2005valence} Copyright 2002 Wiley-VCH and Copyright 2006 Elsevier.
}
\label{Figure_11}
\end{figure*}

Several spectroscopic methodologies, including techniques such as X-ray absorption spectroscopy (XAS), photoelectron spectroscopy (PES), M\"{o}ssbauer spectroscopy, muon spin rotation and relaxation spectroscopy, amongst others, have been judiciously employed to elucidate a plethora of intriguing magnetic and electronic properties exhibited by layered frameworks endowed with metallophilic bilayers.\cite {yoshida2020partially, schreyer2002synthesis, wadati2005valence, sugiyama2006Incommensurate} Layered frameworks, exemplified by $\rm Ag_2{\it M}O_2$ ($M$ = $\rm Mn, Fe, Ni, Cr$), exhibit unusual valency states of $\rm Ag$. At an initial glance, the valency state of the transition metal $M$ in a composition like $\rm Ag_2NiO_2$ may be conventionally perceived as divalent ($\rm Ni^{2+}$), analogous to layered $\rm Ag_2PdO_2$ wherein $\rm Pd$ is in the divalent state ($\rm Pd^{2+}$).\cite {schreyer2001Synthesis} Intriguingly, investigations employing X-ray absorption near edge structure (XANES) measurements have unveiled that the $\rm Ni$ valency state is, in fact, $\rm Ni^{3+}$ rather than $\rm Ni^{2+}$, whilst the $\rm Ag$ valency state assumes $\rm Ag^{1/2+}$ (sub-valent $\rm Ag$).\cite {schreyer2002synthesis} The XANES spectra acquired at the $\rm Ni$ $K$-edge (\textbf {Figure \ref {Figure_11}a}) can be readily explicated utilising the classical chemical shift approach.\cite {bianconi1988xanes, stohr2013nexafs} A comparative analysis of the absorption edge positions between $\rm Ag_2NiO_2$ and the reference compounds ($\rm AgNiO_2$, $\rm NiO$, and $\rm Ni$) evinces a close alignment in the energy positions of $\rm AgNiO_2$ and $\rm Ag_2NiO_2$, thereby suggesting a high similarity in the electronic structure of these compounds and affirming a formal valency exceeding 2+, with 3+ being the most plausible. Furthermore, the initiation of the white line in $\rm Ag_2NiO_2$ manifests at marginally lower energies than that of $\rm AgNiO_2$, and the broadened width of this feature implies the contribution of multiple transitions to its formation.

In the quest for elucidating the electronic structure of $\rm Ag_2NiO_2$, a comprehensive examination of XANES spectra taken at the $\rm Ag$ $\rm {\it L}_{III}$-edge (\textbf {Figures \ref {Figure_11}b and \ref {Figure_11}c}) show conspicuous differences between $\rm Ag_2NiO_2$ and $\rm AgNiO_2$.\cite {schreyer2002synthesis} Notably, the onset of the absorption edge in $\rm Ag_2NiO_2$ occurs at a lower photon energy compared to the reference materials ($\rm AgNiO_2$ and/or $\rm Ag_2O$) wherein silver assumes a monovalent state ($\rm Ag^{1+}$). Furthermore, the intensity of the distinctive pre-edge feature has been correlated with the accessible density of states in the $\rm Ag$ $d$-band,\cite {behrens1999} thereby establishing a connection with the formal valency. Noteworthy is the drastic suppression of this feature within the spectrum of $\rm Ag_2NiO_2$ (\textbf {Figure \ref {Figure_11}b}), signifying, on the one hand, a sub-valent nature of silver, and on the other hand, suggesting that the $d^{10}$-orbital shell can be predominantly regarded as occupied.

Extensive investigations of the electronic structures of $\rm Ag_2NiO_2$, in conjunction with comparative analyses involving $\rm AgNiO_2$, have been undertaken utilising core level and valence band photoelectron spectroscopy. Despite the absence of unequivocal indications concerning disparities in the oxidation states of $\rm Ag$, notable shifts towards lower values in the binding energies of the $\rm Ag$ 3$d$ spectral features in $\rm Ag_2NiO_2$ and $\rm AgNiO_2$ have been unveiled, contrary to expectations when compared to metallic silver. Intriguingly, the binding energy of $\rm Ag_2NiO_2$ exhibits a slightly more positive deviation (0.1 eV) than that observed for $\rm AgNiO_2$.

Comprehensive investigations employing soft X-ray photoemission and absorption spectroscopy have been conducted on $\rm Ag_2NiO_2$ and $\rm Ag_2MnO_2$, leading to insightful findings. The spectroscopic analyses reveal the predominance of trivalent $\rm Ni^{3+}$ in $\rm Ag_2NiO_2$, whilst $\rm Ag_2MnO_2$ exhibits trivalent $\rm Mn^{3+}$ states accompanied by a mixture of $\rm Mn^{2+}$ valency states.\cite {eguchi2010resonant} Furthermore, the resonant photoemission spectra of $M$ 2$p$ - 3$d$ transitions provide evidence for the presence of $M$ 3$d$ states at the Fermi level in both compounds. These observations are rationalised to strongly suggest that the $\rm Mn$ 3$d$ and $\rm Ni$ 3$d$ electronic states at the Fermi level play a crucial role in the notable enhancement of carrier mass, as manifested in the transport properties of $\rm Ag_2MnO_2$ and $\rm Ag_2NiO_2$.\cite {eguchi2010resonant}

\begin{figure*}[!b]
 \centering
 \includegraphics[width=0.9\columnwidth]{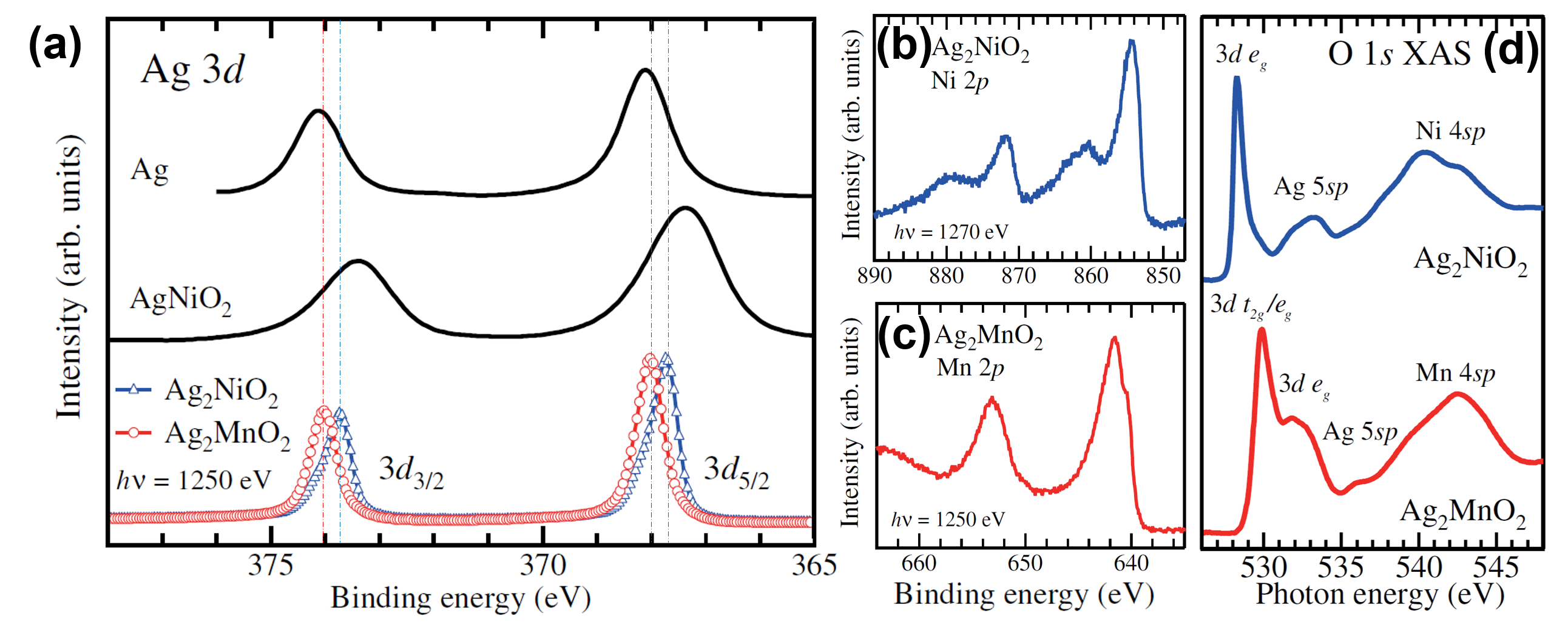}
 \caption{X-ray photoelectron and absorption spectroscopic measurements of $\rm Ag_2MnO_2$ and $\rm Ag_2NiO_2$. (a) Comparison of the $\rm Ag$ 3$d$ core level spectra of $\rm Ag_2MnO_2$ and $\rm Ag_2NiO_2$ with $\rm AgNiO_2$ and $\rm Ag$ metal. The spectra were standardised by normalising them to the peak intensity of the 3$d_{5/2}$ orbital at approximately 368 eV. (b) $\rm Ni$ 2$p$ core level spectrum of $\rm Ag_2NiO_2$. (c) $\rm Mn$ 2$p$ core level spectrum of $\rm Ag_2MnO_2$. (d) Comparison of the $\rm O$ 1$s$ spectra of $\rm Ag_2MnO_2$ and $\rm Ag_2NiO_2$. Figures reproduced with permission.\cite {eguchi2010resonant} Copyright 2010 Japan Physical Society.}
 \label{Figure_12}
\end{figure*}

Soft X-ray absorption and photoemission spectroscopy were conducted on $\rm Ag_2MnO_2$ and $\rm Ag_2NiO_2$ to gain further insights into their electronic structures. \textbf {Figure \ref {Figure_12}a} depicts the $\rm Ag$ 3$d$ core-level spectra of $\rm Ag_2MnO_2$ and $\rm Ag_2NiO_2$, alongside those of $\rm Ag^{1+}NiO_2$ \cite {kang2000photoemission} and $\rm Ag^0$ \cite {bao1996interaction} metal reference compounds. All spectra exhibit spin-orbit splitting corresponding to 3$d_{5/2}$ and 3$d_{3/2}$ orbitals. The $\rm Ag$ 3$d$ binding energies observed in $\rm Ag_2MnO_2$ and $\rm Ag_2NiO_2$ fall between those of $\rm Ag^{1+}NiO_2$ and $\rm Ag^0$ metal spectra, indicating an intermediate state. \textbf {Figures \ref {Figure_12}b and \ref {Figure_12}c} present the $\rm Ni$ 2$p$ and $\rm Mn$ 2$p$ core level spectra, respectively. The $\rm Ni$ 2$p$ spectrum displays main peaks at 2$p_{1/2}$ (~872 eV) and 2$p_{3/2}$ (~854 eV) with broad satellites, resembling the spectra of other $\rm Ni^{3+}$-containing compounds.\cite {wedig2006studies} The $\rm Mn$ 2$p$ spectrum exhibits a similar spectral shape to that of other $\rm Mn^{3+}$ compounds, with a small shoulder feature at approximately 640 eV, possibly arising from $\rm Mn^{2+}$ as observed in the $\rm Mn$ 2$p$ XAS spectra (\textbf {Figure \ref {Figure_12}c}).

Meanwhile, \textbf {Figure \ref {Figure_12}d} illustrates the $\rm O$ 1$s$ X-ray absorption spectroscopy (XAS) spectra of $\rm Ag_2MnO_2$ and $\rm Ag_2NiO_2$. Unlike photoemission spectroscopy (PES), which provides information on the occupied electronic structure, the $\rm O$ 1$s$ XAS spectra capture insights into the unoccupied electronic structure. In the $\rm O$ 1$s$ XAS spectrum of $\rm Ag_2MnO_2$, the sharp peak at the lowest photon energy undergoes a shift towards higher photon energy, indicating the presence of unoccupied $\rm Mn$ 3$d$ states. Additionally, a shoulder feature on the higher-energy side corresponds to unoccupied $e_g$ states. In $\rm Ag_2NiO_2$, the lowest-photon-energy peak corresponds to unoccupied $\rm Ni$ 3$d$ $e_g$ states, exhibiting a spectral shape similar to $\rm AgNiO_2$. The higher-photon-energy region in both compounds corresponds to mixed states of $\rm Mn$ or $\rm Ni$ 4$sp$ and $\rm Ag$ 5$sp$ orbitals.

\begin{figure*}[!b]
 \centering
 \includegraphics[width=0.9\columnwidth]{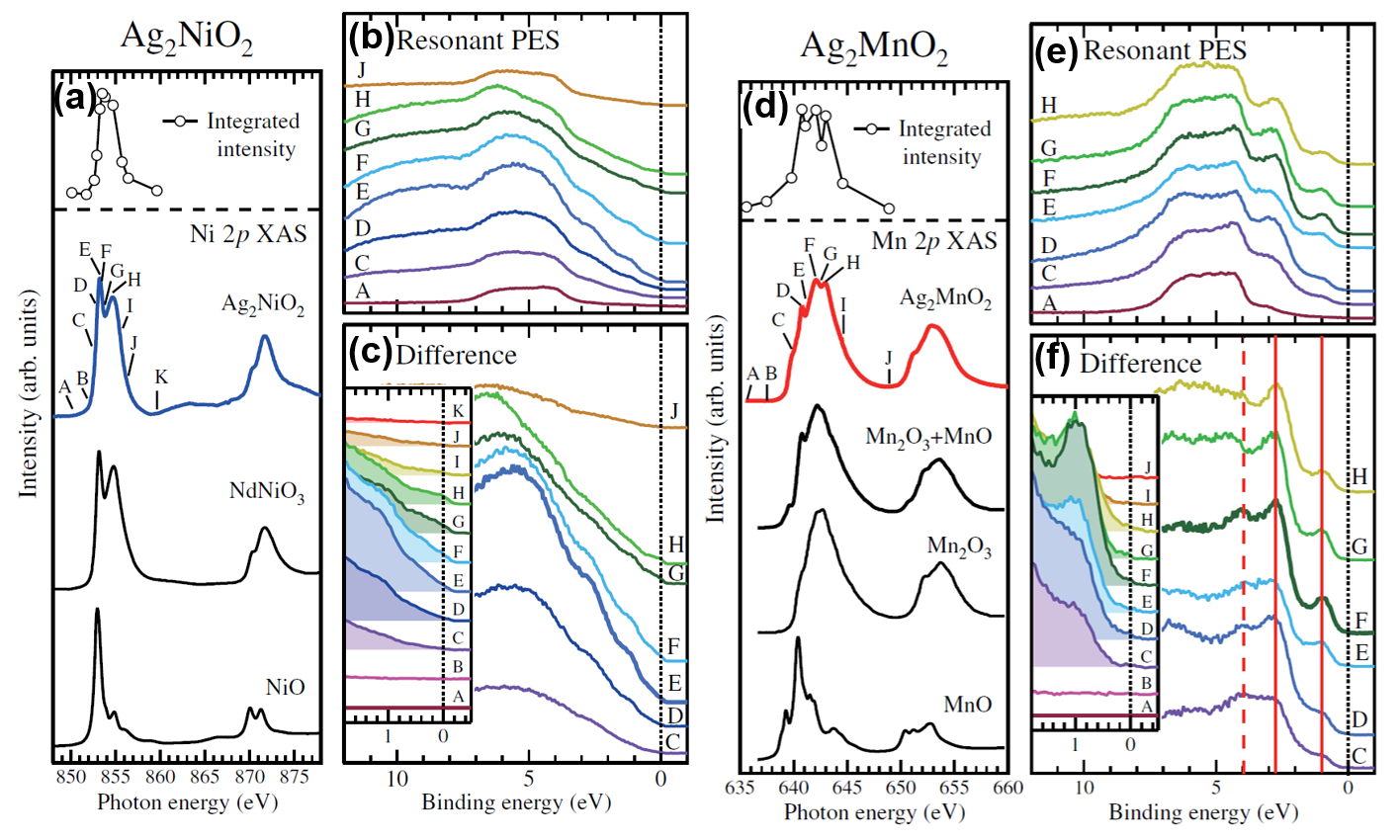}
 \caption{Resonant photoelectron spectroscopic measurements of $\rm Ag_2NiO_2$ and $\rm Ag_2MnO_2$. (a) $\rm Ni$ 2$p$ XAS spectrum of $\rm Ag_2NiO_2$ compared with those of $\rm NiO$ and $\rm NdNiO_3$. The integrated intensity of the resonant PES is plotted on top of the XAS spectra. (b) $\rm Ni$ 2$p$-3$d$ resonant PES spectra and (c) difference spectra between the off-resonance spectrum ` A ' and each spectrum. Inset shows the enlarged view near the Fermi energy ($E_F$). (d) $\rm Mn$ 2$p$ XAS spectrum of $\rm Ag_2MnO_2$ compared with those of $\rm MnO$, $\rm Mn_2O_3$ and an admixture of $\rm MnO$ + $\rm Mn_2O_3$. The integrated intensity of the resonant PES spectra is plotted on top of the XAS spectra. (e) $\rm Mn$ 2$p$-3$d$ resonant PES spectra and (f) difference spectra as that taken in (c). Figures reproduced with permission.\cite {eguchi2010resonant} Copyright 2010 Japan Physical Society.}
 \label{Figure_13}
\end{figure*}

Moreover, \textbf {Figure \ref {Figure_13}} shows the XAS and resonant PES spectra of $\rm Ag_2NiO_2$ and $\rm Ag_2MnO_2$. \textbf {Figures \ref {Figure_13}a and \ref {Figure_13}d} respectively depict the $\rm Ni$ 2$p$ and $\rm Mn$ 2$p$ XAS spectra. The $\rm Ni$ 2$p$ XAS spectrum of $\rm Ag_2NiO_2$ exhibits a strong resemblance to that of $\rm Ni^{3+}$ in $\rm {\it R}NiO_3$ ($R$ = rare earth metals) compounds,\cite {wadati2005valence} thereby indicating the presence of trivalent $\rm Ni^{3+}$ in $\rm Ag_2NiO_2$, as corroborated by first-principles calculations. In contrast, the $\rm Mn$ 2$p$ XAS spectrum of $\rm Ag_2MnO_2$ can predominantly be attributed to $\rm Mn^{3+}$ with the exception of a minor peak at 641 eV (marked as ` D '), which is absent in the $\rm Mn_2O_3$ ($\rm Mn^{3+}$) spectrum. To account for this small peak, the XAS spectrum of $\rm Ag_2MnO_2$ was simulated by combining the experimental $\rm MnO$ ($\rm Mn^{2+}$) and $\rm Mn_2O_3$ ($\rm Mn^{3+}$) XAS spectra. By employing a simulated $\rm MnO$ + $\rm Mn_2O_3$ XAS spectrum comprising 30\% $\rm MnO$ ($\rm Mn^{2+}$) and 70\% $\rm Mn_2O_3$ ($\rm Mn^{3+}$), the experimental spectrum of $\rm Ag_2MnO_2$ was accurately reproduced.\cite {eguchi2010resonant}

Resonant PES spectra, acquired using the photon energies indicated in the $\rm Ni$ and $\rm Mn$ 2$p$ XAS spectra, are presented in \textbf {Figures \ref {Figure_13}b and \ref {Figure_13}e}, respectively. The spectra were normalised with respect to the incident photon current and the number of scan times. \textbf {Figures \ref {Figure_13}c and \ref {Figure_13}f} exhibit the difference spectra obtained by subtracting the `off' resonance spectra (labeled as `A') from each spectrum, aimed at highlighting the $\rm Ni$ and $\rm Mn$ 3$d$ density of states (DOS), respectively. The insets in \textbf {Figures \ref {Figure_13}c and \ref {Figure_13}f} provide an expanded view near the Fermi energy ($E_F$). The difference spectra between the `on' and `off' resonances in $\rm Ag_2NiO_2$ (marked as ` E ' in \textbf {Figure \ref {Figure_13}c}) and $\rm Ag_2MnO_2$ (marked as ` F ' in \textbf {Figure \ref {Figure_13}f}) reflect the partial DOS ($t_{2g}$ and $e_g$) of $\rm Ni$ and $\rm Mn$ 3$d$, respectively. The PES measurements reveal that the DOS at $E_F$ encompasses the $\rm Ni$ and $\rm Mn$ 3$d$ characteristics, as displayed in the enlarged insets in \textbf {Figures \ref {Figure_13}c and \ref {Figure_13}f}. Furthermore, the results suggest a strong correlation between the observed significant mass enhancement at low temperatures in resistivity and specific heat measurements and the hybridised $\rm Ni$ and $\rm Mn$ 3$d$ states at $E_F$. Overall, the core level XAS and PES spectra indicate the presence of $\rm Mn^{3+}$ states mixed with $\rm Mn^{2+}$ states in $\rm Ag_2MnO_2$ and predominantly trivalent $\rm Ni^{3+}$ states in $\rm Ag_2NiO_2$. Based on electroneutrality, the valency states of silver were calculated to be $\rm Ag^{1/2+}$ and $\rm Ag^{0.65+}$ ($\rm Ag^{2/3+}$) in $\rm Ag_2NiO_2$ and $\rm Ag_2MnO_2$, respectively, where the $2/3+$ subvalency under $\rm SU(2)\times U(1)$ is only observed in hybrids (alternating silver monolayered-bilayered materials).\cite{masese2023honeycomb} 

\begin{figure*}[!t]
\centering
\includegraphics[width=0.9\columnwidth]{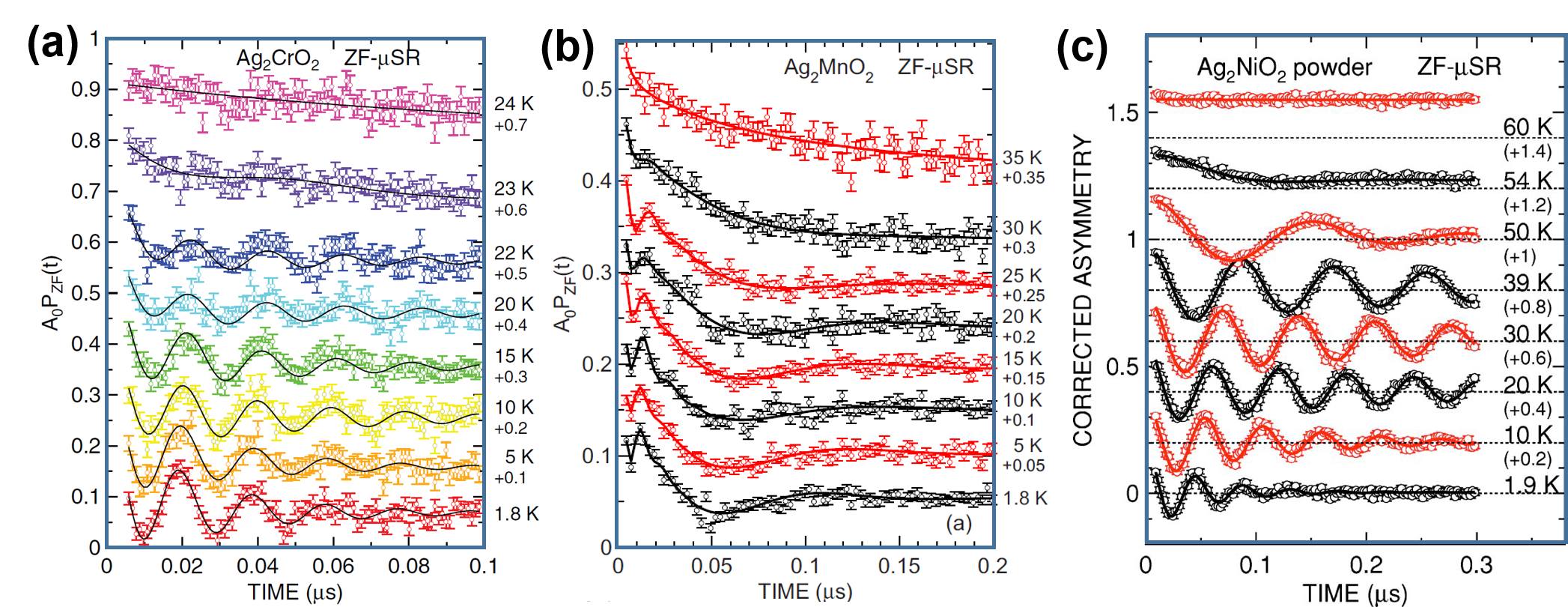}
\caption{Magnetic properties of $\rm Ag_2CrO_2$, $\rm Ag_2MnO_2$ and $\rm Ag_2NiO_2$. ZF $\mu^{+}$ SR spectra attained at various temperatures for (a) $\rm Ag_2CrO_2$, (b) $\rm Ag_2MnO_2$ and (c) $\rm Ag_2NiO_2$. Figures reproduced with permission.\cite {sugiyama2009static, sugiyama2006Incommensurate} Copyright 2014 Elsevier, Copyright 2009 Elsevier, 2013 Royal Society of Chemistry and Copyright 2006 American Physical Society, respectively.}
\label{Figure_14}
\end{figure*}
Utilising X-ray absorption spectroscopy at $\rm Ni$ $\rm {\it K}$, $\rm Ag$ $\rm {\it L}_{III}$, and $\rm O$ $\rm {\it K}$ edges, and employing X-ray photoelectron spectroscopy, with focus on the binding energies of $\rm Ni$ 2$p$ and $\rm Te$ 3$d$ orbitals, discerning insights emerged. These investigations unveiled the specific valency states of the $\rm Ag$ bilayer domains within the global $\rm Ag_2Ni_2TeO_6$ composition to be unequivocally ascribed as $\rm Ag_6^{1/2+}Ni_2^{2+}Te^{5+}O_6$.\cite {masese2023honeycomb}

Positive muon spin rotation and relaxation ($\mu^{+}$SR) spectroscopy has been extensively employed in elucidating the magnetic characteristics of layered $\rm Ag_2{\it M}O_2$ compounds (where $M$ represents $\rm Ni, Mn,$ or $\rm Cr$), as shown in \textbf {Figure \ref {Figure_14}}.\cite {sugiyama2009static} \textbf {Figure \ref {Figure_14}a} shows the $\mu^{+}$SR spectra of $\rm Ag_2CrO_2$, obtained at various temperatures. Below the N\'eel temperature ($T_{\rm N}$ = 24 K), the zero-field spectrum exhibits a conspicuous muon-spin precession signal, thereby providing conclusive evidence for the emergence of static antiferromagnetic order. Furthermore, it was ascertained that the internal field remained temperature invariant, with the exception of the vicinity of $T_{\rm N}$, mirroring the behaviour observed in the susceptibility-temperature curve of $\rm Ag_2CrO_2$. This observation has lent support to the notion that the antiferromagnetic transition is induced by a first-order structural phase transition at $T_{\rm N}$. By merging the outcomes of $\mu^{+}$SR measurements with first-principles calculations, it was deduced that $\rm Ag_2CrO_2$ adopts a partially disordered antiferromagnetic state, which constitutes the most plausible spin structure for this layered compound. Based on the similarity in the ZF spectrum for $\rm Ag_2MnO_2$ and $\rm Ag_2NiO_2$ obtained at the lowest temperature measured (\textbf {Figures \ref {Figure_14}b and \ref {Figure_14}c}), it is postulated that the antiferromagnetic spin arrangement of $\rm Ag_2MnO_2$ bears resemblance to that of $\rm Ag_2NiO_2$.

In addition to investigating the magnetic properties of $\rm Ag_2{\it M}O_2$ compounds, $\mu^{+}$SR spectroscopy has been employed to verify the magnetotransport characteristics of $\rm Ag_2NiO_2$. The $\mu^{+}$SR measurements have substantiated the presence of two precession frequencies, attributed to the existence of a stationary internal magnetic field below the N\'eel temperature ($T_{\rm N}$ = 56 K). By analysing the delay in the initial phase of the precession signal, it was postulated that the ground state of $\rm Ag_2NiO_2$ corresponds to an incommensurate spin density wave state. This proposition represents an alternative scenario for the coexistence of antiferromagnetic order and metallic conductivity.

\subsection{Computational Approaches}
Computational modelling techniques offer unique insights into the intricate atomic-scale mechanisms governing the electronic structure and physicochemical properties of materials, rendering them indispensable tools for materials design. The foremost objective and advantage of computational modelling in robust materials design lie in its capacity to supplement and enhance experimental analyses by unveiling fundamental atomic-scale properties and mechanisms that are challenging to obtain solely through empirical measurements. Notably, atomic-scale computational modelling methods, such as first-principles density functional theory (DFT), have been extensively employed to comprehensively comprehend and predict various properties of layered frameworks that exhibit intriguing metallophilic bilayers.

Density functional theory (DFT) calculations have been employed to validate the valency state of $\rm Co$ ions in $\rm Ag_2CoO_2$, confirming their 2+ state, and to determine that a low-spin state with $S$=1/2 can be achieved by reducing the on-site Coulomb interaction on $\rm Co$.\cite {yoshida2020static} In the case of $\rm Ag_2{\it M}O_2$ ($M$ = $\rm Ni, Mn,$ and $\rm Cr$), the transition metal ions predominantly exhibit a valency state of 3+. If this also holds true for $\rm Ag_2CoO_2$, the magnetic $\rm Co^{3+}$ ions should exhibit high-spin state ($S=2$, $e^2_{g}t^4_{2g}$), intermediate-spin state ($S=1$, $e^1_{g}t^5_{2g}$), or low-spin state ($S=0$, $t^6_{2g}$), depending on the balance between crystal field splitting and the strength of Hund's coupling. Experimental magnetic susceptibility measurements indicate that the effective paramagnetic moment of Co spins closely aligns with the theoretical value for $S=1/2$,\cite {yoshida2020static} a spin state not typically anticipated for the electron configuration of $\rm Co^{3+}$ ions. To gain further insights, DFT calculations have been employed,\cite {yoshida2020static} taking into account the Coulomb interaction ($U$), Hund's coupling, and crystal field splitting. These calculations reveal that $\rm Co^{2+}$ is stabilised in $\rm Ag_2CoO_2$, and the emergence of the low-spin state ($t^6_{2g}e^1_{g}$) on $\rm Co^{2+}$ with a relatively small on-site $U$ explains the observed $S=1/2$ state of $\rm Co$ ions in $\rm Ag_2CoO_2$.

First-principles calculations have been employed to elucidate the underlying factors contributing to the $\rm Ag$ subvalency observed in $\rm Ag_2NiO_2$.\cite {johannes2007formation} These calculations have revealed that the emergence of $\rm Ag$ subvalency ($\rm Ag^{1/2+}$) arises from a robust bonding-antibonding interaction occurring between the $\rm Ag$ bilayers. This interaction effectively pushes a significant portion of the bonding $\rm Ag$ $sp$ bands below both the $\rm Ni$ $d$ orbitals and a majority of the $\rm O$ $p$ orbital states.\cite {johannes2007formation} As a result, the conventional $\rm Ag^{1+}$ valency state is precluded, as the electronic structure indicates the absence of such a configuration. Moreover, the valency state of $\rm Ni$ was affirmed to be in the trivalent state ($\rm Ni^{3+}$), corresponding not to an electronic ground state of $3d^7$, but rather to a $3d^8\underline{L}$ electronic ground state (where $\underline{L}$ represents a ligand hole). Cluster model calculations have affirmed the different $\rm Ni$ bands observed in the core level and valence band photoelectron spectra of $\rm Ag_2NiO_2$ to comprise ground states of $d^7$ and $d^8\underline{L}$. \cite {wedig2006studies}

\begin{figure*}[!t]
 \centering
 \includegraphics[width=0.9\columnwidth]{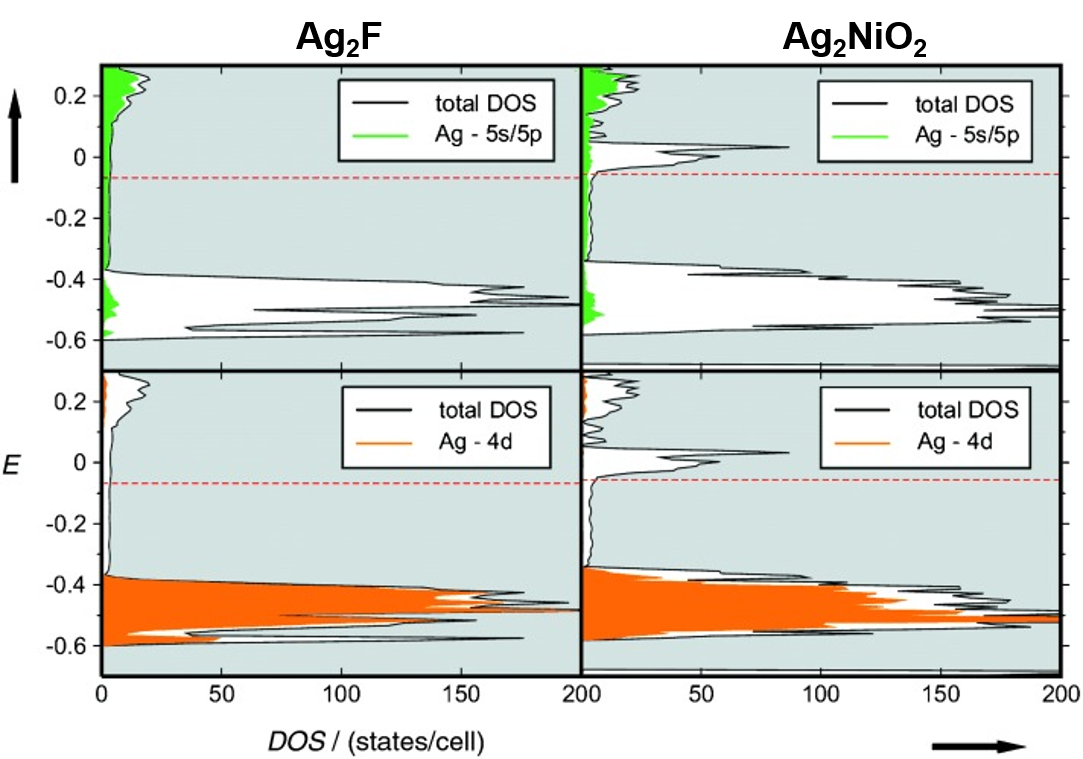}
 \caption{Calculated density of states (DOS) of $\rm Ag_2F$ and $\rm Ag_2NiO_2$. Figures reproduced with permission.\cite {schreyer2002synthesis} Copyright 2002 Wiley-VCH.}
 \label{Figure_15}
\end{figure*}

The atypical distribution of valency states observed in $\rm Ag_2NiO_2$ has been interpreted by considering the stronger ligand-field stabilisation exhibited by low-spin $\rm Ni^{3+}$ centers compared to $\rm Ni^{2+}$ centers in octahedral coordination. Additionally, the presence of low-lying silver 5$s$ and 5$p$ bands, which can accommodate excess electron density, has been reported to further contribute to the observed valency state distribution.\cite {schreyer2002synthesis} The density of states (DOS) analysis, performed at the Hartree-Fock level, reveals similar characteristics between $\rm Ag_2NiO_2$ and $\rm Ag_2F$ (\textbf {Figure \ref {Figure_15}}). Specifically, the DOS projections onto the 4$d$ orbitals of silver, as well as O 2$p$ and Ni 3$d$ contributions, reside significantly below the Fermi level. In the energy range around the Fermi level, dispersed bands are predominantly associated with Ag 5$s$/5$p$ states. Consequently, the DOS and their projections in $\rm Ag_2NiO_2$ closely resemble those observed in $\rm Ag_2F$.\cite {schreyer2002synthesis}

\newpage

\section{\label{Section: Theoretical_Characterisations} Novel Theoretical Characterisations of the Metallophilic Bilayers}

\subsection{Optimised 2D Lattices}

Honeycomb layered oxides with a large and diverse compositional space typically of the form $A_aM_mL_l\rm O_6$ where $A = \rm Ag, Cu, Na, K, Li, {\it etc}.$, $M$ = $s$-block metal atoms \textit {e.g.} Mg or transition metal atoms (\textit {e.g.}, Co, Ni, \textit {etc}.), $L$ = pnictogen, chalcogen or transition metal atoms and $0\leq a \leq4$, $0\leq m \leq2$, $0\leq l \leq 1$ have been shown to either exhibit the Face-Centred Cubic (FCC) or the Hexagonal-Close Packing (HCP) lattices.\cite{kanyolo2021honeycomb, kanyolo2022advances2} These lattices are known to be the only viable options for the most densely-packed lattice structures in three dimensions (3D). In other words, such honeycomb layered oxides meet the criteria for the most densely-packed lattice configurations in the congruent sphere packing problem.\cite{hales2011revision, viazovska2017sphere, zong2008sphere, cohn2017sphere, cohn2009optimality, cohn2014sphere} Moreover, the building blocks for the FCC and HCP lattices are three/two distinct lattice sites of the two dimensional (2D) hexagonal packing superimposed onto each other along the $c$-axis. The fact that the 2D slices are hexagonal lattices means that the congruent sphere packing condition for the most densely packed atomic structures is satisfied by honeycomb layered oxides both in 2D and 3D. Meanwhile, each hexagonal lattice can be labelled by $U$, $V$ or $W$ representing the corresponding atomic lattice sites. For instance, within the aforementioned class of honeycomb layered oxides, the lattice sites corresponding to the sphere centres of atoms in the silver-based honeycomb layered tellurate, ${\rm Ag_2}M_2\rm TeO_6$ ($M = \rm Ni, Co, Cu, Zn, Mg$ \textit{etc}) with $\rm Ag$ linearly/dumbbell 
or prismatically coordinated to $\rm O$ atoms respectively can be represented in the FCC/HCP notation as\cite{kanyolo2022advances2},
\begin{subequations}\label{FCC_HCP_eq}
\begin{align}
   U_{\rm O}V_{(M, M, {\rm Te})}\underbrace{W_{\rm O}W_{\rm Ag}W_{\rm O}}_{\rm linear\,\,coordination}V_{(M, M, {\rm Te})}U_{\rm O},\\
    U_{\rm O}V_{(M, M, {\rm Te})}\underbrace{W_{\rm O}V_{\rm Ag}W_{\rm O}}_{\rm prismatic\,\,coordination}V_{(M, M, {\rm Te})}U_{\rm O},
\end{align}
\end{subequations}
thus showcasing their lattice optimisation. Here, the subscripts are the respective atoms in the sites. It is prudent to note that each lattice site is comprised of a triplet of hexagonal sub-lattice sites $u$, $v$ or $w$ written in the notation as $U_{u, v, w}$, $V_{u, v, w}$ and $V_{u, v, w}$, which allows for the identification of bipartite atomic lattice sites such as $V_{M, M, {\rm Te}}$ in eq. (\ref{FCC_HCP_eq}) where $u = M$ and $v = M$ form the bipartite honeycomb lattice comprised of $M$ atoms.

From a mathematical physics point of view, valence bond theory (indispensable in computational approaches to understanding the various morphologies and topologies exhibited by honeycomb layered oxides) only serves to determine the precise relative distances between atoms constrained by strict sphere packing optimisation conditions. Provided we are only interested in understanding geometric and topological features in honeycomb layered oxides, valence bond theoretic approaches become less relevant. Moreover, focusing on the 2D slices, it turns out that the partition function for the congruent sphere packing problem in honeycomb layered oxides can be written as the linear programming bound partition function for the spinless modular bootstrap for conformal field theories (CFTs) under the algebra ${\rm U(1)}^c$ given by\cite{afkhami2020high, hartman2019sphere},
\begin{align}\label{Z_eq}
    Z_{\Lambda}(\beta) = \sum_{\vec{r} \in \Lambda}\frac{\exp(-\pi \beta r^2)}{\eta^{2c}(i\beta)},
\end{align}
where $\beta$ is a complex parameter, $c$ is the central charge related to the spatial dimensions by $d = 2c$,
\begin{align}
    \eta(i\beta) = \prod_{N = 1}^{\infty}\frac{\exp(-\pi\beta/12)}{1 - \exp(-\beta 2\pi N)} = q^{\frac{1}{24}}(\tau)\prod_{N = 1}^{\infty}(1 - q^N(\tau)) = \eta(\tau),
\end{align}
is the Dedekind eta function and the sum is over appropriately normalised displacement vectors $\vec{r} \in \Lambda$ connecting sphere centres within the lattice $\Lambda$ with $r^2 = |\vec{r}|^2$ the norm of the vectors. After a redefinition $\tau = i\beta$ and $q(\tau) = \exp(2\pi i\tau)$, we identify the famous Dedekind eta function $\eta(\tau)$. The Dedekind eta function, $\eta(\tau)$ and the lattice theta function, $\theta_{\Lambda}(\tau) = Z_{\Lambda}(\tau)\eta^{2c}(\tau)$ respectively transform as\cite{siegel1954simple, hartman2019sphere, kanyolo2022advances2},
\begin{align}\label{S_eq}
    \eta(-1/\tau) = \sqrt{-i\tau}\eta(\tau),\,\,\,\,\,\theta_{\Lambda^*}(-1/\tau) = |\det(\Omega)|\sqrt{-i\tau}\theta_{\Lambda}(\tau). 
\end{align}
under the group of modular symmetries, $S, T \in SL_2(\mathbb{Z})$, where $|\det(\Omega)|$ is the volume of the fundamental domain (primitive/unit cell) of $\Lambda$ written here as the determinant of the Gram matrix $\Omega$ constructed from the unit basis vectors,  $\Lambda^*$ is the dual lattice and\cite{kanyolo2022cationic},
\begin{align}\label{gen_eq}
S = \begin{pmatrix}
0 & -1\\ 
1 & 0
\end{pmatrix},\,\, T = \begin{pmatrix}
1 & 1\\ 
0 & 1
\end{pmatrix},
\end{align}
are the generators of the modular group (\textit{i.e.} the special linear group of matrices with integer entries $\mathbb{Z}$ and determinant $\pm 1$. The determinant is chosen to be $+1$, which selects the projective special linear group $PSL_2(\mathbb{Z}) \equiv SL_2(\mathbb{Z})/\{\pm 1\}$, thus restricting $\tau$ to the upper-half of the complex plane: $\tau \in \mathbb{H}_+$). Since dynamics of a physical system remain invariant when the partition function is multiplied by a constant (corresponding to shifting the vacuum energy by a constant), we shall take \textit{a priori} $\det(\Omega) = \pm 1$. Strictly, this assumption is not valid due to fact that even unimodular lattices (\textit{i.e.} lattices where $\det{\Omega} = \pm 1$ and the norm is even, $r^2/2 = n \in \mathbb{Z}_+$) only exist in dimensions divisible by $8$.\cite{kanyolo2022advances2} Nonetheless, the vacuum energy argument emancipates our approach from this constraint. Thus, using eq. (\ref{S_eq}) in eq. (\ref{Z_eq}) we find $Z_{\Lambda}(1/\beta) = Z_{\Lambda^*}(\beta)$, and taking the lattice $\Lambda$ to be even, we can obtain $Z_{\Lambda}(\beta - 24i/d) = Z_{\Lambda}(\beta)$ where these two transformations are generated by $S$ and $T^d$ matrices respectively. 

Notably, linear programming bounds using the partition function in eq. (\ref{Z_eq}) have been computed, which yield the hexagonal lattice ($\Lambda = hx$) as the only optimised lattice in $d = 2$ dimensions, with other significant cases in $d = 8$ and $d = 24$ dimensions corresponding to the $E_8$ and Leech lattices respectively.\cite{hartman2019sphere} Thus, whenever the lattice is bipartite (\textit{e.g.} the honeycomb lattice ($hc$)), one possible construction for the partition function is given by the algebra ${\rm U(1)}^c\times {\rm U(1)}^c$,
\begin{align}\label{bosonic_Z_eq}
    Z_{|\Lambda|^2}(\beta, \overline{\beta}) = Z_{\Lambda}(\beta)\times Z_{\overline{\Lambda}}(\overline{\beta}) = \sum_{\vec{r} \in \Lambda}\frac{\exp(-\pi \beta r^2)}{\eta^{2c}(i\beta)} \times\sum_{\vec{\overline{r}} \in \overline{\Lambda}}\frac{\exp(-\pi\overline{\beta}\overline{r}^2)}{\eta^{2c}(-i\overline{\beta})},
\end{align}
where $|\Lambda|^2 = \Lambda \times \overline{\Lambda}$ is a useful notation for the bipartite lattice constructed from its constituent lattices $\Lambda$ and $\overline{\Lambda}$ with $\vec{e}_i \in\Lambda$ and $\vec{\overline{e}}_i \in \overline{\Lambda}$ the unit basis vectors of the respective lattices. In the definition of $Z_{\overline{\Lambda}}(\overline{\beta})$, we have used, 
\begin{align}
    \overline{\eta}(\overline{\tau}) = \overline{q}^{\frac{1}{24}}(\tau)\prod_{N = 1}^{\infty}(1 - \overline{q}^N(\overline{\tau})) = \eta(-i\overline{\beta}), 
\end{align}
where $\overline{q}(\overline{\tau}) = \exp(-i\overline{\tau})$ and $\overline{\tau} = i\overline{\beta}$. The Gram matrices $\Omega$ and $\overline{\Omega}$ defining the volume of the fundamental domain by their determinants are respectively given by, 
\begin{align}
    \Omega = \left (\prod_{i = 1}^{d}\vec{e}_i \right ) = (\vec{e}_1, \cdots, \vec{e}_{i = d}), \,\,\, \overline{\Omega} = \left (\prod_{i = 1}^{d}\vec{\overline{e}}_i\right ) = (\vec{\overline{e}}_1, \cdots, \vec{\overline{e}}_{i = d}).
\end{align}
For instance, in $d = 2c = 2$ dimensions, the unit basis vectors of the hexagonal lattices ($\Lambda = hx, \overline{\Lambda} = \overline{hx}$) comprising the bipartite honeycomb lattice $|\Lambda|^2 = hc$ correspond to, 
\begin{align}
    \vec{e}_1 = (1/2, -\sqrt{3}/2)^{\rm T},\,\, \vec{e}_2 = (1/2, +\sqrt{3}/2)^{\rm T},\,\,\,\, \vec{\overline{e}}_1 = (1/2, -\sqrt{3}/2)^{\rm T},\,\, \vec{\overline{e}}_2 = (1/2, +\sqrt{3}/2)^{\rm T},
\end{align}
which yield the identical Gram matrices, 
\begin{align}\label{Omega_eq}
\Omega = \begin{pmatrix}
1/2 & 1/2\\ 
-\sqrt{3}/2 & +\sqrt{3}/2
\end{pmatrix},\,\,\,\, 
\overline{\Omega} = \begin{pmatrix}
1/2 & 1/2\\ 
-\sqrt{3}/2 & +\sqrt{3}/2
\end{pmatrix},
\end{align}
with determinant, $\det(\Omega) = \det(\overline{\Omega}) = \sqrt{3}/2$. Thus, we can re-scale each unit basis vector by the factor $\sqrt{2/\sqrt{3}}$ in order to guarantee  $\det(\Omega) = \det(\overline{\Omega}) = 1$.

Moreover, since the hexagonal lattice is now self-dual ($hx = hx^*$), we find a useful operation under the $S$ transformation or its inverse $S^{-1}$ given by $Z_{hx}(1/\beta) = Z_{hx^*}(\beta) = Z_{hx}(\beta)$ and $Z_{\overline{hx}}(1/\overline{\beta}) = Z_{\overline{hx^*}}(\overline{\beta}) = Z_{\overline{hx}}(\overline{\beta})$ respectively. Moreover, under the $T$ transformation or its inverse, $T^{-1}$, we obtain $Z_{hx}(\beta \pm i) = \exp(\pm i\pi/12)Z_{hx}(\beta)$ and $Z_{\overline{hx}}(\overline{\beta} \pm 1) = \exp(\mp\pi/12)Z_{\overline{hx}}(\overline{\beta})$ (we have used $r^2/2 = n$ and $\overline{r}^2/2 = \overline{n}$ with $\overline{n}, n \in \mathbb{Z}_+^2$ given by positive integers). Since the two phases $\exp(\pm \pi/12)$ and $\exp(\mp \pi/12)$ cancel each other out, the partition function for the bipartite honeycomb lattice is invariant under both $S$ (or $S^{-1}$) and $T$ (or $T^{-1}$) transformations. It was found that, such modular transformations on the honeycomb lattice can be employed to define a geometric spin degree of freedom identified as pseudo-spin.\cite{kanyolo2022cationic, kanyolo2022advances2} We shall refer to the construction of the partition function in  eq. (\ref{bosonic_Z_eq}) as bosonic, which reflects the fact of this modular invariance. On the contrary, the partition function of a fermionic lattice is defined by\cite{kanyolo2022advances2}, 
\begin{align}\label{fermionic_Z_eq}
    Z(\beta) = Z_{\Lambda}(\beta) + Z_{\overline{\Lambda}}(\overline{\beta}) = \sum_{\vec{r} \in \Lambda} \frac{\exp(-\pi \beta r^2) + \exp(\pi \beta r^2)}{\eta^{2c}(i\beta)} \equiv \sum_{n = 0}^{\infty}2b_n\exp(\beta J(\beta))\cosh(2\pi\beta n) \equiv \sum_{n = 0}^{\infty} b_nZ(n, \beta),
\end{align}
where we have set $\beta + \overline{\beta} = 0$, $\Lambda = \overline{\Lambda}$ and $r^2 = \overline{r}^2$, and introduced the functions,
\begin{align}
    Z(n, \beta) = 2\exp(\beta J(\beta))\cosh(2\pi\beta n),\,\,\,\,J(\beta) = -2c\ln (\eta(i\beta))/\beta,
\end{align}
with $b_n$ the number of vectors of norm $r^2/2 = n$ in the lattice $\Lambda$. For the hexagonal lattice, $\Lambda = hx$ corresponding to $d = 2$ dimensions, $b_n$ is given by the number of integral solutions of the Diophantine equation\cite{Sloane1964Theta}, 
\begin{align}\label{Diophantine_eq}
   |x + y\exp(\pi i/3)|^2 = x^2 + y^2 - xy = n.
\end{align}
This fact can be appreciated by checking that,
\begin{align}
    r^2 = 2|x\vec{e}_1 - y\vec{e}_2|^2 = 
2\left |\begin{pmatrix}
1/2 & 1/2\\ 
-\sqrt{3}/2 & \sqrt{3}/2
\end{pmatrix}\vec{R}\right |^2 = 
    2|\Omega \vec{R}|^2 = 2x^2 + 2y^2 - 2xy = x^2 + y^2 + z^2= 2n,
\end{align}
where $\vec{R} = (x, -y)^{\rm T}$ and $z = x - y$. Evidently, the $S$ transformation $Z(1/\beta) = Z(\beta)$ is also manifest on the fermionic lattice, whereas the $T$ transformation is not. Since the sum $\beta + \overline{\beta} = 0$ is set to vanish, $\beta$ and $\overline{\beta}$ must transform by $S, T$ and $S^{-1}, T^{-1}$ respectively as $\beta \rightarrow 1/\beta, \beta \rightarrow \beta - i24/d$ and $\overline{\beta} \rightarrow 1/\overline{\beta}, \overline{\beta} \rightarrow \overline{\beta} + i24/d$, which identifies the transformation of $\eta(\tau)$ in the denominator under $T$ as the problematic term for $d \neq 24n$ ($n = \mathbb{Z}_+$). In mathematics, the connection between the coefficients of complex functions such as $Z(\beta)$ with modular symmetry and the number of solutions of polynomials (over finite number fields) such as eq. (\ref{Diophantine_eq}) is a subject of active research under the Langlands program.\cite{lmfdb} A famous example is the \textit{modularity theorem}, which implies \textit{in part} that a modular form $f(-1/\tau) = \tau^k f(\tau)$ ($k \in \mathbb{Z}_+$ is a positive integer known as the weight of the modular form) can always be defined with $k = 2$ for Elliptic curves (\textit{e.g.} the polynomial of the Weierstra\ss\,\,form, $y^2 = 4x^3 - g_2x - g_3$, with constants $g_2$ and $g_3$, and a non-vanishing discriminant, $\Delta = g_2^3 - 27g_3^2 \neq 0$), provided one considers solutions over the field of rational numbers \textit{modulo} prime numbers and/or integers.\cite{breuil2001modularity} 

Proceeding, we can now introduce the pseudo-spin model, where the problematic term $J(\beta)$ corresponds to the off-diagonal elements,
\begin{align}\label{J_eq}
J_{ij}(\beta) = \frac{1}{2}
\begin{pmatrix}
0 & J(\beta)\\ 
J(\beta) & 0
\end{pmatrix},
\end{align}
of a pseudo-spin correlation term in the two-spin Ising model\cite{kanyolo2022advances2}, 
\begin{align}\label{Ising_H_eq}
    H_{\rm Ising}(h(n)) = -\sum_{i,j}J_{ij}(\beta)S_i^zS_j^z - h(n)\sum_{i = 1}^2S_i^z, 
\end{align}
where $S_i^z, S_j^z$ are the $z$-component pseudo-spin operators, $h(n) = 2\pi n$ is the pseudo-magnetic flux and $J(\beta)$ is the Heisenberg term that depends on the parameter $\beta$, now interpreted to be the inverse temperature. The exact solution for the partition functions $Z(n, \beta)$ correspond to, 
\begin{align}\label{Tr_exacT_sol_eq}
    {\rm Tr}\left \{\exp\left (-\beta H_{\rm Ising}(n) \right )\right \} = Z(n, \beta) = \lambda_+(n, \beta) + \lambda_-(n, \beta) = \exp(\beta J(\beta))\cosh(2\pi\beta n),
\end{align}
where $\lambda_1, \lambda_2$ are the eigenvalues\cite{kanyolo2022cationic},
\begin{align}
    \lambda_{\pm}(n, \beta) = \exp(\beta J(\beta))\cosh(2\pi\beta n)\pm \sqrt{\left (\exp(\beta J(\beta))(\sinh(2\pi\beta n))^2 + \exp(-2\beta J(\beta))\right )},
\end{align}
and the trace ${\rm Tr}$ is performed over the pseudo-spins $S_1^z$ and $S_2^z$. It is clear that each partition function $Z(n, \beta)$ must subsequently be weighted by $b_n$ in a final trace over the positive integers $n$ as in eq. (\ref{fermionic_Z_eq}) to recover $Z(\beta)$. Finally, since the fluxes in the pseudo-spin model are integers, we can introduce the concept of Chern numbers\cite{kanyolo2021partition}, 
\begin{align}\label{old_condition_eq}
    2\pi n = \int_0^{\beta} dt_{\rm E}\,\,\langle \Psi|i\partial_{t_{\rm E}}|\Psi \rangle = \oint d\vec{x}\cdot\langle \Psi|i\vec{\nabla}_{\vec{x}}|\Psi \rangle = \int d^2x\,\,\vec{n}\cdot\vec{\mathcal{B}}(\vec{x}),
\end{align}
where $\mathcal{B}(\vec{x}) = \vec{\nabla}_{\vec{x}}\times\langle \Psi|i\vec{\nabla}_{\vec{x}}|\Psi \rangle$ is the pseudo-magnetic field, $\vec{n}$ is the vector normal to the surface of integration, $t_{\rm E} = -it$ is Euclidean time obtained by Wick rotation, the trajectories $\vec{x}(t_{\rm E} + \beta) = \vec{x}(t_{\rm E})$ are considered periodic in $\beta$ and $d^2x\,\,\vec{n}$ is the area element perpendicular to the pseudo-magnetic field (in the direction of the unit vector $\vec{n}$). The nature of such a pseudo-magnetic field and the wave function $\Psi$ is considered in the subsequent subsections.

\subsection{Two-dimensional (2D) Liouville Conformal Field Theory}

The 2D slices of $A$ cations comprising the $ab$ planes of honeycomb layered oxides suchlike $A_aM_mL_l\rm O_6$, where $A = \rm Ag, Cu, Na$, $\rm Li, K$, \textit{ etc.}, $M$ = $s$-block metal atoms \textit {e.g.} Mg or transition metal atoms (\textit {e.g.}, Co,Ni, \textit {etc}.), $L$ = pnictogen, chalcogen or transition metal atoms and $0\leq a \leq4$, $0\leq m \leq2$, $0\leq l \leq 1$ have the property that $A$ cations can be rendered mobile when a sufficient electric field, $\vec{E}$ is applied supplying the activation energy. This process of extraction/re-insertion (deintercalation/intercalation) can be exploited in cathode technologies for secondary battery applications.\cite{kanyolo2020idealised, kanyolo2021honeycomb, kanyolo2022advances2, kanyolo2022advances, kanyolo2022cationic} In the theoretical description of the (de)intercalation process, the diffusion dynamics of the $A$ cations in a single crystal of $A_aM_mL_l\rm O_6$ are described by a Langevin equation, 
\begin{align}
    \vec{n}\times\frac{d^2\vec{x}}{dt^2} = -2\vec{p} + \vec{n}\times\vec{\nabla}\Phi,
\end{align}
where $\vec{n} = (0, 0, 1)$ is the unit vector normal to the $ab$ plane, $\vec{p} = \langle \Psi|i\vec{\nabla}|\Psi \rangle$ is the momentum of a diffusing $A$ cation with its wave function $\Psi$ appearing in eq. (\ref{old_condition_eq}), $\Phi$ is the geometric potential energy in the material governing the diffusion process whose gradient is proportional to the electric field, $\vec{\nabla}\Phi \propto \vec{E}$ and the subscript in the gradient, $\vec{\nabla}_{\vec{x}}$ is dropped for convenience. We can integrate over a closed diffusion trajectory to yield, 
\begin{align}\label{int_Langevin_eq}
    \int d\vec{x}\cdot\vec{n}\times\frac{d^2\vec{x}}{dt^2} - \int d\vec{x}\cdot\vec{n}\times\vec{\nabla}\Phi = \int dt\,k_{\rm g} + \int d^2x\,\vec{\nabla}^2\Phi = -2\int d\vec{x}\cdot\vec{p} = -4\pi n,
\end{align}
where $k_{\rm g} = (d\vec{x}/dt)\cdot\vec{n}\times d^2\vec{x}/dt^2$ is the geodesic curvature and $\oint d\vec{x}\cdot\vec{p} = 2\pi n$ corresponding to eq. (\ref{old_condition_eq}) ($\vec{\mathcal{B}} = \vec{\nabla}\times\vec{p}$) can also be assumed to correspond to the so-called old quantum condition, where $n \in \mathbb{Z}_+$ takes on positive integers.\cite{ishiwara2017universal} Here, we have used Stoke's/Green's theorem to obtain the Laplacian $\nabla^2\Phi$. It is clear that, we can introduce a conformal metric, 
\begin{align}
    dt^2 = \exp(2\Phi(x,y))(dx^2 + dy^2) = g_{ab}dx^adx^b = \exp(2\Phi(x,y))\delta_{ab}dx^adx^b,
\end{align}
defining the manifold $\mathcal{A}$ enclosed by the trajectories $\partial \mathcal{A}$, and equipped with the conformal metric tensor $g_{ab} = \exp(2\Phi)\delta_{ab}$ and $\delta_{ab}$ the Kronecker delta. This means that the Langevin equation can be viewed as an It$\rm \hat{o}$ processes with the Fokker-Planck equation corresponding to the Liouville equation, 
\begin{align}
    \nabla^2\Phi = -K\exp(2\Phi) = -K\sqrt{\det(g_{ab}(\Phi))}, 
\end{align}
where $K$ is the Gaussian curvature of $\mathcal{A}$. Thus, the integrated Langevin equation (eq. (\ref{int_Langevin_eq})) yields the Gauss-Bonnet theorem,
\begin{align}\label{Gauss_Bonnet_eq}
    \int_{\partial \mathcal{A}} dt\,k_{\rm g} + \int_{\mathcal{A}} d^2x\,\sqrt{\det(g_{ab})}\,K = 2\pi \mathcal{E}(\mathcal{A}), 
\end{align}
with $\mathcal{E}(\mathcal{A}) = 2 - 2g = -2n$ the Euler characteristic of $\mathcal{A}$ ($g$ is the genus of $\mathcal{A}$). Thus, since $\Phi$ is related to the electric field $\vec{E}$, we can identify $n = g - 1$ as the number of diffusing cations or equivalently $g$ is the number of vacancies created by the electric field, $\vec{E}$. 

The famous \textit{Theorema Egregium} (which states that the \textit {Gaussian curvature of the surface is intrinsic to that surface and hence invariant under isometries}) alongside the Gauss-Bonnet theorem (eq. (\ref{Gauss_Bonnet_eq})) imply that Liouville theory defined by the Gaussian curvature does not depend on the embedding ($R^2$ surface) of the lattice theory of diffusing cations. Rather, both theories are related at the topological level by the Gauss-Bonnet theorem which yields $g = n + 1$, with $g$ the genus of the surface and $n$ the number of diffusing charges. This introduces a duality between cation diffusion on lattices in the $ab$ plane ($R^2$) and Liouville theory (interpreted as a theory of vacancies).\cite{kanyolo2022advances2} A close analogue of the Liouville equation is the 3D Newtonian gravity,
\begin{align}\label{Newton_eq}
    \nabla^2\Phi(x,y,z) = -2\pi \ell\rho(x,y,z),
\end{align}
where $\rho_{\rm 3D}(x,y,z) = -\frac{\ell}{2}\partial\rho(x,y,z)/\partial z$ and $\frac{\ell}{2}\rho(x,y,0)$ can be interpreted as the 3D vacancy number density and 2D pair-correlation function respectively, and $\ell$ the integration cut-off length along the $z$ direction where $-\frac{\ell}{2}\int_{\mathcal{A}}d^{\,2}x\,\,\rho(x,y,\ell) = 1$, with the derivative making it clear that we are dealing with a 2D slice via a 3D-to-2D projection. Consequently, the analogy follows from the well-known normalisation $\frac{\ell}{2}\int_{\mathcal{A}}d^{\,2}x\,\,\rho(x,y,0) = g - 1$ of the pair-correlation function\cite{kanyolo2022advances2} which yields, 
\begin{subequations}\label{pair_corr_eq}
\begin{align}
    \int_{\mathcal{M} \subset R^3}d^{\,3}x\,\,\rho_{\rm 3D}(x,y,z) = -\frac{\ell}{2}\int_{\mathcal{A}}d^{\,2}x\int_0^{\ell} dz\,\,\frac{\partial \rho(x,y,z)}{\partial z} = \frac{\ell}{2}\int_{\mathcal{A}}d^{\,2}x\,\,\rho(x,y,0) - \frac{\ell}{2}\int_{\mathcal{A}}d^{\,2}x\,\,\rho(x,y,\ell) = g,
\end{align}
where $\mathcal{M} = \mathcal{A}\times\ell$ is a manifold in $R^3$, $\mathcal{A}$ is the 2D manifold enclosed by a diffusion trajectory and $\ell$ is the 1D manifold along the $z$-axis within the interval $[0, \ell]$. Thus, we can identify the 2D vacancy number density to be given by,
\begin{align}
    \rho_{\ell}(x,y) \equiv \frac{\ell}{2}\rho(x,y,0) - \frac{\ell}{2}\rho(x,y,\ell),
\end{align}
\end{subequations}
where $\rho(x,y,z)$ serves as an interpolation function. Due to eq. (\ref{Newton_eq}), we can instantly write down its relativistic counterpart, which has certainly been considered elsewhere.\cite{kanyolo2020idealised, kanyolo2022advances2} Given the intriguing similarities and connections between the cations and their vacancies, it is prudent to make a formal connection between the two formalisms. In particular, we note that Liouville theory is essentially a 2D gravity theory, with $g_{ab}$ and $\Phi$ the dynamical fields, whereas the cationic theory is a topological 2D field theory of charged particles. The quantum versions of these two descriptions correspond to 2D quantum gravity and topological 2D quantum field theories respectively.\cite{kanyolo2020idealised, kanyolo2021honeycomb, kanyolo2022advances2, kanyolo2022cationic} 

\begin{figure*}[!t]
 \centering
 \includegraphics[width=0.9\columnwidth]{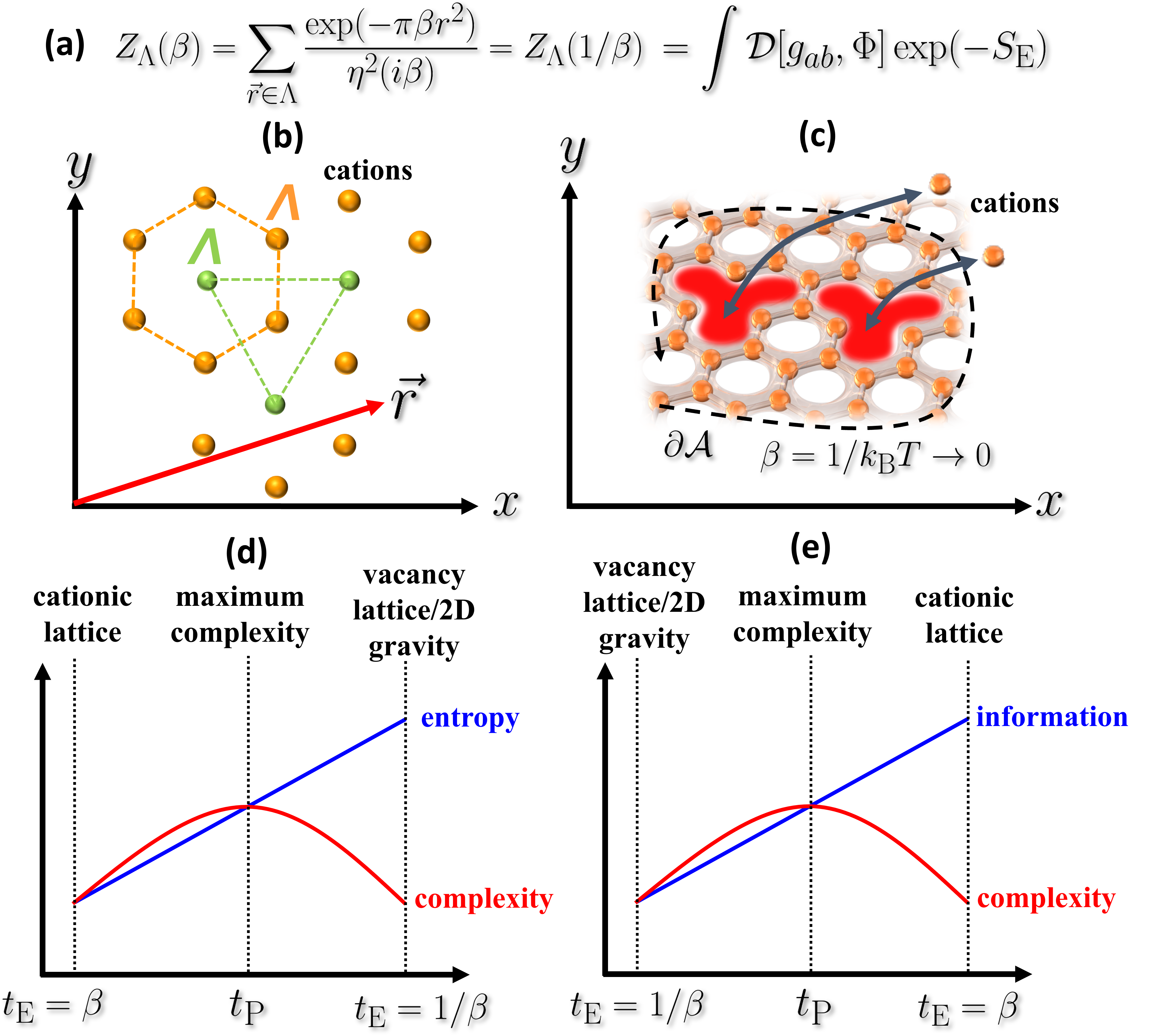}
 \caption{A depiction of the assumptions employed to bootstrap the path integral in eq. (\ref{QG_eq2}). (a) The equation suggesting the equivalence of the lattice partition function $Z_{\Lambda}(\beta) = Z_{\Lambda*}(1/\beta)$ and the emergent 2D gravity path integral given in eq. (\ref{QG_eq2}) with $\Lambda \equiv \Lambda^*$; (b) A depiction of the hexagonal (green) and honeycomb (orange) lattices of cations labelled by $\Lambda$. The cations must be assumed static and immobile from their relative positions to calculate $Z_{\lambda}(\beta)$; (c) The extraction of cations by raising the temperature $T = 1/k_{\rm B}\beta \rightarrow \infty$, depicting the boundary $\partial\mathcal{A}$ of the manifold $\mathcal{A}$; (d, e) A heuristic schematic of the periodicity given in eq. (\ref{Thermal_periodicity_eq}) of the thermal extraction (de-intercalation, (d)) and re-insertion (interaclation, (e)) processes as pertaining the entropy/information (blue) and the complexity measure (red), with $t_{\rm E} = it$ Euclidean time related to ordinary time $t$ by a Wick rotation and the `Page time' $t_{\rm P} = \frac{1}{2}(\beta + 1/\beta)$ labelling the Euclidean time it takes to achieve maximum complexity/minimum simplicity, where the sum of \textit{information} $+$ \textit{entropy} is assumed a conserved quantity.}
 \label{Figure_16}
\end{figure*}

Focusing on the 2D quantum gravity theory, we consider the partition function $Z_{\rm L}(\beta)$ given by the Euclidean path integral\cite{kanyolo2022cationic}, 
\begin{align}\label{QG_eq}
    Z_{\rm L}(\beta) = \int \mathcal{D}[\Phi, g_{ab}(\Phi)]\exp(-S_{\rm E}(\Phi, g_{ab}(\Phi))),
\end{align}
where $S_{\rm E}(\Phi, g_{ab})$ is the Euclidean action for the Liouville theory. Recalling that $\beta$ in eq. (\ref{Z_eq}) is also assumed to be the inverse temperature, $S$ transformation given by $Z_{\rm L}(\beta) = Z_{\rm L}(1/\beta)$ near absolute zero, $\beta \rightarrow \infty$ (ground state of the system) renders the system approximately periodic in $\beta$, 
\begin{align}\label{Thermal_periodicity_eq}
    \lim_{\beta \rightarrow \infty} Z_{\rm L}(t_{\rm E} + \beta) =  \lim_{\beta \rightarrow \infty} Z_{\rm L}(t_{\rm E} + 1/\beta) = Z_{\rm L}(t_{\rm E}),
\end{align}
with $t_{\rm E} = -it$ the Euclidean time. This means that $\beta = -i\tau$ can also be treated as a Wick rotation. Unfortunately, since the path integral over the geometries $\int \mathcal{D}[g_{ab}]$ is famously ill-defined in quantum gravity, one results to Bootstrap methods.\cite{kanyolo2021partition} In our case, we first consider the idealised case of a 2D lattice of static $A$ cations in a cathode material at absolute temperature, $\beta = 1/k_{\rm B}T \rightarrow \infty$ where $k_{\rm B}$ is the Boltzmann constant. At this stage, the measure of \textit{complexity} is at its minimum since the lattice is completely ordered. As the temperature is increased, $\beta$ becomes finite and lattice vibrations begin to slowly form (phonons begin to be excited from the vacuum). As the temperature is increased just beyond a certain threshold, $k_{\rm B}T \geq E_{\rm a}$ given by the activation energy $E_{\rm a}$, lattice vacancies adiabatically form as the cations begin to diffuse. Further raising the temperature leads to a greater measure of \textit{complexity}, as more complex structures such as quasi-particles are excited from the lattice vacuum. In this picture, the entropy of the system of cations will grow with this measure of \textit{complexity}. However, beyond a certain time scale $t_{\rm P}$ (`Page time'\cite{page1993information}, in analogy with black hole thermodynamics), the lattice description must fail since more cations will have been dislodged from their lattice positions as diffusion increasingly sets in. Thus, it will no longer make sense to theoretically keep track of the lattice of $A$ cations. Instead, a theory of diffusion for the $A$ cations described by the Langevin equation suffices, which in turn corresponds to Liouville theory and by extension the 2D quantum gravity theory. A schematic has been availed in \textbf{Figure \ref{Figure_16}}, where the sum of \textit{information} $+$ \textit{entropy} is assumed a conserved quantity. 

Nonetheless, we can insist on a lattice description provided the 2D quantum gravity theory is interpreted as a lattice theory of the vacancies created as $1/\beta = k_{\rm B}T \rightarrow \infty$. Thus, as $\beta \rightarrow 0$, the measure of \textit{complexity} decouples from entropy which continues to increase, as \textit{complexity} decreases, tracing a Page curve.\cite{page1993information} Thus, the partition function for the 2D quantum gravity theory must be related by $S$ transformation, $Z_{\rm \Lambda}(1/\beta) = Z_{\rm L}(\beta) = Z_{\Lambda^*}(\beta)$ to the lattice theory of cations. This realisation is powerful, since it defines the Euclidean path integral,
\begin{align}\label{QG_eq2}
     Z_{\rm L}(\beta) = \int \mathcal{D}[\Phi, g_{ab}(\Phi)]\exp(-S_{\rm E}(\Phi, g_{ab}(\Phi))) = Z_{\Lambda}(1/\beta) = \sum_{\vec{r^*} \in \Lambda^*}\frac{\exp(-\pi\beta r^{*2})}{\eta^{2c}(i\beta)} = Z_{\Lambda^*}(\beta),
\end{align}
under some reasonable assumptions which transform the Euclidean path integral into a lattice summation. Since the Euclidean action for the Liouville theory is given by, 
\begin{align}
    S_{\rm E}(\Phi, g_{ab}(\Phi)) = \int_0^{\beta} dt_{\rm E}\int_{\mathcal{A}} d^{\,2}x\,\,\frac{1}{2}\left ( (\vec{\nabla}\Phi(x,y))^2 + K\sqrt{\det(g_{ab}(\Phi))}\right ) = S_{\rm E}(\Phi) + S_{\rm E}(g_{ab}(\Phi)),
\end{align}
where $\sqrt{\det(g_{ab}(\Phi))} = \exp(2\Phi)$. Here, we shall the consider the manifold, $\mathcal{A}$ to be compact (without boundary), which can be achieved by setting, $k_{\rm g} = 0$ corresponding to the equilibrium condition for the Langevin equation. This means that, in eq. (\ref{QG_eq}), we should make the following transformations, 
\begin{align}
   \int \mathcal{D}[g_{ab}(\Phi)]\times \leftrightarrow \sum_{n = 0}^{\infty}b_n^*\times,\,\,\, \int \mathcal{D}[\Phi]\exp(-S_{\rm E}(\Phi)) = \left (\prod_{N = 1}^{\infty}\frac{\exp(2\pi N/2)}{1 - \exp(-2\pi N\beta)}\right )^d \leftrightarrow \eta^{2c}(i\beta),
\end{align}
where we have introduced infinite harmonic oscillator modes of integer frequency $N$ for the Liouville field $\Phi$ in each dimension of space $d = 2c = 2$ and employed a zeta function regularisation of the vacuum energy of the form, $1 + 2 + 3 \cdots = \sum_{n = 1}^{\infty}1/N^{-1} \rightarrow \zeta(-1) = -1/12$. The Gauss Bonnet theorem introduces $n = g - 1$ in the exponent of the integrand provided $k_{\rm g} = 0$, thus making the connection to eq. (\ref{old_condition_eq}) as expected. Finally, the dual lattice $\Lambda^*$ is introduced by identifying $b_n^*$ as the number of vectors of even norm, $r^{*2} = 2n$ where $\vec{r^*} \in \Lambda^*$.

\subsection{Monolayer-Bilayer Phase Transition}

A peculiar feature of the Gauss-Bonnet theorem is the genus $g = n + 1 = 0$ state, which appears to be forbidden by the fact that the vacancy number $n$ has to be non-negative. To understand the significance of such a peculiarity, suppose we can find physical processes on a 2D lattice $\Lambda$ that manifest at genus $g = n + 1 \neq 0$ when some order parameter defined by the anti-commutator, 
\begin{align}\label{Delta_eq}
   \left \{\phi_{\Delta}, \phi_{\Delta}^{\dagger} \right \} \equiv \left \{\sum_{w \neq 0}\phi_{\hat{s}}(w), \sum_{\overline{w} \neq 0}\phi_{\hat{s}^{\dagger}}^{\dagger}(\overline{w})\right \} \coloneqq \frac{1}{C}\sum_{\vec{r} \in \Lambda, \vec{r} \neq 0}\frac{1}{(r^2/2)^{s}} = -2\Delta,
\end{align}
is vanishing, where $s = \hat{s} + \hat{s}^{\dagger}$ are some  operators to be determined, $C$ is a proportionality constant and $\{\phi_{\Delta}, \phi_{\Delta}^{\dagger}\} = \phi_{\Delta}\phi_{\Delta}^{\dagger} + \phi_{\Delta}^{\dagger}\phi_{\Delta}$ is the anti-commutator. The subscripts are treated as conformal dimensions of the complex fields $\phi_{\hat{s}^{\dagger}}^{\dagger}(\overline{w}), \phi_{\hat{s}}(w)$ where $\overline{w} = x - iy$ and $w = x + iy$ are complex variables and $r^2 = \overline{w}w = x^2 + y^2$. We first observe that, $\hat{s}^{\dagger}, \hat{s}$ are the scaling dimensions of the fields by checking that,
\begin{align}
    \phi_{\hat{s}^{\dagger}}^*(\lambda \overline{w}) = \lambda^{-\hat{s}^{\dagger}}\phi_{\hat{s}^{\dagger}}^*(\overline{w})\,\,\,,\,\,\,\phi_{\hat{s}}(\lambda \overline{w}) = \lambda^{-\hat{s}}\phi_{\hat{s}}(\overline{w}),
\end{align}
also appropriately re-scales the order parameter with $\lambda$ a scaling parameter. Assuming that $\Lambda$ is even, meaning $r^2/2 = n \in \mathbb{Z}_+ \neq 0$, the order parameter takes the form, 
\begin{align}\label{L_eq}
     2\Delta = -\frac{1}{C}\sum_{n = 1}^{\infty}\frac{b_n}{n^s} = -L(s)/C,
\end{align}
with $b_n$ the number of vectors of norms $r^2 = 2n$. Thus, suppose the lattice can be characterised by the Mellin transform,
\begin{align}
    \frac{1}{\Gamma(s)}\int_0^{\infty} \frac{d\beta}{\beta}(2\pi\beta)^sf(i\beta) = L_{\rm E}(s) = -2C\Delta,
\end{align}
of ``the lattice theta function'' corresponding to a cusp (modular) form of some elliptic curve, $E: y^2 = 4x^3 - g_2x - g_3$ over the field of rational numbers satisfying the (Fricke) involution\cite{zagier1984series},
\begin{align}
    f_{\rm E}(-1/N\tau) = \sum_{n = 1}^{\infty}b_nq^n(-1/N\tau) = -wN\tau^kf_{\rm E}(\tau), 
\end{align}
with weight $k = 2$ and $\tau = i\beta$ where, $q(\tau) = \exp(2\pi i\tau)$ is (the square of) the nome and the level $N = N_{\rm E}$ is equal to the conductor $N_{\rm E}$ of the elliptic curve over the rationals with $w = \pm 1$ the (global) root number. Here, $\Gamma(s) = \int_0^{\infty} d\beta\beta^{s - 1}\exp(-\beta)$ is the Gamma function and one also takes $f_{\rm E}(\tau + 1) = f_{\rm E}(\tau)$ (provided the congruence subgroup such as $\Gamma_c(N)$ with $c = 0, 1$ contains matrix, $S$ given in eq. (\ref{gen_eq})). It follows that the order parameter $\Delta$ will be proportional to 
$L_{\rm E}(s) = \sum_{n = 1}^{\infty}b_n/n^s$, where the cusp form is the generating function for the $b_n$.\cite{koblitz2012introduction} Note that, in the case of the hexagonal lattice, the coefficients of the lattice theta function $b_n$ are instead given by the number of integral solutions of eq. (\ref{Diophantine_eq}) and not the cusp form, \textit{i.e.} (to the best of our knowledge) the hexagonal lattice theta function ($\vec{r} \neq 0$) does not correspond to a cusp form, and hence $\Delta$ is not proportional to an $L$-function (technically, $L$ functions must have an analytic continuation to the entire complex plane obtained by the Mellin transform of their cusp form, excluding possibly their poles).

Instead, it suffices to take the continuum limit $b_n \rightarrow 2C > 0$ which now defines the constant.\cite{kanyolo2023pseudo} In this limit, the features of the lattice $\Lambda$ and hence the modularity of $\Delta$ become unimportant, with the order parameter approximated by the Riemann zeta function, $\zeta(s) = \sum_{n = 1}^{\infty}(1/n^s) = -\Delta$ which has the analytic continuation,
\begin{align}
    \zeta(s) = 2^s\pi^{s - 1}\sin (\pi s/2)\Gamma(1 - s)\zeta(1 - s),
\end{align}
excluding the pole at $s = 1$. Under our construction, the critical point of the phase transition, where a conformal field theory exists is located at $\Delta = 0$. Now, recall that, we are interested in the bipartite honeycomb lattice $\Lambda = hc$ ($\vec{r} \in hc$) comprised of two hexagonal lattices $hx, \overline{hx}$ ($w \in hx, \overline{w} \in \overline{hx}$). Defining the scaling dimensions,
\begin{align}\label{scaling_dims_eq}
    \hat{s} = -\hat{a}\hat{a}^{\dagger} + i\hat{\gamma}_{\hat{a}\hat{a}^{\dagger}},\,\,\,\hat{s}^{\dagger} = -\hat{a}\hat{a}^{\dagger} - i\hat{\gamma}_{\hat{a}\hat{a}^{\dagger}},
\end{align}
where $\hat{\gamma}_{\hat{a}\hat{a}^{\dagger}}^{\dagger} = \hat{\gamma}_{\hat{a}\hat{a}^{\dagger}}$ is a Hermitian operator, $[a, a^{\dagger}] = 1$ is the commutation relation for creation and annihilation operators acting on the Hilbert space of a single quantum harmonic operator and $\hat{n} = \hat{a}^{\dagger}\hat{a}$, $s = -2\hat{a}\hat{a}^{\dagger}$ is proportional to the pseudo-flux obtaining $s = -2g$, where $g = n + 1 = 1, 2, \cdots \in \mathbb{Z}_+$ corresponds to the trivial zeros of $\zeta(s)$. Meanwhile, a phase transition occurs at $n = -1$ corresponding to the two-sphere with genus $g = 0$, where $2\Delta = 1$. This phase transition can be understood via the treatment of the fields as quantum operators of the form, 
\begin{align}
    \phi_{\Delta}^{\dagger} = \sum_{\overline{w} \neq 0}\frac{1}{(|\overline{w}|^2/2)^{\hat{s}^{\dagger}}} 
    \rightarrow \sum_{n = 1}^{\infty}\frac{c_n^{\dagger}}{n^{\hat{s}^{\dagger}}},\,\,\,\,\phi_{\Delta} = \sum_{w \neq 0}\frac{1}{(|w|^2/2)^{\hat{s}}} \rightarrow \sum_{n = 1}^{\infty}\frac{c_n}{n^{\hat{s}}},
\end{align}
where $c_n^{\dagger}$, $c_n$ are raising and lowering operators with the commutation relation, $\left\{ c_n, c_m^{\dagger}\right \} = 2\delta_{nm}$ which yields the result in eq. (\ref{L_eq}) by performing the anti-commutation. 

Thus, a non-vanishing order parameter, $\Delta \neq 0$ introduces quantum correlations between $\phi_{\Delta}^{\dagger}$ and $\phi_{\Delta}$ leading to the non-commutivity of eq. (\ref{Delta_eq}) at $g = 0$. Consequently, our idealised model implies that atoms in $hx$ must interact with atoms in $\overline{hx}$ introducing quantum correlations between them. It is these quantum correlations at $g = 0$ that are responsible for bonds between atoms in across the lattices such as argentophilicity emerging between atoms in opposite hexagonal lattices across large distances. In this picture, a monolayer-bilayer phase transition is expected in this idealised model at $g = 0$, where the bipartite lattice bifurcates into its constitute lattices introducing quantum correlations between them. The quantum correlations can be interpreted geometrically as a pseudo-magnetic degree of freedom, 
\begin{align}\label{old_condition_eq2}
    -2\pi(\Delta (g) + 1) = \left \langle \int_{\mathcal{A}_g} d^{\,2}x\,\,\vec{n}\cdot\mathcal{B}_{\Delta} \right \rangle, 
\end{align}
where $\langle \cdots \rangle$ is a quantum averaging and $\mathcal{A}_g$ is the manifold of genus $g$. Comparing eq. (\ref{old_condition_eq}) and eq. (\ref{old_condition_eq2}), we obtain,
\begin{align}\label{q_average_eq}
    -\Delta (g) = \left \langle n(\beta) \right\rangle + 1 = \frac{1}{\Gamma(s)}\int_0^{\infty} \frac{d\beta}{\beta}\beta^sn(\beta) = \zeta(s),
\end{align}
where $s = -2g$ and we have used $a^{\dagger}a \rightarrow n(\beta) = 1/(\exp(\beta \omega) - 1)$ under the assumption of a thermal state $\omega = 1$. Thus, the averaging is performed by a partition function,
\begin{align}
    \langle \cdots \rangle \coloneqq \frac{1}{\Gamma(s)}\int \frac{d\beta}{\beta}\beta^s\exp(-\beta)(\cdots),
\end{align}
given by the Gamma function, $\Gamma(s)$.

Moreover, one can check that acting on the vacuum states of each lattice with $\hat{a}^{\dagger}$ and $\hat{a}^{'\dagger}$ respectively,
\begin{align}
\frac{(\hat{a}^{\dagger})^{g - 1}(\hat{a}'^{\dagger})^{g - 1}}{g!}|0\rangle\,|0\rangle' = |n\rangle\,|n\rangle' \equiv |\mathcal{A}_g\rangle,
\end{align}
extracts in total $n = g - 1$ pairs of cations equivalent to the statement of the Gauss-Bonnet theorem,
\begin{align}\label{normalisation_eq1}
    \mathcal{E}(\mathcal{A}_g) = \langle\mathcal{A}_g|\hat{\mathcal{E}}|\mathcal{A}_g\rangle  = \frac{1}{2\pi}\int_{\mathcal{A}_g}d^{\,2}x\,\,K\exp(2\Phi(x,y,0)),
\end{align}
where $\hat{\mathcal{E}} = -\hat{a}^{\dagger}\hat{a} - \hat{a}^{'\dagger}\hat{a}'$, $K$ is the Gaussian curvature and $\Phi$ is the Liouville field. This suggests the quantum state of the honeycomb lattice is a Thermofield Double (TFD) state\cite{kanyolo2022local} of entangled hexagonal lattice atoms, 
\begin{align}
    |{\rm TFD} \rangle  = \frac{1}{\sqrt{Z_{\rm TFD}(\beta)}}\sum_{g = 1}^{\infty}\exp\left (\beta\hat{\mathcal{E}_g}/4\right )|\mathcal{A}_g\rangle = \frac{1}{\sqrt{Z_{\rm TFD}(\beta)}}\sum_{n = 0}^{\infty}\exp\left(-\beta n\right)|n\rangle|n\rangle', 
\end{align}
where $Z_{\rm TFD}(\beta) = 1/(1 - \exp(-\beta))$ is the partition function which guarantees the normalisation $\langle {\rm TFD}|{\rm TDF} \rangle = 1$. Thus, transferring well-known properties of the TFD state to the honeycomb lattice, we find that the honeycomb lattice state $|\rm TFD\rangle$ is in a pure state with a density matrix $\hat{\rho} = |{\rm TFD}\rangle \langle {\rm TFD}|$ satisfying ${\rm Tr}_{\rm hc}(\hat{\rho}) = {\rm Tr}_{\rm hc}(\hat{\rho}^2) = 1$ whereas individually each hexagonal lattice is in a thermal state, ${\rm Tr}_{\rm hx}(\hat{\rho}^2) = {\rm Tr}_{\rm \overline{hx}}(\hat{\rho}^2) \neq 1$ with ${\rm Tr}_{\rm hx}(\hat{\rho}) = {\rm Tr}_{\rm \overline{hx}}(\hat{\rho}) = 1$. Although not further explored herein, this construction finally justifies the analogy with black hole thermodynamics displayed in \textbf{Figure \ref{Figure_16}}. 

Considering the quantum averaging of the order parameter $\langle \Delta \rangle$, we find that it tends to infinity, 
\begin{align}
    2\langle \Delta \rangle = -\langle \{\phi_{\Delta}^{\dagger}, \phi_{\Delta}\}\rangle = -2\sum_{n = 1}^{\infty}\frac{1}{n^{\langle s \rangle = 1}} \rightarrow -\infty,
\end{align}
which in number theory reflects the infinitude of primes. Here, $\langle s \rangle = 1$ is obtained from eq. (\ref{scaling_dims_eq}) with the definition for the quantum averaging given by eq. (\ref{q_average_eq}). Hence, unlike for the other values of $s$, there is no means to perform a Riemann zeta function regularisation scheme since the Riemann zeta function is not analytic at $s = 1$. Nonetheless, it turns out that this is a blessing in disguise since it will allow us to make use of the Hermitian operator $\hat{\gamma}_{\hat{n}}^{\dagger} = \hat{\gamma}_{\hat{n}}$ in eq. (\ref{scaling_dims_eq}). Physically, since $\Delta$ cannot tend to infinity, another CFT must exist at $\phi_{\Delta = 1/2} = 0$.  To see this, consider the coherent states of the operator $\phi_{\Delta}$. Referring to $c_n^{\dagger}, c_n$ in the previous definition as Dirac operators ($c_n^{\dagger}\neq c_n$) due to their anti-commutation rule $\{ c_n^{\dagger}, c_n\} = 2\delta_{nm}$, we shall introduce instead the Majorana operators, $c_n^{\dagger} = c_n$. The eigenstates of the Majorana operators are simultaneous quantum states of both $\phi_{\Delta}^{\dagger}$ and $\phi_{\Delta}$, unlike the Dirac case. Defining new scaling dimensions, 
\begin{align}\label{scaling_dims_eq2}
    \tilde{s} = -\left\langle \hat{a}\hat{a}^{\dagger}\right\rangle + i\hat{\gamma}_{\hat{a}\hat{a}^{\dagger}},\,\,\,\tilde{s}^{\dagger} = -\left\langle \hat{a}\hat{a}^{\dagger}\right\rangle - i\hat{\gamma}_{\hat{a}\hat{a}^{\dagger}},
\end{align}
we require that $\phi_{\Delta}(\tilde{s}) = 0$ and $\phi_{\Delta}(\tilde{s}^{\dagger}) = 0$. Since we are calculating with (Majorana) fermions, $c_n = c_n^{\dagger} = 1$ for fully occupied coherent states, implying the eigenvalues of $\hat{\gamma}_{\hat{a}\hat{a}^{\dagger}}$ will be at the essential zeros $\phi_{\Delta}^{\dagger} = \zeta(\tilde{s}^{\dagger}) = 0, \phi_{\Delta} = \zeta(\tilde{s}) = 0$ of the Riemann zeta function on the real line $-\left\langle \hat{a}\hat{a}^{\dagger}\right\rangle = 1/2$. Unfortunately, the chances of establishing the explicit form of $\hat{\gamma}_{\hat{a}\hat{a}^{\dagger}}$ within this Review are close to none, since this is equivalent to proving the Riemann hypothesis under the Hilbert-P\'{o}lya conjecture!.\cite{conrey2015riemann} Nonetheless, similar to $g = n + 1 \neq 0$, physical considerations require that these states be monolayered, implying complementarity between the essential and trivial zeros.\cite{kanyolo2023pseudo} 

Quantum mechanically, such a Bohr complementarity can be expressed in the form, $[n, \gamma] = i\kappa$, where $\kappa$ is an unimportant constant re-scaling $\gamma$ and the Hilbert-P\'{o}lya operator $\gamma$ is also the sought after Hermitian quantum phase operator.\cite{susskind1964quantum} Heuristic evidence of this claim lies, for instance, in certain results in Random Matrix Theories (RMT) such as the Circular Unitary Ensemble, where the statistics of the Riemann zeros behave like the statistics of a random quantum phase.\cite{keating2000random} Further discussions on RMT are beyond the intended scope of this {\it Review}. Additional progress towards understanding important features of this coherent state in bilayered structures can be made by introducing a bootstrap for the normalisation of the 2D vacancy number density (eq. (\ref{pair_corr_eq}),

\begin{align}\label{normalisation_eq2}
    \int_{\mathcal{A}}d^{\,2}x\,\,\rho_z(x, y) \coloneqq \lim_{X \rightarrow \infty}\frac{\hat{a}\hat{a}^{\dagger}}{(\ln X)^{r(z)}},
\end{align}
where $r(z) = (\ell - z)/\ell = 0, 1 \geq 0$ are the binary integers defined at $z = \ell$ and $z = 0$ respectively, and $\ell$ is the bifurcation distance measured from the origin ($z = 0$) along the unit normal $\vec{n} = (0, 0, 1)$. For instance, take $\ell = \pi/k_{\rm F}$, where $k_{\rm F} = |\vec{k}_{\rm F}|$ and $\vec{k}_{\rm F}$ is the Fermi wave vector appearing in the 3D Ruderman-Kittel-Kasuya-Yosida (RKKY) interaction $J_{\rm RKKY}^{\rm 3D}(z) \propto \cos(2k_{\rm F}z)/z^3 - \sin(2k_{\rm F}z)/z^4 = 1/z^3$, which is pathological at $z = 0$.\cite{kanyolo2023pseudo}. Clearly, since $g = \langle n| \hat{a}\hat{a}^{\dagger}|n \rangle = n + 1 \neq 0$ and $z$, $z = \ell$ corresponds to the normalisation in eq. (\ref{pair_corr_eq}) in the limit $\lim_{X \rightarrow \infty} 1/(\ln X)^{r(z = \ell)} = \lim_{X \rightarrow \infty} 1/(\ln X)^0 = 1$, whereas the pathology at $z = 0$ corresponds to a vanishing vacancy number since $\lim_{X \rightarrow \infty} 1/(\ln X)^{r(z = 0)} = \lim_{X \rightarrow \infty} 1/(\ln X)^1 \rightarrow \infty$, regardless of the value of $g \neq 0$ signifying a large degeneracy for the monolayered structure $z = 0$. This degeneracy must be lifted by a monolayer-bilayer phase transition, avoiding the pathology. In this way, the coordinate $z$ acts an order parameter. Unfortunately, $X$ is a mathematical artefact which we are presently unable to link to a physical variable. Nonetheless, eq. (\ref{normalisation_eq2}) is useful since we can consider the semi-classical analogue of eq. (\ref{L_eq}) at $s = 1$ to be equivalent to $\rho_z(x,y) = \langle n |\hat{\phi}^{\dagger}(x,y,z)\hat{\phi}(x,y,z)|n \rangle$ with $g = n + 1$, where $\hat{\phi}(x,y,z) = (a^{\dagger}/\sqrt{C})\sum_{n = 1}^{\infty} \phi_n(z)\exp(i2\pi (x + iy))$ and, 
\begin{multline}
   \int_{-1/2}^{+1/2}\int_0^{\infty}dxdy\,\,\hat{\phi}^{\dagger}(x, y, z)\hat{\phi}(x, y, z)
    = \frac{\hat{a}\hat{a}^{\dagger}}{C}\sum_{n = 1}^{\infty}\sum_{m = 1}^{\infty}\frac{\overline{\phi}_n(z)\phi_m(z)}{2\pi (n + m)}\frac{\sin(\pi(n - m))}{\pi(n - m)} 
    = \frac{\hat{a}\hat{a}^{\dagger}}{C}\sum_{n = 1}^{\infty}\frac{\overline{\phi}_n(z)\phi_n(z)}{4\pi n} = \frac{\hat{a}\hat{a}^{\dagger}}{C}\sum_{n = 1}^{\infty}\frac{b_n(z)}{n} \\
    = \frac{\hat{a}\hat{a}^{\dagger}}{C} L_{\rm E}(s = 1) \rightarrow \frac{\hat{a}\hat{a}^{\dagger}}{C}\lim_{X \rightarrow \infty} \frac{C_{\rm E}}{(\ln X)^r} = \lim_{X \rightarrow \infty} \frac{\hat{a}\hat{a}^{\dagger}}{(\ln X)^r}, 
\end{multline}
with $b_n \coloneqq \overline{\phi}_n\phi_n/4\pi$ the real-valued coefficients of a cusp form of weight 2 and level $N$ generated by some elliptic curve $E$ over the rationals, and the 2D manifold $\mathcal{A}_g/g$ defining a single primitive cell ($\mathcal{A}_g/g \coloneqq D$: $0 \leq y \leq +\infty, -1/2 \leq x \leq +1/2$), $r = r(z)$ is the so-called arithmetic Rank of the elliptic curve $E$ of conductor $N_{\rm E} = N$ and $C_{\rm E}\ln X$ arises from the Euler product form of the (incomplete) $L_{\rm E}(s)$ function of the elliptic curve $E$ at $s = 1$ for primes less than or equal to $X$.\cite{breuil2001modularity, lmfdb, koblitz2012introduction} Here, the arrow ($\rightarrow$) assumes the conjecture of Birch and Swinnerton-Dyer, with the proportionality constant given by $C = C_{\rm E}$.\cite{birch1965notes, wiles2006birch} 

By this treatment, we can repackage the scaling dimensions as a complex-valued integral of the form,
\begin{align}
    \hat{s} \coloneqq \ell\int_{\mathcal{M}} d^{\,3}x\,\,\Psi^*\partial_z\Psi,\,\,\,\hat{s}^{\dagger}  \coloneqq \ell\int_{\mathcal{M}} d^{\,3}x\,\,\Psi\partial_z\Psi^*,
\end{align}
where $\Psi = \sqrt{\rho}\exp\left (i\int \vec{dx}\cdot\vec{m}/\ell \right )$ with $\vec{m}$ a magnetisation to be defined, $\Psi^* = \sqrt{\rho}\exp\left (-i\int \vec{dx}\cdot\vec{m}/\ell \right )$ is the complex conjugate of $\Psi$, and $\mathcal{M}$ and $\rho$ are the 3D manifold and the interpolation function appearing in eq. (\ref{pair_corr_eq}) respectively. Now introduce the complex-valued tensor equation reminiscent of the tensor field equations proposed in an idealised model of cation diffusion in honeycomb layered materials\cite{kanyolo2023pseudo}, 
\begin{align}\label{tensor_eq}
    \partial_uK_{uv} = -4\pi \ell\Psi^*\partial_v\Psi,\,\,\,K_{uv} = \partial_u\partial_v\Phi + 2ei\varepsilon_{uvw}A_w/\ell,
\end{align}
where $\partial_u = \partial/\partial x^u, \partial_v = \partial/\partial x^v$ are partial derivatives in 3D Euclidean space, $\varepsilon_{uvw}$ is the Levi-Civita symbol, $A_u \equiv \vec{A} = (A_x, A_y, A_z)$ is the electromagnetic vector potential and $e$ is the elementary charge (for simplicity, we henceforth set $e = 1$). Clearly, the average magnetisation vector is given by the imaginary part of eq. (\ref{tensor_eq}), 
\begin{align}
    \vec{\gamma} \coloneqq \int_{\mathcal{M}} d^{\,3}x\,\,\vec{\Gamma} \coloneqq \langle\,\,\vec{m}\,\,\rangle = \int_{\mathcal{M}} d^{\,3}x\,\,\rho\vec{m} = \frac{1}{2\pi \ell} \int_{\mathcal{M}} d^{\,3}x\,\,\vec{\nabla}\times\vec{A},  
\end{align}
where $\vec{\Gamma} = \rho\vec{m} = (\vec{\nabla}\times\vec{A})/2\pi \ell$ is the magnetisation density and $\langle \cdots \rangle \coloneqq \int_{\mathcal{M}} d^{\,3}x\,\,|\Psi|^2(\cdots)$. Thus, by the normalisation conditions in eq. (\ref{normalisation_eq1}) and eq. (\ref{normalisation_eq2}), we find,
\begin{align}
    \hat{s} \coloneqq \ell\int_{\mathcal{M}} d^{\,3}x\,\,\Psi^*\partial_z\Psi = -g + i\gamma_z,\,\,\,\hat{s}^{\dagger}  \coloneqq \ell\int_{\mathcal{M}} d^{\,3}x\,\,\Psi\partial_z\Psi^* = -g - i\gamma_z,
\end{align} 
with $g = -aa^{\dagger}$, which predicts that bilayered materials can be modulated into monolayers by the $z$ component of a magnetic field where the magnetisation values occur at the essential zeros $\tilde{s} = 1/2 + i\gamma_z$ or $\tilde{s}^{\dagger} = 1/2 - i\gamma_z$. Finally, for a magnetic field along the $z$ direction independent of $z$ coordinate, since $\int_{\mathcal{M}} d^{\,3}x \equiv \int_0^{\ell}\int_{\mathcal{A}} d^{\,2}x$, we obtain the fluxes, 
\begin{align}\label{flux_zeros_eq}
    \gamma_z(g) = \frac{1}{2\pi}\int_{\mathcal{A}_g}d^{\,2}x\,\,(\vec{\nabla}\times\vec{A})_z,
\end{align}
which are magnetic fluxes through the surface of manifolds $\mathcal{A}_g$. Since the trivial zeros are given by the cationic charges $g = n + 1$, this lends further credence to the Bohr complementarity proposed above, since $[n, \gamma] = i$ now corresponds to the well-known commutation relation between fluxes and charges.\cite{kanyolo2021renormalization} Typical flux quantisation where $\gamma_z/2\pi = N \in \mathbb{Z}$ are integer values correspond to winding numbers (monopole numbers) on a two-sphere ($\mathcal{A} = S^2$). However, in eq. (\ref{flux_zeros_eq}), the topology of genus $g = aa^{\dagger} \neq 0$ surfaces $\mathcal{A}$ imply the atypical nature of the winding numbers. Moreover, since the charges are restricted to positive integers ($aa^{\dagger} \geq 1$), the fluxes indeed correspond to the ill-defined quantum phase operator of the harmonic oscillator, whose Hermicity is known to be questionable.\cite{susskind1964quantum} Nonetheless, should we believe these expressions, this realisation suggests the essential zeros correspond to winding numbers along geodesics on closed Riemann surfaces $\mathcal{A}$. Another interesting observation can be made within the context of Lee-Yang theorem\cite{knauf1999number, conrey2015riemann}, whereby the monolayer-bilayer phase transition is modulated or tuned by an external imaginary magnetic flux\cite{kanyolo2023pseudo} of the form $\varphi(\gamma_z) \coloneqq i\gamma_z/2\pi$ within the context of a Heisenberg spin model partition function, $\Xi(T,\varphi(\gamma_z))$. In particular, the partition function $\Xi(0, \varphi(\gamma_z))$ in the ground state ($T = 1/\beta = 0$) is defined as the Taylor expansion,   
\begin{align}
    \Xi(0, \varphi(\gamma_z)) = \lim_{\beta \rightarrow \infty} \int_0^{\beta}\,\,Z(u,\varphi(\gamma_z))du = \lim_{\beta \rightarrow \infty}\int_0^{\beta}\,\,{\rm Tr}\left \{ \exp(-uH_{\rm Ising}(\varphi(\gamma_z)) \right \}du = \lim_{\beta \rightarrow \infty}\sum_{n = 0}^{\infty}\frac{\Xi_n(\beta)}{2n!}(-1)^n\gamma_z^{2n},
\end{align}
where the functional form of $H_{\rm Ising}(\varphi)$ is given in eq. (\ref{Ising_H_eq}), $S_j, S_i$ are spin operators, $J_{ij}$ is the Heisenberg exchange term of the form given in eq. (\ref{J_eq}), the functional form of the partition functions $Z(u, \varphi) = \exp(uJ(u))\cosh(2\pi u\varphi)$ is given in eq. (\ref{Tr_exacT_sol_eq}) with the definition $J(u) \coloneqq \ln (2\Phi(u))/u$, the constants are given by $\Xi_n(\beta) = 2\int_0^{\beta}\Phi(u) u^{2n}du$ and,
\begin{align}
   \Phi(u) \coloneqq q^{1/2}(u)\sum_{n = 1}^{\infty}\left (2\pi^2q^4(u)n^4 - 3\pi q^2(u)n^2 \right )\exp(-\pi n^2q^2(u)) = \Phi(-u),
\end{align}
is the super-exponentially decaying function\cite{rodgers2020bruijn} with $q(u) = \exp(2u)$ interpreted as a Boltzmann factor (compare the role of $\Phi(u)$ with $\eta(i\beta)$ in eq. (\ref{fermionic_Z_eq})). Consequently, the ground state partition function becomes equivalent to the Riemann Xi function\cite{conrey2015riemann}, $\xi(1/2 + \varphi) \equiv \Xi(0, \varphi)$ with its zeros $\Xi(0, \varphi_g) = 0$ at $\beta \rightarrow +\infty$ occurring at imaginary values $\varphi_g = i\gamma_z(g)/2\pi$ satisfying the Lee-Yang theorem and hence the Riemann hypothesis.\cite{knauf1999number, conrey2015riemann} Finally, one can certainly re-define the Hermitian tensor given in eq. (\ref{tensor_eq}) as $K_{uv} \coloneqq R_{uv} + i\frac{2e}{\ell}\varepsilon_{uvw}A_w$, where $R_{uv} \coloneqq \mu_{uabv}\varepsilon_{ab}/\ell$ introduces possible flexoelectric effects in the monolayer-bilayer phase transition, with $\varepsilon_{ab}$ the second order strain tensor and $\mu_{uabv}$ the so-called fourth-rank polar tensor (flexoelectric coefficient).\cite{surmenev2023influence} Additional ingredients beyond the scope of this work are necessary to completely appreciate the extra features of such modulation of the monolayer-bilayer phase transition.\cite{kanyolo2023pseudo} Moreover, there will certainly be questions pertaining the manifestation of the monolayer-bilayer phase transition from a bond theoretic point of view. This will be tackled in the subsequent sections, where we argue that silver metallophilicity and precisely argentophilic bilayers are as a result of such a monolayer-bilayer phase transition.

\subsection{Valence Bond Theoretic Approaches}

The determination of a metal's valency state pertains to the enumeration of valence electrons participating in chemical bonding within a compound. Numerous elements possess an established valency state associated with their placement in the periodic table, owing to compliance with the octet rule. Conversely, certain elements manifest multivalency due to the comparably analogous energy levels of their valence electrons. As an illustrative example, the silver ion, unlike the divalent magnesium cation ($\rm Mg^{2+}$), may exhibit a range of oxidation states such as +3, +2, and even $\rm -1$ (as shown in \textbf {Table \ref {Table_2}}). Typically, oxidation states are denoted by integers encompassing positive, zero, or negative values. 
On this front, the condensed matter physics and chemistry inherent in silver-based layered materials with anomalous fractional valency states $1/2+, 2/3+$, \textit{etc} (herein referred to as sub-valency/sub-valent states) has remained a great mystery to the scientific community since their discovery. As such, progress towards understanding the origin of the silver sub-valency in layered materials has been limited, with the problem compounded by the discovery of unconventional silver-silver interactions constituting so-called \textit{metallophilic} (in the case of silver, argentophilic) bonds in the aforementioned Ag-based materials, as well as other silver cluster materials.\cite{schmidbaur2015argentophilic} The connection between argentophilicity and sub-valency appears to be non-trivial, since other materials with metallophilic bonds such as copper and gold (cuprophilic and aurophilic interactions, respectively), have been envisaged, \textit{albeit} appearing presently not to manifest sub-valency. Moreover, thallophilic interactions have also been found in thallium-based layered oxides such as $\rm Tl_2MnTeO_6$ reported to be bilayered \textit{albeit} appear to be without sub-valency.\cite {nalbandyan2019Tl2MnTeO6} 

Particularly, argentophilicity in bilayered structures manifests uncharacteristically shorter silver-silver bonds compared to conventional silver metallic bonds, suggesting a different bonding mechanism is needed to explain the stability of such bilayers. Moreover, the low temperature superconductivity reported in $\rm Ag_2^{1/2+}F^{1-}$ remains unexplained to date.\cite{andres1966superconductivity} Incidentally, localised valence electrons in other silver bilayered materials such as $\rm Ag_{16}^{1/2+}B_4^{3+}O_{10}^{2-}$ which preclude metallic conductivity in favour of semi-conductivity or insulator behaviour, alongside other peculiarities indicate the physics and chemistry of silver especially in silver materials with bilayers is poorly understood. Meanwhile, after the advent of powerful computational tools in the many-body valence theory such as Molecular Dynamics (MD) and Density Functional Theory (DFT), focus has shifted towards understanding the many body physics and chemistry of such metallophilic bilayers seeking to accurately simulate their ground and excited state behaviour, believed to capture most of the essential properties of these materials such as magnetic susceptibility, thermal and electrical conductivity amongst others, \textit{albeit} with limited success. Computational modelling, frequently entailing conflicting perspectives and engendering extensive scholarly discourse, have evolved over numerous decades in connection with understanding the origin and nature of the silver sub-valency. Recent noteworthy efforts towards understanding these discrepancies in the properties of silver-based bilayered materials include employing electron localisation functions, the crystal overlap Hamiltonian population and non-covalent interaction (NCI) scalar fields in DFT computations.\cite{kovalevskiy2020uncommon, lobato2021comment, yin2021reply}  Particularly, the distinctive distribution of valency states observed for instance in $\rm Ag_2^{1/2+}Ni^{3+}O_2$ has been interpreted through the lens of an intensified ligand-field stabilisation effect experienced by the low-spin $\rm Ni^{3+}$ species, in contrast to the $\rm Ni^{2+}$ species residing within an octahedral coordination environment.\cite{schreyer2002synthesis} Furthermore, this intricate phenomenon has been attributed to the presence of energetically favourable silver $5s$ and $5p$ bands, enabling the accommodation of excess electron density.\cite{schreyer2002synthesis} First-principles calculations have been employed to discern the underlying factors contributing to the $\rm Ag$ subvalency observed in $\rm Ag_2NiO_2$. \cite {johannes2007formation} These calculations peg the $\rm Ag$ subvalency ($\rm Ag^{1/2+}$) to a robust bonding-antibonding interaction occurring between the $\rm Ag$ bilayers. This interaction effectively impels a substantial fraction of the bonding $\rm Ag$ {\it sp} bands beneath both the $\rm Ni$ $d$ orbitals and a majority of the $\rm O$ $p$ orbital states. \cite {johannes2007formation} Consequently, the conventional $\rm Ag^{1+}$ valency state is categorically precluded, as the electronic structure indicates the absence of such a configuration. However, it is not evident whether such treatments effectively translate to the entire class of silver-based metallophilic bilayered materials found in literature.

\begin{table*}
\caption{Anomalous oxidation states of silver exhibited in Ag-based compounds.\cite{masese2023honeycomb, schreyer2002synthesis, susloparova2022, yoshida2020static, taniguchi2020butterfly, yoshida2011novel, matsuda2012partially, yoshida2008unique, yoshida2006spin, sorgel2007ag3ni2o4, tong2011, argay1966redetermination, wang1991, ikezawa1985, fujita1974, andres1966superconductivity, ido1988, freed1940magnetic, yoshida2020partially, sugiyama2009static, ji2010orbital, bailey1887xliii, ahlert2003ag13oso6, kovalevskiy2020uncommon, derzsi2021ag, minamikawa2022electron, ho1990photoelectron, dixon1996photoelectron, schneider2005unusual, beesk1981x, kurzydlowski2021fluorides, grzelak2017metal, jansen1992ag5geo4, standke1985, ramirez2006, santoyo2010, ramirez2020, ott1928, mcmillan1960, wyckoff1922, molleman2015surface} Materials that (are expected to) exhibit metallophilic bilayers or monolayer-bilayer hybrid are comprised of $\rm Ag$ cations with oxidation state of $+1/2$ to $+2/3$.} The theoretically predicted $\rm Ag{\it X}_2$ ($X$ = $\rm Cl, Br, I $) and $\rm Ag_6Cl_4$ have been included, for the sake of completeness.\\
\label{Table_2}
\begin{center}
\scalebox{0.85}{
\begin{tabular}{llll} 
\hline
\textbf{Compound} & \textbf{Oxidation state of Ag} & \textbf{Compound} & \textbf{Oxidation state of Ag}\\ 

\hline\hline
$\rm Ag_2O_3$\cite{standke1985} & +3 & $\rm Ag_3OH$\cite{molleman2015surface} & +1/3\\
$\rm Ag{\it X}_2$ ($X$ = Cl, Br, I)\cite{ramirez2006, santoyo2010, ramirez2020} & +2 & $\rm Ag_2{\it M}O_2$ ($\rm {\it M} = Co, Cr, Co, Ni$ and $\rm Fe$)\cite{yoshida2006spin, schreyer2002synthesis, yoshida2008unique, sugiyama2009static, ji2010orbital} & +1/2\\
$\rm AgO$\cite{mcmillan1960} & +2 & $\rm Ag_{16}B_4O_{10}$\cite {kovalevskiy2020uncommon} & +1/2\\
$\rm AgF_2$\cite{kurzydlowski2021fluorides, grzelak2017metal} & +2 & $\rm Ag_6{\it M}_2TeO_6$ ($M$ = Ni, Co, Cu, Mg, Zn)\cite{masese2023honeycomb} & +1/2\\
$\rm Ag_2O$\cite{wyckoff1922} & +1 & $\rm Ag_4O$\cite{bailey1887xliii} & +1/2\\
$\rm Ag_5SiO_4$\cite{linke1994subvalent} & +4/5 & $\rm Ag_2F$\cite {ott1928, wang1991, ikezawa1985, fujita1974, andres1966superconductivity, ido1988, williams1989neutron, freed1940magnetic} & +1/2\\
$\rm Ag_5GeO_4$\cite{jansen1992ag5geo4} & +4/5 & $\rm Ag_{13}OsO_6$\cite {ahlert2003ag13oso6} & +4/13\\
$\rm Ag_3O$\cite{beesk1981x} & +2/3 & $\rm Ag_{13}(OH)_4$\cite{molleman2015surface} & +4/13\\
$\rm Ag_6Cl_4$\cite {derzsi2021ag} & +2/3 & ${\rm Ag}_NM^{+}$\cite{minamikawa2022electron, ho1990photoelectron, dixon1996photoelectron, schneider2005unusual} & $\rm -1$\\
\hline
\end{tabular}}
\end{center}
\end{table*}

\begin{figure*}[!t]
 \centering
 \includegraphics[width=0.85\columnwidth]{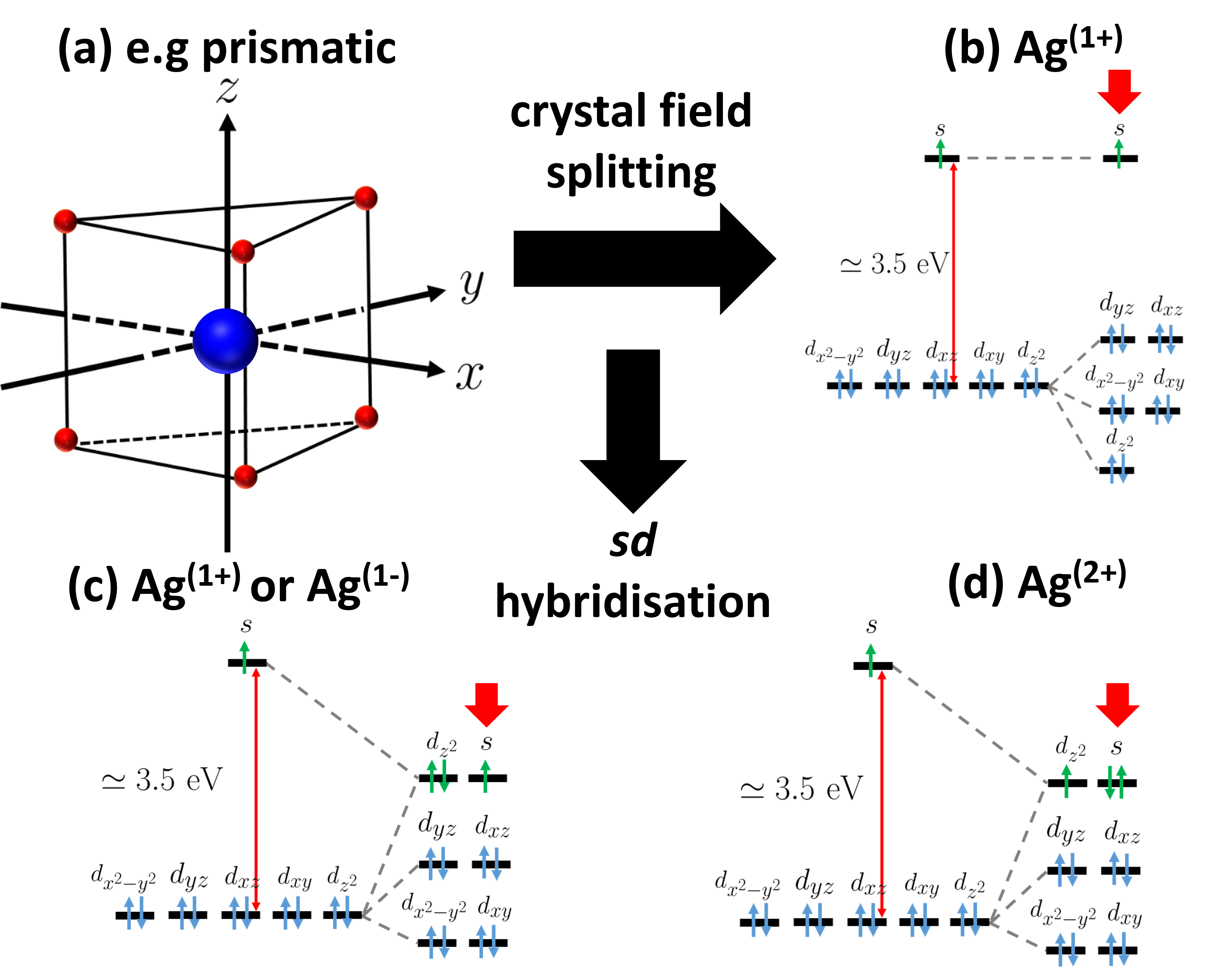}
 \caption{Influence of (exemplar) prismatic coordination of $\rm Ag$ atoms with oxygen or fluoride atoms on $\rm Ag$ electronic configuration, leading to various degenerate states in metallophilic bilayers. (a) Exemplar prismatic coordination of silver atoms (blue) to oxygen or fluoride (red) atoms in monolayered or bilayered honeycomb layered frameworks. (b) Typical crystal field splitting of the Ag $4d^{10}$ orbitals in the exemplar prismatic environment, depicting the electronic configuration [Kr]$4d^{10}5s^1$ ($\rm Ag^{(1+)}$) where the bracket denotes the valence state. The energy gap between $4d$ and $5s$ orbitals before $sd$ hybridisation (indicated by the thin upward red arrows) is assumed to be $3.5$ eV\cite{blades2017evolution}; (c) Crystal field splitting of the Ag $4d^{10}$ orbitals in a prismatic environment and subsequent $sd_{z^2}$ hybridisation, depicting the electronic configuration [Kr]$4d^{10}5s^1$ ($\rm Ag^{(1+)}$ or $\rm Ag^{(1-)}$) where the bracket denotes the degenerate Ag valence states $1+$ and $1-$; (d) Likewise, the crystal field splitting of the Ag $4d^9$ orbitals in a prismatic environment and subsequent $sd_{\rm z^2}$ hybridisation, depicting the electronic configuration [Kr]$4d^95s^2$ ($\rm Ag^{(2+)}$ or $\rm Ag^{(1-)}$) where the bracket denotes the degenerate Ag valence states $2+$ and $1-$. Note that, the thick red downward arrow indicates the valence electrons ($5s$ orbital) assigned by $\rm SU(2)\times U(1)$ symmetry breaking (eq. (\ref{SU2U1_eq})), whose effect is to slightly lower the energy of the $4d_{z^2}$ orbital relative to the $5s$ orbital via a structural phase transition analogous to orbital ordering of $e_g$ orbitals.\cite{pavarini2016Quantum} Figure reproduced with permission.\cite{masese2023honeycomb} Copyright 2023 Wiley-VCH.}
 \label{Figure_17}
\end{figure*}

On the other hand, significant progress in the theoretical front was recently reported\cite{masese2023honeycomb} whereby, degenerate valence states of silver arranged in a honeycomb lattice in 2D interact under non-Abelian electromagnetic interactions exhibiting a broken $\rm SU(2)\times U(1)$ symmetry, analogous to electron-neutrino interactions in the standard model of particle physics.\cite{weinberg1967model} Particularly, due to the energy proximity of silver $4d^{10}$ and $5s^1$ electrons and other environmental factors such as crystal field splitting of the $4d$ orbitals encouraging $sd$ hybridisation as shown in \textbf{Figure \ref{Figure_17}}, the electronic configuration of silver atom is expected to be in either one of two degenerate states, namely the typical [Kr]$4d^{10}5s^1$ configuration responsible for the usual oxidation state $\rm Ag^{1+}$ or the proposed [Kr]$4d^95s^2$ configuration. This suggests two other oxidation states exist, namely $\rm Ag^{2+}$ and $\rm Ag^{1-}$, obtained by designating the electrons in either the $5s^2$ or $4d^9$ orbitals respectively as the valence electrons (or both, considering $\rm Ag^{3+}$). The non-Abelian gauge interactions between these degenerate silver states admits the Lagrangian\cite{masese2023honeycomb}, 
\begin{multline}\label{SU2U1_eq}
    \mathcal{L} = -\int d^{\,3}x\,\,\frac{1}{4}\left({\rm Tr}\left (\partial_{\mu}W_{\nu} - \partial_{\nu}W_{\mu} + 2q_{\rm w}[W_{\mu}, W_{\nu}] \right )^2 + \left(\partial_{\mu}A_{\mu} - \partial_{\nu}A_{\mu}\right)^2 \right)\\
    + \int d^{\,d}x\,\left (\overline{\psi}_{\rm L}\gamma^{\mu}(i\partial_{\mu} + 2q_{\rm w}W_{\mu} + \frac{Y}{2}q_{\rm e}A_{\mu})\psi_{\rm L} + \overline{\psi}_{\rm R}\gamma^{\mu}(i\partial_{\mu} + \frac{Y}{2}q_{\rm e}A_{\mu})\psi_{\rm R}\right)\\
    - \int d^{\,d}x\,\left(\frac{1}{2}\left |\left (\partial_{\mu} + iq_{\rm w}W_{\mu} + i\frac{1}{2}q_{\rm e}A_{\mu}\right )\phi_{\Delta} \right|^2 + y_{\rm c}\overline{\psi}_{\rm R}\phi_{\Delta}^{\dagger}\psi_{\rm L} + y_{\rm c}\overline{\psi}_{\rm L}\phi_{\Delta}\psi_{\rm R} + \cdots\right ),
\end{multline}
interacting under $\rm SU(2)\times U(1)$ gauge interaction. Here, $\vec{W}^{\mu} = \vec{W}^{\mu}\cdot\vec{\sigma}/2 = (W_1^{\mu}, W_2^{\mu}, W_3^{\mu})\cdot(\sigma_1, \sigma_2, \sigma_3)/2$ and $A_{\mu}$ are the $\rm SU(2)$ and $\rm U(1)$ gauge fields, $\vec{\sigma} = (\sigma_1, \sigma_2, \sigma_3)$ are the Pauli vectors, $\psi_{\rm L} = \rm (Ag^{(1+)}, Ag^{(1-)})^{\rm T}$ and $\psi_{\rm R} = \rm Ag^{2+}$ and their Hermitian conjugate counterparts $\overline{\psi}_{\rm L} = \psi_{\rm L}^{*{\rm T}}\gamma^0, \overline{\psi}_{\rm R} = \psi_{\rm R}^{*{\rm T}}\gamma^0$ are 2D/3D Dirac spinors of the silver atoms [Kr]$4d^{10}5s^1$ (left-handed) and [Kr]$4d^95s^0$ (right-handed) respectively where electrons in the $5s$ orbital are the designated valence electrons of the Ag atoms commensurate with the valence states (bracketed) $(1+), (1-)$ and the oxidation state (not bracketed) $2+$, $d = 2,3$ is the dimensionality space, $q_{\rm e}, q_{\rm w}$ are the corresponding charge couplings of the gauge fields to the Ag Dirac fermions, $\partial_{\mu} = (\partial/\partial t, \partial/\partial x, \partial/\partial y, 2\Delta(d)\partial/\partial z)$ are space-time partial derivatives, $\Delta(d) = (d - 2)/2$ is the conformal dimension, $\gamma^{\mu} = (\gamma^0, \gamma^1, \gamma^2, 2\Delta(d)\gamma^3)$ are the Gamma matrices, $\phi_{\Delta} = \nu\exp(i\theta\gamma_{\Delta}^5)(0, \Delta(d))^{\rm T}$ (or its Hermitian conjugate $\phi_{\Delta}^{\dagger} = \nu\exp(-i\theta\gamma_{\Delta}^5)(0, 2\Delta(d))$) is a complex scalar field suchlike in eq. (\ref{Delta_eq}) arising from the lattice geometry/topology with $\Delta(d) = (d - 2)/2$ the conformal dimension, $\theta$ a real phase, $\nu$ a constant with mass dimensions and $\gamma_{\Delta}^5 = i\gamma^1\gamma^2(\gamma^3)^{2\Delta}$ the fifth gamma matrix, and $y_{\rm c}$ is a dimensionless (Yukawa) coupling. Clearly, $q_{\rm e} \simeq 1.602\times 10^{-19}$ C is the elementary charge. Unfortunately, we do not know \textit{a priori} the value of the $\rm SU(2)$ coupling, $q_{\rm w}$, \textit{albeit} a reasonable assumption below will set it to be $q_{\rm w} = q_{\rm e}$. Throughout, we have used Einstein summation convention and in the Lagrangian we have set the Fermi speed $v_{\rm F} = 1$ to simplify relativistic expressions.

Considering the mixing of the gauge fields by their corresponding couplings, 
\begin{align}\label{Weinberg_eq}
    \begin{pmatrix}
A^{\mu}\\ W_3^{\mu}
\end{pmatrix}
= 
\frac{1}{\sqrt{q_{\rm e}^2 + q_{\rm w}^2}}
\begin{pmatrix}
q_{\rm w} & q_{\rm e}\\ 
-q_{\rm e} & q_{\rm w}
\end{pmatrix}
\begin{pmatrix}
p^{\mu}\\ Z^{\mu}
\end{pmatrix},
\end{align}
into new superposition gauge fields $p_{\mu}$ and $Z_{\mu}$. This mixing is necessary to derive the relationship between the valence states and the oxidation states, thus treating them on equal footing. We should consider the linear combination of the generators corresponding to the Gellmann-Nishijima formula\cite{masese2023honeycomb}, $Q = 2I + Y$ in eq. (\ref{SU2U1_eq}) where $I = \sigma_3/2$ is the isospin with eigenvalues $\pm 1/2$ with $2I$ corresponding to the valency, $Y$ is the hypercharge corresponding to the initial oxidation state and $Q$ is effective oxidation state after $\rm SU(2)\times U(1)$ symmetry breaking. Thus, [Kr]$4d^{10}5s^1$ ($\psi_{\rm L} = \rm (Ag^{(1+)}, Ag^{(1-)})^{\rm T}$) is in valence state $2I = \pm 1$ \textit{albeit} oxidation state $Y = 0$ whereas [Kr]$4d^95s^0$ ($\psi_{\rm R} = \rm Ag^{2+}$) is in oxidation state $Y = +2$ \textit{albeit} valence state $2I = 0$. Thus, the Gellmann-Nishijima formula yields the effective oxidation states, $Q = +1, -1$ and $+2$ for the left-handed and right-handed Ag fermions respectively. Moreover, in the superposition, the mass-less photon $A_{\mu}$ which only couples to [Kr]$4d^95s^0$ is transformed into the mass-less $p_{\mu}$ photon which now couples to all the fermions with the number of effective charges $q_{\rm eff} = q_{\rm w}q_{\rm e}/\sqrt{q_{\rm e}^2 + q_{\rm w}^2}$ given by their effective oxidation ($Q$) values. 

\begin{figure*}[!t]
 \centering
 \includegraphics[width=\columnwidth]{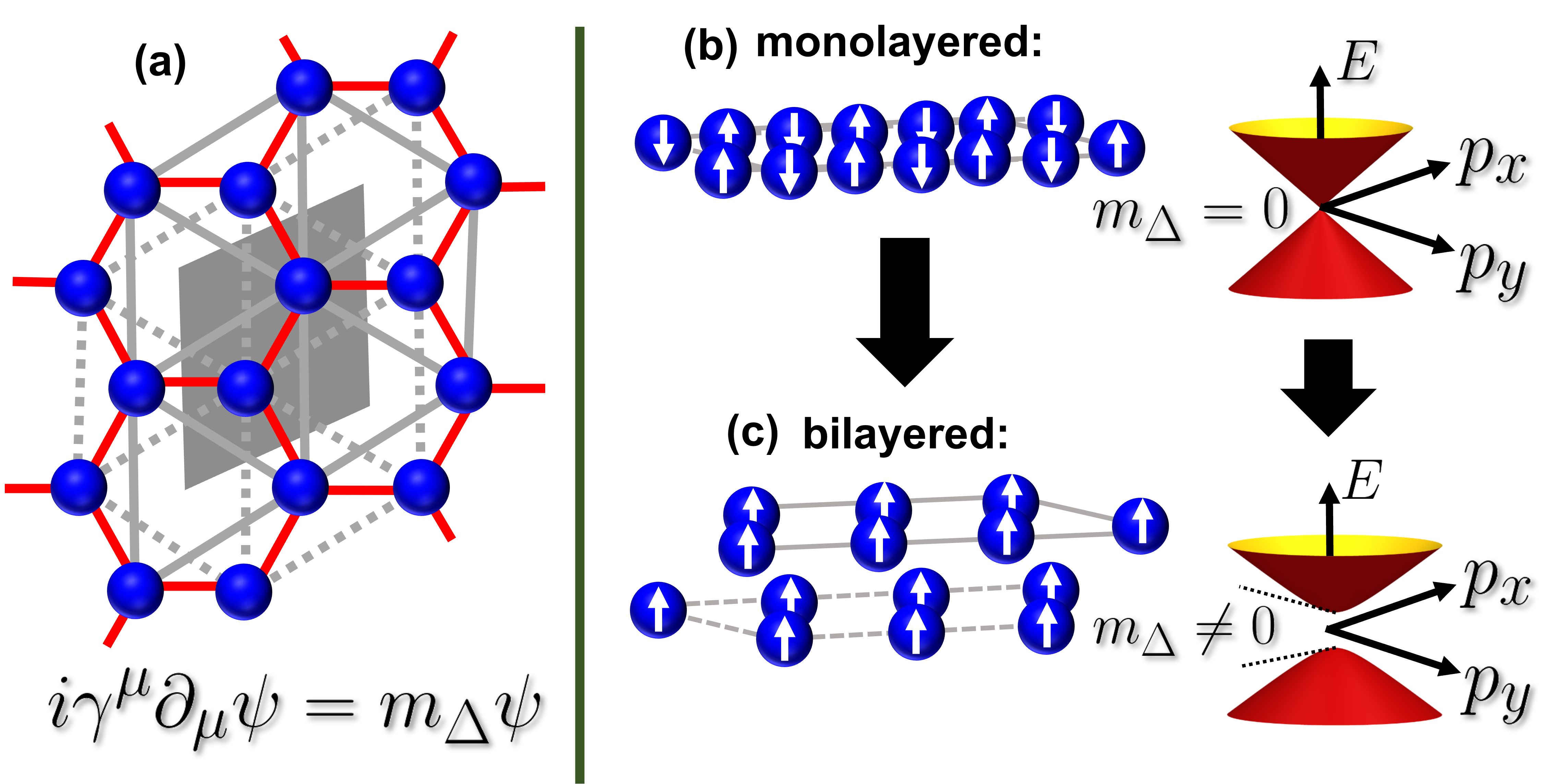}
 \caption{A rendition of the silver cations on the bipartite honeycomb lattice showcasing $\rm SU(2)\times U(1)$ and pseudo-spin models. (a) The bipartite honeycomb lattice (red solid lines) comprised of a pair of hexagonal lattices (grey solid and dashed lines) with Ag atoms at the vertices. The honeycomb lattice primitive cell (shaded in grey) is comprised of left- and right- chiral silver fermions $\psi = (\rm Ag^{2+}, Ag^{1-})^{\rm T} \coloneqq Ag^{1/2+}$ assumed to obey the massive Dirac equation $i\gamma^{\mu}\partial_{\mu}\psi = m_{\Delta}\psi$ at the vicinity of the honeycomb vertex, during (de)-intercalation processes; (b) A monolayer of Ag cations whereby anti-parallel pseudo-spin up ($\uparrow$) and down ($\downarrow$) degrees of freedom have been introduced on the honeycomb lattice. By analogy with Zeeman splitting, this configuration is plausible only for a vanishing pseudo-magnetic field proportional to the generated mass, $m_{\Delta} = 2m\Delta(d = 2) = 0$, where $\Delta(d) = (d - 2)/2$ is the conformal dimension. The energy ($E$)-momentum ($\vec{p}$) dispersion relation for 2D mobile cations near a honeycomb vertex is $E = v_{\rm F}\vec{p}$, where $\vec{v}_{\rm F}$ is the Fermi velocity and $\vec{p} = (p_x, p_y)$; (c) Emergence of Ag metallophilic interactions between the two constituent hexagonal lattices, which lead to the bifurcation of the honeycomb lattice stabilising the bilayered structure. Within the pseudo-spin model, this manifests as the parallel alignment of the pseudo-spins due to a spontaneous emergence of a pseudo-magnetic field proportional to the Dirac mass term, $m_{\Delta} = 2m\Delta(d = 3) = m \neq 0$. Thus, the 2D dispersion relation for mobile cations diffusing in 2D near a honeycomb vertex is $E = \sqrt{|\vec{p}|^2v_{\rm F}^2 + m^2v_{\rm F}^4}$, representing an energy gap in the energy spectrum near the Fermi energy. Similar condensed matter physics is expected for conduction electrons on the honeycomb lattice of Ag cations, expected to lead to a conductor/semi-conductor/insulator phase transition.\cite{kanyolo2022advances2, kanyolo2023pseudo}}
 \label{Figure_18}
\end{figure*}

\begin{figure*}[!t]
 \centering
 \includegraphics[width=\columnwidth]{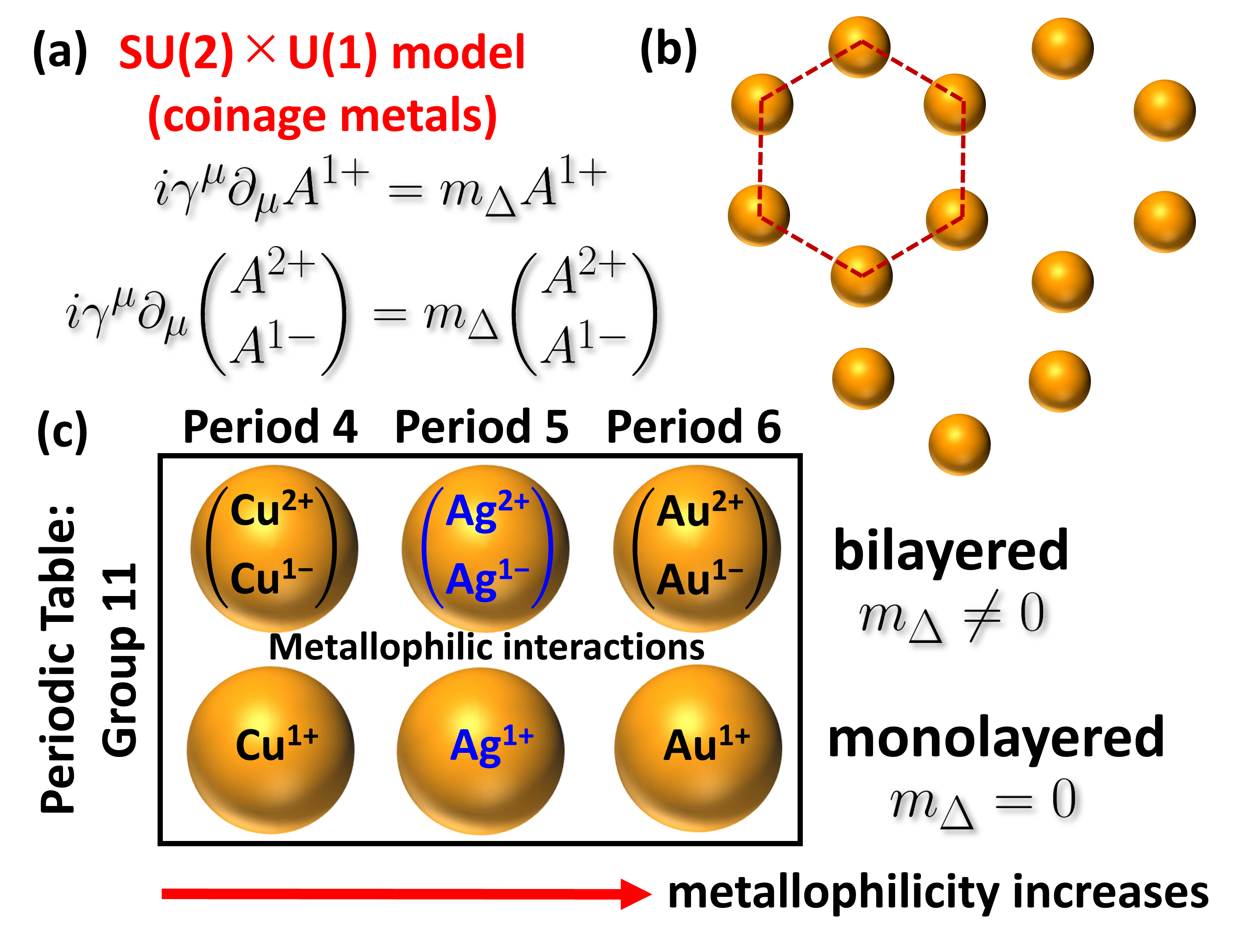}
 \caption{Similar $\rm SU(2)\times U(1)$ physics and chemistry of $A = \rm Cu, Ag, Au$ along group 11 elements of the periodic table (coinage metal atoms). (a) The Dirac equation in the $\rm SU(2)\times U(1)$ model for $A^{\alpha}$, where $\alpha = 1+, 2+, 1-$ is the valence/oxidation state of the coinage metal atom, valid only on the honeycomb lattice given in (b), where a coinage metal atom is located at a honeycomb vertex; (b) A honeycomb lattice of coinage metal atoms; (c) A depiction of the group 11 atoms in the periodic table of elements showing the massive and mass-less cations in the $\rm SU(2)\times U(1)$ model. Here, $A = \rm Cu, Au$ are in black and $\rm Ag$ is in blue. Metallophilicity increases with atomic size as shown by the red arrow, suggesting the generated mass $m_{\Delta}$ would increase in a similar fashion. Thus, this depiction is analogous to the 3 generations of leptons in the standard model.\cite{weinberg1967model}}
 \label{Figure_19}
\end{figure*}

Proceeding, given the elaborate nature of this Lagrangian, we shall focus on specific aspects related to the valence states and the emergence of metallophilic bilayers. In particular, applying the mixing in eq. (\ref{Weinberg_eq}) and the appropriate hypercharges, and only keeping the Ag fermion terms and alongside all the mass terms generated in the Lagrangian in eq. (\ref{SU2U1_eq}), we obtain,
\begin{multline}\label{SU2U1_eq2}
    \mathcal{L} = \int d^{\,d}x\,\left (\overline{\psi}\gamma^{\mu}i\partial_{\mu}\psi + \sqrt{2}q_{\rm w}\overline{\rm Ag^{1-}}\gamma^{\mu}{\rm Ag^{1+}}W_{\mu}^- + \sqrt{2}q_{\rm w}\overline{\rm Ag^{1+}}\gamma^{\mu}{\rm Ag^{1-}}W_{\mu}^+ + q_{\rm w}\overline{\rm Ag^{1+}}\gamma^{\mu}{\rm Ag^{1+}}W_{\mu}^3 - q_{\rm w}\overline{\rm Ag^{'1-}}\gamma^{\mu}{\rm Ag^{'1-}}W_{\mu}^3\right )\\
    + \int d^{\,d}x\,\left (q_{\rm e}\overline{\rm Ag^{2+}}\gamma^{\mu}{\rm Ag^{2+}}A_{\mu} -m_{\Delta}\left (\overline{\rm Ag^{2+}}{\rm Ag^{'1-}} + \overline{{\rm Ag^{'1-}}}{\rm Ag^{2+}}\right ) - \frac{1}{2}m_{\rm W}^2W^{+\mu}W_{\mu}^- - \frac{1}{2}m_{\rm Z}^2Z^{\mu}Z_{\mu}\right )\dots\\
    = \int d^{\,d}x\,\left ( \overline{\rm Ag^{1+}}\gamma^{\mu}i\partial_{\mu}{\rm Ag^{1+}} + \sqrt{2}q_{\rm w}\overline{\rm Ag^{'1-}}\gamma^{\mu}{\rm Ag^{1+}}W_{\mu}^- + \sqrt{2}q_{\rm w}\overline{\rm Ag^{1+}}\gamma^{\mu}{\rm Ag^{'1-}}W_{\mu}^+ + q_{\rm w}\overline{\rm Ag^{1+}}\gamma^{\mu}{\rm Ag^{1+}}W_{\mu}^3 \right )\\
    + \int d^{\,d}x\,\left ( \overline{\rm Ag^{1/2+}}\gamma^{\mu}i\partial_{\mu}{\rm Ag^{1/2+}} + q_{\rm eff}\overline{\rm Ag^{1/2+}}\gamma^{\mu}{\rm Ag^{1/2+}}p_{\mu} + \frac{(q_{\rm e}^2 - q_{\rm w}^2)}{\sqrt{q_{\rm w}^2 + q_{\rm e}^2}}\overline{\rm Ag^{1/2+}}\gamma^{\mu}{\rm Ag^{1/2+}}Z_{\mu} \right )\\
    + \int d^{\,d}x \left ( - m_{\Delta}\overline{\rm Ag^{1/2+}}{\rm Ag^{1/2+}} - \frac{1}{2}m_{\rm W}^2W^{+\mu}W_{\mu}^- - \frac{1}{2}m_{\rm Z}^2Z^{\mu}Z_{\mu} \right )\dots, 
\end{multline}
where $\psi = ({\rm Ag^{1+}, Ag^{'1-}, Ag^{2+}})^{\rm T}$, ${\rm Ag^{1/2+}} = ({\rm Ag^{2+}, Ag^{'1-}})^{\rm T}$, $\overline{{\rm Ag^{1/2+}}} = {\rm Ag^{1/2+}}^{\dagger}\gamma^0$, $W_{\mu}^{\pm} = \frac{1}{\sqrt{2}}(W_{\mu}^1 \pm iW_{\mu}^2)$, ${\rm Ag^{'1-}} = \exp(-i\theta\gamma_{\Delta}^5){\rm Ag^{1-}}$ is a rotation by a chiral phase factor, $m_{\Delta} = y_{\rm c}\nu\Delta(d)$ is the Dirac mass finite only in $d = 3$ dimensions, $m_{\rm Z} = m_{\rm W}\sqrt{1 + q_{\rm e}^2/q_{\rm w}^2}$ and $m_{\rm W} = q_{\rm W}\nu\Delta(d)$ are the masses of the gauge fields after the mixing. Note that, we have dropped the brackets in the superscript since all the fermions are now considered to be in their oxidation states with respect to the gauge fields. Essentially, this Lagrangian explains the origin of silver sub-valency by considering combinations such as $\rm Ag_2^{1/2+} = Ag^{2+}Ag^{1-}$ and $\rm Ag_3^{2/3+} = Ag^{2+}Ag^{1-}Ag^{1+}$, which also implies interactions between these degenerate states constituting the argentophilic interactions. Based on this, it is prudent to consider the following re-scaling in eq. (\ref{SU2U1_eq2}): $p^{\mu} \rightarrow \sqrt{2}p^{\mu}$; $\rm Ag^{1/2+} \rightarrow Ag^{1/2+}/\sqrt{2}$ (hence $\overline{\rm Ag^{1/2+}} \rightarrow \overline{\rm Ag^{1/2+}}/\sqrt{2}$); and $\partial_{\mu} \rightarrow 2\partial_{\mu}$, which as a consequence re-scales the square-root term as $\sqrt{2}q_{\rm w} \rightarrow q_{\rm w}$ and more importantly the $p_{\mu}$ coupling term as,
\begin{align}\label{q_eff_dash_eq}
    q_{\rm eff} = \frac{q_{\rm e}q_{\rm w}}{\sqrt{q_{\rm e}^2 + q_{\rm w}^2}} \rightarrow q'_{\rm eff} = \frac{q_{\rm e}q_{\rm w}}{\sqrt{2q_{\rm e}^2 + 2q_{\rm w}^2}}.
\end{align}
Thus, we find the re-scaled coupling to $p_{\mu}$ to be given by $q'_{\rm eff} = q_{\rm e}q_{\rm w}/\sqrt{2q_{\rm e}^2 + 2q_{\rm w}^2} = q_{\rm e}/2$, and thus $q_{\rm e} = q_{\rm w}$, which also eliminates interactions of $\rm Ag^{1/2+}$ with the $Z_{\mu}$ electromagnetic interaction (boson). Interactions involving $Z_{\mu}$ and $W_{\mu}^{\pm}$ bosons are screened over length scales $h/2\pi m_{\rm Z}v_{\rm F}$ and $h/2\pi m_{\rm W}v_{\rm F}$ respectively where $v_{\rm F}$ is the Fermi speed of the cation on the honeycomb lattice and $h \simeq 6.626\times 10^{-34}$ Js is the Planck constant. Amongst other non-trivialities, this implies transitions such as $\sqrt{2}q_{\rm w}\overline{\rm Ag^{1-}}\gamma^{\mu}{\rm Ag^{1+}}W_{\mu}^-$ have a finite energy cost, of order $3.5 {\rm eV} \sim m_{\rm W}v_{\rm F}^2$ (see eq. (\ref{mass_WZ_eq})). Moreover, the Lagrangian predicts $W_{\mu}^3$ interactions with $\rm Ag^{1+}$ on the honeycomb lattice are long range since, like $p_{\mu}$, $W_{\mu}^3$ is also mass-less. A depiction of the honeycomb lattice bifurcation into a metallophilic bilayer is availed in \textbf{Figure \ref{Figure_18}}. Similar considerations for group 11 atoms, $\rm Cu$ and $\rm Au$ (\textbf{Figure \ref{Figure_19}}) predict metallophilic bilayers in coinage metal atom-based honeycomb layered frameworks. Finally, the theory makes limited use of the $Z_{\mu}$ boson, which is entirely decoupled from $\rm Ag$ interactions.

Moreover, de-localisation of the usually itinerant $\rm 5s^1$ valence electron in [Kr]$4d^{10}5s^1$ is impeded by the closed-shell pairing in the $5s^2$ state of [Kr]$4d^95s^2$, which offers a straightforward localisation scheme. Thus, this work makes a giant analytic leap towards understanding the origin of sub-valency and argentophilicity, without employing any \textit{ab initio} DFT computational methods. Particularly, \textit{ab initio} DFT computations (including but not limited to the most up-to-date packages such as The Vienna \textit{Ab initio} Simulation Package (VASP) and CRYSTAL23,\cite{hafner2008ab, kresse1993ab, kresse1994ab, erba2022crystal23}) employ a Hamiltonian diagonalisation scheme such as the Hatree-Fock approximation and the Kohn-Sham formalism for the many-electron Schr\"{o}dinger equation designating the electrons that do not take part in bonding interactions as core and the rest as valence electrons, thus finding the minimum energy (ground state) configuration of the system designating it as the appropriate valence state configuration of the ions in the bond. Thus, it is prudent to consider where the limitations with such diagonalisation schemes of many-body Schr\"{o}dinger equation lie, especially in predicting a degenerate silver ground state. 

\begin{figure*}[!t]
 \centering
 \includegraphics[width=0.8\columnwidth]{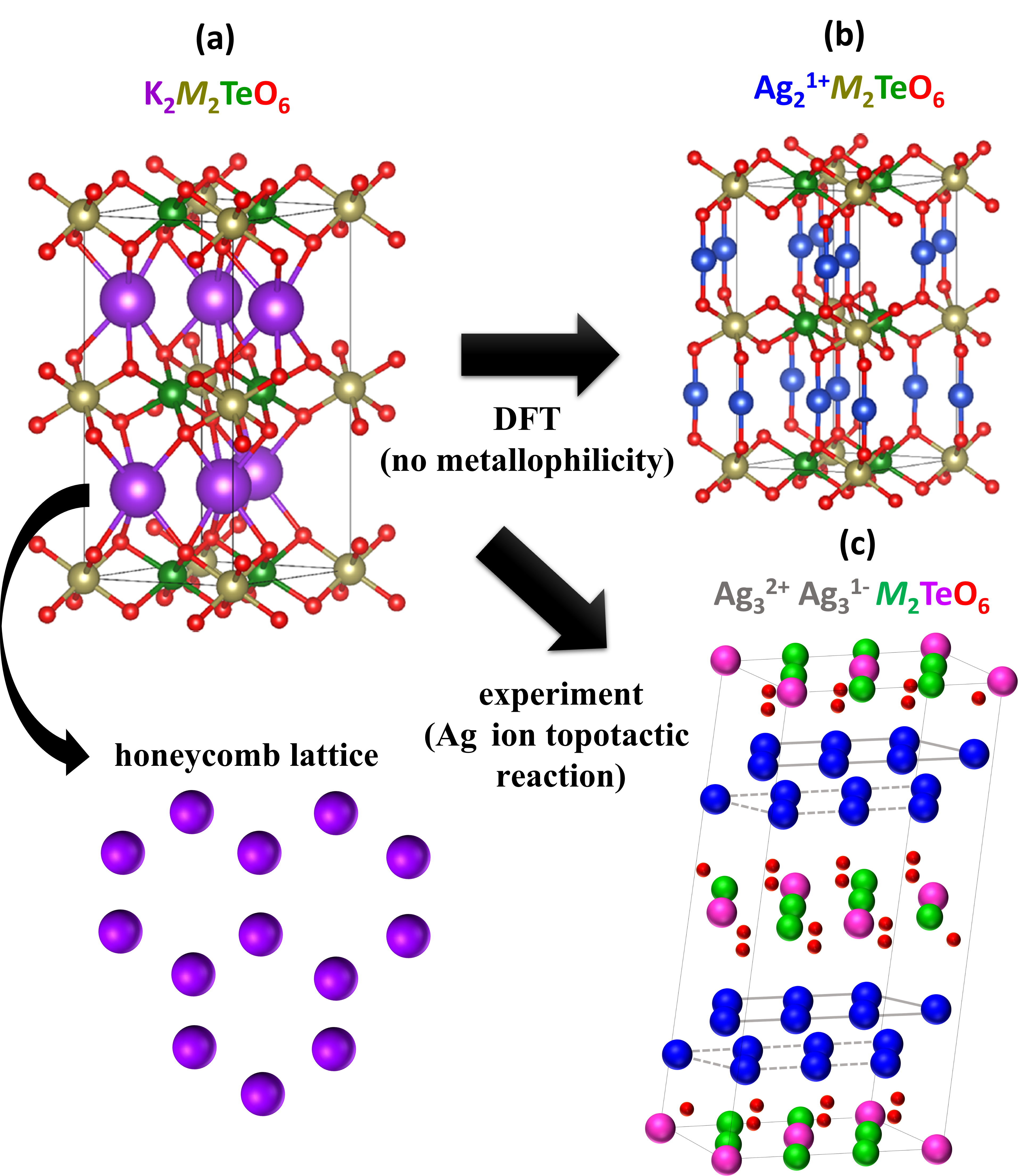}
 \caption{Differing Density Functional Theory (DFT) results (using the Vienna \textit{Ab initio} Simulation Package (VASP) program, particularly the Kohn-Sham formalism) from the experimentally-obtained bilayered material ${\rm Ag}_6M_2\rm TeO_6$ ($M = \rm Ni, Co, Mg$ \textit{etc}).\cite{masese2023honeycomb, tada2022implications} (a) The structure of the honeycomb layered tellurate $\rm K_2Ni_2TeO_6$ exhibiting a monolayered honeycomb lattice of $\rm K$ cations (inset: below); (b) The simulated honeycomb layered tellurate ${\rm Ag_2^{1+}}M_2\rm TeO_6$ using the VASP program, exhibiting a honeycomb lattice of $\rm Ag^{1+}$ (monolayered). The simulation package employs the Kohn-Sham formalism which prsently does not make distinctions between the degenerate valency states of $\rm Ag$, thus predicting only (non-)metallophilic monolayered structures\cite{tada2022implications}; (c) The experimentally observed ${\rm Ag_3^{2+}Ag_3^{1-}}M_2{\rm TeO_6}$ with $\rm Ag$ metallophilic bilayers, obtained from the precursor materials ${\rm K_2}M_2\rm TeO_6$ ($M = \rm Ni, Co, Mg$ \textit{etc}) via topotactic ion exchange.\cite{masese2023honeycomb}}
 \label{Figure_20}
\end{figure*}

We shall consider a simplistic argument that highlights the success of the $\rm SU(2)\times U(1)$ model relative to typical Hamiltonian diagnalisation schemes such as the Kohn-Sham formalism employed in \textit{ab initio} DFT calculations.\cite{tada2022implications} In particular, the characteristic polynomial, $\det(H - xI)$ of a three-fold degenerate Hamiltonian, $H$ on a given 2D lattice, where $I$ is the identity matrix, can be expressed as a cubic equation of the form,
\begin{align}\label{degeneracy_eq}
    \det(H - xI) = (x - E)(x - E)(x - E) = 0,
\end{align}
which only has one solution given by $x = E$ corresponding to the degenerate ground state energy, $E$. Thus, in Hamiltonian diagonalisation schemes in DFT computations, these degenerate states are indistinguishable. On the other hand, according to a theorem by Peierls applied to 2D\cite{peierls1991more}, physical systems \textit{abhor} such degeneracies which means crystalline distortions, \textit{e.g.} in the form of a phase transition, must arise in such a way as to lift the degeneracy. In the $\rm SU(2)\times U(1)$ model\cite{masese2023honeycomb}, these degenerate states can transform into each other via electron-photon and electron-gauge interactions giving rise to a continuous $\rm SU(2)\times U(1)$ gauge symmetries on the lattice. Peierls theorem\cite{peierls1991more} offers us a path towards analytically breaking these continuous symmetries, thus obtaining a unique ground state with a lower energy than the ground state obtained in DFT with $\rm Ag^{1+}$ oxidation state. The manifestation of this new physics will primarily depend on whether the 2D lattice of silver is honeycomb or hexagonal. In particular, for the hexagonal case, the revised characteristic polynomial instead takes the form,
\begin{align}
    \det(H - xI) = (x - E_{-1})(x - E_{+1})(x - E_{+2}) = 0,
\end{align}
where the eigenvalues $x = E_{+1}, E_{-1}, E_{+2}$ are the three lowest energy solutions corresponding to the three silver valence states $\rm Ag^{1+}$, $\rm Ag^{1-}$ and $\rm Ag^{2+}$ with $0 < E_{+1} < E_{-1} < E_{+2}$, implying the unique and non-degenerate ground state on the hexagonal lattice $E = E_{+1}$ corresponds to $\rm Ag^{1+}$. The situation differs greatly on the bipartite honeycomb lattice comprising a pair of hexagonal lattices. A cation in one hexagonal lattice interacts with another in the other via the non-Abelian $\rm SU(2)\times U(1)$ gauge fields which introduces the three fold degeneracy displayed in eq. (\ref{degeneracy_eq}). This means that this vacuum degeneracy must be broken by a Peierls instability corresponding to the mixing of eigenvalues of the form, $x = E_{+1}, (E_{-1} + E_{+2})/2, (E_{-1} - E_{+2})/2$, where the non-degenerate ground state is given by $x = (E_{-1} - E_{+2})/2 = E < 0$ (which can be verified using $0 < E_{+1} < E_{-1} < E_{+2}$). In the $\rm SU(2)\times U(1)$ model, this mixing of energy eigenvalues also leads to cationic valence electron mixing via $sd$ hybridisation leading to the emergence of the sub-valent state $\rm Ag_2^{1/2+} = Ag^{2+}Ag^{1-}$. Unlike the individual degenerate states that are massless on the honeycomb lattice, this sub-valent state is a massive mode which manifests as a bifurcated (3D) honeycomb lattice stabilised by argentophilic bonds. Incidentally, this argument explains the lack of metallophilic bilayers predicted by Density Functional Theory (DFT) calculations under the Kohn-Sham diagonalisation scheme, as depicted in \textbf{Figure \ref{Figure_20}}. 

Moreover, analogous to graphene where a pseudo-spin degree of freedom originates from the dynamics of the carbon $p_z^1$ electrons, it is possible to introduce pseudo-spins on the bipartite honeycomb lattice inherited from the behaviour of the half-filled $4d_{z^2}^1$ orbitals of silver, provided it is energetically isolated from the rest of $4d^9$ orbitals by crystal field splitting linked to the particular coordination chemistry of silver to oxygen or halide anions.\cite{masese2023honeycomb} Thus, under additional assumptions\cite{kanyolo2022advances2}, the Dirac mass term can be transformed into the form, 
\begin{multline}\label{H_Dirac_eq}
    \mathcal{L}_{\rm Dirac\,\,mass} \simeq  -\int d^{\,d}x\,\,m_{\Delta}\overline{\rm Ag^{1/2+}}{\rm Ag^{1/2+}}
    \propto \int d^{\,d}x\,\,\left ( -  T|A|^2\left (\sum_{\ell} G(i\omega_{\ell}, \vec{x})\right )^2S_1S_2 -\mathcal{B}_{\Delta}(S_1 + S_2)\right )\\
    = \int d^{\,d}x\,\,\left (-J_{\rm RKKY}S_1S_2 - \mathcal{B}_{\Delta}(S_1 + S_2)\right ),
\end{multline}
where $G(i\omega_{\ell}, \vec{x})$ is the fermionic Green function\cite{aristov1997indirect}, $|A|^2$ is the crystal field splitting term, $S_1, S_2$ are pseudo-spin degrees of freedom, $\mathcal{B}_{\Delta}$ is the pseudo-magnetic term (such as in eq. (\ref{old_condition_eq}) and eq. (\ref{old_condition_eq2})) representing a structural phase transition (bifurcation) of the honeycomb lattice responsible for `orbital ordering' of the degenerate $4d_{z^2}$ and $5s$ orbitals depicted in \textbf{Figure \ref{Figure_17}(c)} and \textbf{Figure \ref{Figure_17}(d)} introducing an energy gap which renders the $5s$ orbital at a slightly higher energy designating its electrons as the valence electrons, and $\omega_{\ell} = \pi T(2\ell + 1)$ are fermionic matsubara frequencies. This offers a way to incorporate the $\rm SU(2)\times U(1)$ model in computations, whereby simulations can be performed in a similar manner to ferromagnetic/anti-ferromagnetic calculations. Note that, this dual description makes use of the fact that the honeycomb lattice is bipartite where $S_1, S_2$ is the $z$-component of pseudo-spin in the honeycomb unit cell. The first (Heisenberg) term is the $sd$-hybridisation term also responsible for degenerate states, which corresponds to the Ruderman-Kittel-Kasuya-Yosida (RKKY) interaction\cite{kanyolo2022advances, kanyolo2022advances2, aristov1997indirect},
\begin{align}
    J_{\rm RKKY} = |A|^2T\left (\sum_{\ell} G(i\omega_{\ell},  \vec{x})\right )^2 \propto \left (\cos(2k_{\rm F}R)/R^3 - \sin{(2k_{\rm F}R)}/R^4\right ),
\end{align}
with $k_{\rm F}$ the Fermi wave vector of the $5s^2$ conduction electrons and $R = |\vec{x}|$. Within this pseudo-spin picture, the bifurcation of the bipartite honeycomb lattice into its constituent hexagonal lattice forming argentophilic bilayers is a form of `Zeeman effect' due to a finite pseudo-magnetic field $\mathcal{B}_{\Delta} \propto m_{\Delta}/R^2$ proportional to the generated Dirac mass, arising spontaneously on the honeycomb lattice as depicted in \textbf{Figure \ref{Figure_18}}. 

Finally, whilst the precise nature of this bifurcation is underexplored, evidently the 2D-to-3D crossover/ monolayer-bilayer phase transition avoids violating the Coleman-Mermin-Wagner theorem which states that, spontaneous breaking of a continuous symmetry (in this case $\rm SU(2)\times U(1)$) is impossible in 2D.\cite{zee2010quantum} In addition, since the honeycomb lattice plays an important role in the stability of bilayered structures, this lends credence to the deployment of geometry theories to augment descriptions of phenomena such as Peierls instability not readily treatable by valence bond theory. It is worth noting that argentophilicity is a blanket word for the unconventional silver-silver interactions in materials, with the interaction potential for such bonds potentially taking different forms as already appears to be the case in elemental silver (longer bonds of order $2.89$ \AA) compared to the present class of honeycomb layered oxides with shorter argentophilic bonds ($\leq 2.83$ \AA). Within the $\rm SU(2)\times U(1)$ model, it appears one could consider, $\rm Ag_2^0 = Ag^{1+}Ag^{1-}$ interactions in the silver metal, instead of just $\rm Ag^0$, whereby the argentophilic interaction is $\rm Ag^{1+} - Ag^{1-}$ instead of $\rm Ag^{2+} - Ag^{1-}$ which arises in $\rm Ag^{1/2+}_2 = Ag^{2+}Ag^{1-}$ in the case of argentophilic bilayers. As a result of this construction, the bond lengths are bound to differ, and more crucially no sub-valency is required in the silver metal. Alternatively, it is conceivable that the $\rm Ag^+ - Ag^+$ interactions in elemental silver are exemplars of metallic interactions, which are thought to be mediated by itinerant $5s^1$ electrons. We also note that, chemical reactions such as eq. (\ref{AgF_Ag_eq}) may be indicative of a larger gauge symmetry such as $\rm SU(3)\times U(1)$ involving $\psi_{\rm L} = (\rm Ag^{(0)}, Ag^{(1+)} Ag^{(1-)})$ and $\psi_{\rm R} = \rm Ag^{2+}$ not considered in this {\it Review}. 
In other words, these considerations imply that testing the mechanisms for argentophilic interactions in some other analytical or computational framework is likely not to translate straightforwardly into testing the $\rm SU(2)\times U(1)$ model, which already yields promising analytic results.

\newpage

\section{\label{Section: Prospects} Frontier Directions and Conclusion}

\subsection{Synthetic Approaches, Novel Design Strategies and Exploring Uncharted Compositional Spaces}

Within the domain of honeycomb layered frameworks featuring metallophilic bilayers, the potential for further advancement remains abundant. A significant milestone in this trajectory was exemplified by the successful synthesis of $\rm Ag_2NiO_2$ through the electrochemical $\rm Ag$-ion insertion/intercalation of delafossite $\rm AgNiO_2$ within a solid-state $\rm Ag$-ion battery configuration (as shown in \textbf {Figure \ref{Figure_5}}).\cite{liu2007} This achievement showcases the feasibility of the electrochemical $\rm Ag$-ion insertion/intercalation technique as a propitious strategy to diverse $\rm Ag$-based layered compositions. Within this broad class, the electrochemical $\rm Ag$-ion insertion of delafossite structures (such as $\rm Ag{\it M}O_2$ (where $M$ denotes $\rm Fe, Al, Ga, In, Sc, Co$), $\rm Ag{\it M}S_2$ ($M$ = $\rm Ni, Cr$), $\rm Ag{\it M}Te_2$ ($M$ = $\rm Cr, Ni$) and $\rm Ag{\it M}Se_2$ ($M$ = $\rm Cr, Ni$))\cite {shannon1971chemistrya, shannon1971chemistryb, okamoto1972, bongers1968, aliev1981} and the honeycomb layered materials including $\rm Ag_3{\it M}_2SbO_6$ (where $M$ represents $\rm Ni, Co, Cu, Zn, Mg$),\cite {politaev2010, zvereva2016} $\rm Ag_3{\it M}_2BiO_6$,\cite {berthelot2012} and $\rm Ag_2{\it M}_2TeO_6$ \cite {masese2023honeycomb} (as summarised in \textbf {Table \ref {Table_14}}), would emerge as a gateway to accessing novel layered frameworks manifesting $\rm Ag$ bilayers. These pioneering endeavours will not only expand the frontiers of scientific inquiry but also hold transformative potential for the development of functional materials with enhanced properties and multifaceted applications. 

\begin{figure*}[!t]
 \centering
 \includegraphics[width=0.65\columnwidth]{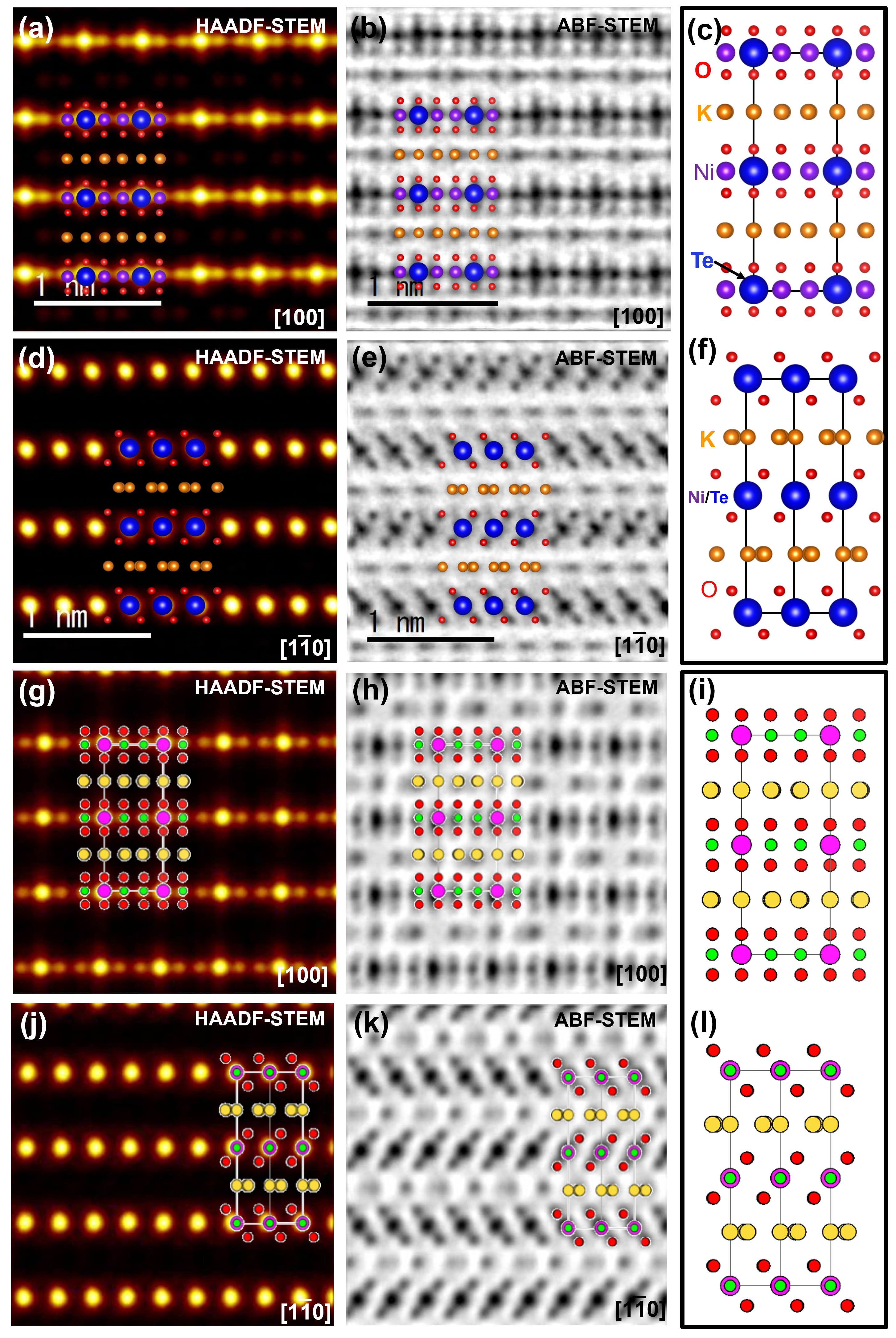}
 \caption{Exemplar layered material manifesting honeycomb arrangement of mobile cations. (a) HAADF-STEM image of $\rm K_2Ni_2TeO_6$ taken along [100] zone axis showing the ordering sequence of $\rm Ni$ and $\rm Te$ atoms. Inset shows a projected model of the crystal structure, for clarity. (b) Corresponding ABF-STEM image taken along [100] zone axis displaying the arrangement of potassium atoms occupying a honeycomb lattice. (c) A rendition of the crystal structure of $\rm K_2Ni_2TeO_6$ along the [100] direction. (d) HAADF-STEM visualisation along the [1$\overline{1}$0] zone axis, and (e) Corresponding ABF-STEM image. (f) Projection of the crystal structure along [1$\overline{1}$0], as shown in d and e. (g) HAADF-STEM and (h) ABF-STEM images of $\rm Na_2Ni_2TeO_6$ taken along [100] zone axis. (i) A rendition of the crystal structure of $\rm Na_2Ni_2TeO_6$ along the [100] direction. (j) HAADF-STEM and (k) ABF-STEM visualisation along the [1$\overline{1}$0] zone axis. (l) Projection of the crystal structure along [1$\overline{1}$0], as shown in j and k. Figures reproduced with permission.\cite {masese2021unveiling, masese2021topological} Copyright 2021 American Chemical Society and Copyright 2021 Elsevier.}
 \label{Figure_21}
\end{figure*}

In addition, the fortuitous discovery of topochemical ion-exchange in layered materials with a honeycomb arrangement of cations, exemplified by compounds like $\rm Na_2Ni_2TeO_6$ and $\rm K_2Ni_2TeO_6$,\cite {masese2021unveiling, masese2021topological} has emerged as a pivotal approach for accessing honeycomb layered materials featuring a bilayer arrangement of coinage metal atoms, such as $\rm Ag$ in $\rm Ag_2{\it M}_2TeO_6$ (where $M$ encompasses $\rm Co, Cu, Ni, Cu, Zn$).\cite {masese2023honeycomb} This discovery holds the potential to invigorate not only the ongoing advancements in compositional tuneability but also the structural versatility within the realm of honeycomb layered oxides. The investigation of the extensive array of layered compounds featuring a honeycomb lattice structure comprised of mobile cations, and the assessment of their potential for topochemical ion exchange utilising metallophilic cations, presents a highly promising avenue for the synthesis of a diverse range of layered materials characterised by the presence of metallophilic bilayers. Within this context, \textbf {Figure \ref {Figure_21}} depicts transmission electron microscopy images showcasing the honeycomb layered configuration of $\rm {\it A}_2Ni_2TeO_6$ ($\rm {\it A}$ = $\rm Na, K$),\cite {masese2021unveiling, masese2021topological} wherein both the arrangement of transition metals and the positioning of mobile cations exhibit a honeycomb pattern. By employing topochemical Ag-ion exchange, these honeycomb layered materials underwent transformation (\textit {i.e.}, bifurcation of the honeycomb lattice in the presence of argentophilic cations), leading to the emergence of novel layered structures featuring $\rm Ag$ bilayers (\textbf{Figure \ref{Figure_2}}).\cite {masese2023honeycomb} Related layered materials that can be explored via topochemical ion-exchange are also shown in \textbf {Table \ref {Table_14}}. As we venture deeper into the intricacies of topochemical ion-exchange, the intricate interplay between the metallophilic coinage metal atoms and the honeycomb lattice in which they reside unravels tantalising prospects for tailoring the properties and functionalities of these compounds. This interplay offers an avenue for precise control and manipulation, opening up new avenues for the design and development of materials with tailored characteristics and desirable applications.

Meanwhile, high-pressure synthetic pathways present untapped potential for the design of novel structures, as exemplified by the successful endeavours undertaken with $\rm Ag_2{\it M}O_2$ materials\cite{schreyer2002synthesis, yoshida2020static, taniguchi2020butterfly, yoshida2011novel, matsuda2012partially, yoshida2008unique, yoshida2006spin, sorgel2007ag3ni2o4, yoshida2020partially, sugiyama2009static, liu2007, Yoshida2007impurity, Yoshida2007impurity, taniguchi2018fabrication} (where $M$ encompasses $\rm Fe, Co, Ni, Cr, Mn, {\it etc.}$). Additionally, the realm of high-temperature metathetic routes remains relatively unexplored, holding promise as a frontier ripe for exploitation in the quest for new materials. Furthermore, the power of computational tools can greatly enhance exploratory synthetic routes. From density functional theory (DFT) calculations to machine learning algorithms, these computational approaches can enable expedited {\it in silico} design of new layered framework compositions featuring the bilayer arrangement of cations. The integration of experimental exploration with computational methodologies will not only accelerate the discovery process but also expand the design frontiers of such layered frameworks.

\begin{table*}
\caption{Feasible synthesis techniques for designing new layered frameworks manifesting metallophilic bilayers experimentally not yet discovered, as presently surveyed.}\label{Table_14}
\begin{center}
\scalebox{0.55}{
{\color{black}\begin{tabular}{lcl} 
\hline
\textbf{Synthesis Technique} & \textbf{Precursors} & \textbf{Expected Compound} \\ 
\hline\hline
& & \\
\textbf{Solid-State Reaction } (high temperature/high pressure) & $\rm Ag$, $\rm Ag_2O$, $\rm Ga_2O_3$ & $\rm Ag_2GaO_2$\\
 & $\rm Ag$, $\rm Ag_2O$, $\rm In_2O_3$ & $\rm Ag_2InO_2$\\
 & $\rm Ag$, $\rm Ag_2O$, $\rm Sc_2O_3$ & $\rm Ag_2ScO_2$\\
 & $\rm Ag$, $\rm Ag_2O$, $\rm Al_2O_3$ & $\rm Ag_2AlO_2$\\
 & $\rm Ag$, $\rm Ag_2O$, $\rm In_2O_3$, $\rm {\it M}_2O_3$ ($M$ = Cr, Fe, Mn, Co)  & $\rm Ag_2{\it M}_{1-{\it x}}In_{\it x}O_2$\\
 & $\rm Ag$, $\rm Ag_2O$, $\rm {\it M}O$ ($M$ = Ni, Co, Zn, Cu, Mg) & $\rm Ag_6{\it M}TeO_6$\\
 & & \\
\textbf{Electrochemical Ag-ion intercalation} (All-solid-state battery setup) & delafossite $\rm Ag{\it M}O_2$ ($M$ = Fe, Al, Ga, In, Sc, Co)  & $\rm Ag_2{\it M}O_2$ ($M$ = Fe, Al, Ga, In, Sc, Co) \\
 & $\rm Ag{\it M}S_2$ ($M$ = Ni, Cr) & $\rm Ag_2{\it M}S_2$ \\
 & $\rm Ag{\it M}Se_2$ ($M$ = Ni, Cr) & $\rm Ag_2{\it M}Se_2$ \\
 & $\rm Ag{\it M}Te_2$ ($M$ = Ni, Cr) & $\rm Ag_2{\it M}Te_2$ \\
 & $Ag_3M_2D\rm O_6$ ($M = {\rm Ni, Co, Mg, Zn, Mn}$; $D = \rm Bi, Sb, Ru$) & $Ag_6M_2D\rm O_6$ \\
 & &  \\
\textbf{Topochemical Ag ion-exchange} (molten Ag salts) &  $\rm Na_{2/3}\,(Mg_{1/3}Mn_{2/3})O_2$, $\rm Na_{2/3}\,(Ni_{1/3}Mn_{2/3})O_2$,  & Ag-based compounds with bilayer domains \\
& $\rm Na_{2/3}Ni_{1/6}Mg_{1/6}Ti_{2/3}O_2$, ${\rm Na}_x{\rm Ni}_{x/2}{\rm Mn}_{1-x/2}\rm O_2$ ($2/3 \leq x < 4/5$), &  \\
& $\rm Na_{0.76}Cu_{0.22}Fe_{0.11}Mn_{0.67}O_2$, $\rm Na_{2/3}Fe_{2/3}Mn_{1/3}O_2$, $\rm Na_{2/3}Fe_{1/2}Mn_{1/2}O_2$, &  \\
 & $\rm NaNi_2BiO_{6-\delta}$, $\rm Na_{2/3}Mg_{0.28}Mn_{0.72}O_2$, $\rm Na_{2/3}Mg_{1/3}Ti_{2/3}O_2$, $\rm Na_{2/3}Fe_{1/3}Mn_{2/3}O_2$, &  \\
  & $\rm Na_{2/3}CoO_2$, $\rm Na_{2/3}Mg_{1/4}Mn_{3/4}O_2$, $\rm Na_{7/9}Ni_{0.20}Co_{0.38}Mn_{0.42}O_2$, $\rm Na_{2/3}Mn_{8/9}Al_{1/9}O_2$, &  \\
  & $\rm Na_{0.76}Cu_{0.22}Fe_{0.11}Mn_{0.67}O_2$, $\rm Na_{7/9}Ni_{0.15}Fe_{0.48}Mn_{0.37}O_2$, $\rm Na_{0.7}Ni_{0.35}Sn_{0.65}O_2$, &  \\
   & $\rm Na_{2/3}Ni_{2/3}Te_{1/3}O_2$, $\rm Na_{0.66}NbO_2$, $\rm Na_{0.51}MnO_{0.93}$, $\rm Na_{0.628}Fe_{0.03}Co_{0.97}O_2$, $\rm Na_{0.74}Ni_{0.58}Sb_{0.42}O_2$,  &  \\
    & $\rm Na_{2/3}Ni_{1/3}Ti_{2/3}O_2$, $\rm Na_{2/3}Mn_{1/3}Fe_{1/3}Co_{1/3}O_2$, $\rm Na_{0.6}MnO_2$, $\rm Na_{0.6}Cr_{0.6}Ti_{0.4}O_2$,  &  \\
     & $\rm Na_{0.6}Mn_{0.65}Ni_{0.25}Co_{0.1}O_2$, $\rm Na_{0.653}Mn_{0.929}O_2$, $\rm Na_{2/3}Mn_{0.65}Ni_{0.2}Co_{0.15}O_2$, $\rm Na_{2/3}MnO_2$,  &  \\
& $\rm Na_{2/3}Mg_{0.28}Mn_{0.72}O_2$, $\rm Na_{2/3}Mn_{7/9}Zn_{2/9}O_2$, $\rm Na_{2/3}CoO_2$, $\rm Na_{2/3}Co_{1/2}Mn_{1/2}O_2$,  &  \\
 & $\rm Na_{2/3}Co_{2/3}Mn_{2/9}Ni_{1/9}O_2$, $\rm Na_{2/3}Co_{0.25}Mn_{0.65}Cr_{0.10}O_2$, $\rm Na_{2/3}Ni_{0.1}Cu_{0.2}Mn_{0.7}O_2$,  &  \\
  & $\rm Na_{2/3}Ni_{1/3}Mn_{1/2}Ti_{1/6}O_2$, $\rm Na_{2/3}Ni_{0.22}Cu_{0.11}Mn_{0.56}Ti_{0.11}O_2$, $\rm Na_{2/3}Fe_{1/5}Mn_{4/5}O_2$, &  \\
   & $\rm Na_{2/3}Mn_{4/5}Fe_{1/10}Ti_{1/10}O_2$, $\rm Na_{2/3}Fe_{0.20}Ni_{0.15}Mn_{0.65}O_2$, $\rm Na_{2/3}Ni_{0.26}Zn_{0.07}Mn_{2/3}O_2$, &  \\
 & $\rm Na_{2/3}Ni_{0.23}Mg_{0.1}Mn_{2/3}O_2$, $\rm Na_{2/3}Mn_{1/2}Fe_{1/4}Co_{1/4}O_2$, $\rm Na_{2/3}Mn_{0.65}Co_{0.2}Ni_{0.15}O_2$,  &  \\
  & $\rm Na_{2/3}Ni_{2/5}Co_{1/5}Mn_{2/5}O_2$, $\rm Na_{2/3}Cr_{1/3}Ti_{2/3}O_2$, $\rm Na_{0.7}MnO_2$, $\rm Na_{0.75}CoO_2$, &  \\
   & $\rm Na_{0.70}Mn_{3/5}Ni_{3/10}Co_{1/10}O_2$, $\rm Na_{0.76}Cu_{0.22}Fe_{0.11}Mn_{0.67}O_2$, $\rm Na_{4/5}Co_{4/5}Ti_{1/5}O_2$, &  \\
 & $\rm Na_{7/9}Ni_{0.20}Co_{0.38}Mn_{0.42}O_2$, $\rm Na_{0.7}RhO_2$, $\rm Na_{0.7}Ni_{0.35}Sn_{0.65}O_2$, &  \\   
  & $\rm K_{0.41}CoO_2$, $\rm K_{0.5}MnO_2$, $\rm K_{0.6}CoO_2$, $\rm K_{0.6}Cr_{0.6}Ti_{0.4}O_2$, $\rm K_{0.5}Mn_{0.7}Ni_{0.3}O_2$, &  \\
  & $\rm K_{0.59}Mg_{0.53}Sb_{0.47}O_2$, $\rm K_{0.5}Ni_{0.5}Sb_{0.5}O_2$, $\rm K_{0.5}Co_{0.5}Sb_{0.5}O_2$, $\rm K_{0.56}Ni_{0.52}Sb_{0.48}O_2$, &  \\
  & $\rm K_2Ni_2TeO_6$ ($\rm K_{2/3}Ni_{2/3}Te_{1/3}O_2$), $\rm K_{0.6}CoO_2$, $\rm K_{0.5}Mn_{0.7}Fe_{0.2}Ti_{0.1}O_2$, $\rm K_{0.6}Mn_{0.8}Ni_{0.1}Ti_{0.1}O_2$, &  \\ 
& $\rm K_{2/3}Ni_{1/3}Co_{1/3}Te_{1/3}O_2$ ($\rm K_2NiCoTeO_6$), $\rm K_{0.75}[Ni_{1/3}Mn_{2/3}]O_2$, $\rm K_{0.75}[Mn_{0.8}Ni_{0.1}Fe_{0.1}]O_2$, &  \\ 
& $\rm K_{0.83}[Ni_{0.05}Mn_{0.95}]O_2$, $\rm K_{0.21}MnO_2$, $\rm K_{0.7}[Cr_{0.85}Sb_{0.15}]O_2$, $\rm K_{0.76}Fe_{0.2}Mg_{0.1}Mn_{0.7}O_2$, &  \\
& $\rm K_{5/9}Mn_{7/9}Ti_{2/9}O_2$, $\rm K_{0.44}Ni_{0.22}Mn_{0.78}O_2$, $\rm K_{0.75}[Ni_{1/3}Mn_{2/3}]O_2$, $\rm K_{0.83}[Ni_{0.05}Mn_{0.95}]O_2$, &  \\
& $\rm K_{0.6}Co_{0.67}Mn_{0.33}O_2$, $\rm K_{0.6}Co_{0.66}Mn_{0.17}Ni_{0.17}O_2$, $\rm K_{0.41}CoO_2$, $\rm K_{0.65}Fe_{0.5}Mn_{0.5}O_2$, &  \\
& $\rm K_{2/3}Mg_{2/3}Te_{1/3}O_2$, $\rm K_{0.5}Mn_{0.8}Co_{0.1}Ni_{0.1}O_2$, $\rm K_{0.5}Mn_{0.7}Ni_{0.3}O_2$, $\rm K_{0.6}Cr_{0.6}Ti_{0.4}O_2$, $\rm K_{0.45}Mn_{0.5}Co_{0.5}O_2$ &  \\ 
& $\rm K_{0.70}[Cr_{0.86}Sb_{0.14}]O_2$, $\rm K_{0.5}MnO_2$, $\rm K_{0.69}CrO_2$, $\rm K_{0.54}[Co_{0.5}Mn_{0.5}]O_2$, $\rm K_{0.5}[Mn_{0.8}Fe_{0.1}Ni_{0.1}]O_2$, &  \\
& $\rm K_{1/2}Mn_{5/6}Mg_{1/12}Ni_{1/12}O_2$, $\rm K_{0.45}Mn_{1-{\it x}}Fe_{\it x}O_2$ ($x$ = $\rm 0, 0.1, 0.2, 0.3, 0.4$ and $\rm 0.5$), $\rm K_{0.8}CrO_2$, &  \\ 
& $\rm K_{0.5}[Ni_{0.1}Mn_{0.9}]O_2$,  $\rm K_{0.5}[Mn_{0.85}Ni_{0.1}Co_{0.05}]O_2$, $\rm K_{0.4}Fe_{0.1}Mn_{0.8}Ti_{0.1}O_2$, $\rm K_{0.45}Co_{1/12}Mg_{1/12}Mn_{5/6}O_2$, &  \\
& $\rm K_{0.45}Mn_{0.9}Mg_{0.1}O_2$, $\rm K_{0.39}CrO_2$, $\rm K_{0.23}MnO_2$, $\rm K_{0.54}Mn_{0.78}Mg_{0.22}O_{2}$, $\rm K_{0.45}Ni_{0.1}Fe_{0.1}Mn_{0.8}O_2$, &  \\
&  $\rm K_{\it x}Mn_{0.7}Ni_{0.3}O_{2}$ ($x$ = $\rm 0.4-0.7$), $\rm K_{0.4}Fe_{0.1}Mn_{0.8}Ti_{0.1}O_2$, $\rm K_{0.5}Mn_{0.7}Co_{0.2}Fe_{0.1}O_2$, &  \\
& $\rm K_{0.5}Mn_{0.72}Ni_{0.15}Co_{0.13}O_2$, $\rm K_{0.45}Mn_{0.9}Al_{0.1}O_2$, $\rm K_{0.67}Mn_{0.83}Ni_{0.17}O_2$, $\rm K_{0.5}Mg_{0.15}[Mn_{0.8}Mg_{0.05}]O_2$, &  \\
& $\rm K_{0.45}Ni_{0.1}Co_{0.1}Mg_{0.05}Mn_{0.75}O_2$, $\rm K_{0.45}Ni_{0.1}Co_{0.1}Al_{0.05}Mn_{0.75}O_2$, $\rm K_{0.48}Mn_{0.4}Co_{0.6}O_2$ &  \\
& $\rm K_{0.45}Rb_{0.05}Mn_{0.85}Mg_{0.15}O_2$, $\rm K_{0.35}Mn_{0.8}Fe_{0.1}Cu_{0.1}O_2$, $\rm K_{0.5}Mn_{0.8}Fe_{0.2}O_2$, $\rm K_{0.3}Mn_{0.95}Co_{0.05}O_2$ &  \\  
 & &  \\
\textbf{Topochemical Au ion-exchange} (molten Au salts) &  $\rm Na_2{\it M}_2TeO_6$ ($M = \rm Mg, Zn, Co, Ni, Cu$), $\rm K_2{\it M}_2TeO_6$  & $\rm Au_6{\it M}TeO_6$ \\
\textbf{Topochemical Cu ion-exchange} (molten Cu salts) &  $\rm Na_2{\it M}_2TeO_6$ ($M = \rm Mg, Zn, Co, Ni, Cu$), $\rm K_2{\it M}_2TeO_6$  & $\rm Cu_6{\it M}TeO_6$ \\
\textbf{Topochemical Tl ion-exchange} (molten Tl salts) &  $\rm Na_2{\it M}_2TeO_6$ ($M = \rm Mg, Zn, Co, Ni, Cu$), $\rm K_2{\it M}_2TeO_6$  & $\rm Tl_6{\it M}TeO_6$ \\
& &  \\ 
\hline
\end{tabular}}}
\end{center}
\end{table*}

The increasing application of artificial intelligence (AI) in the field of exploratory synthesis provides a compelling view into the future possibilities of materials preparation.\cite {raccuglia2016, coley2018, butler2018} By combining methodologies involving AI and machine learning, it becomes possible to accelerate the synthesis procedures associated with diverse honeycomb layered frameworks that feature metallophilic bilayers. AI algorithms can be trained to analyse vast amounts of data and predict optimal synthesis conditions, leading to more efficient and targeted fabrication of such layered frameworks. Moreover, the integration of machine learning techniques can enable the discovery of {\it de novo} honeycomb layered frameworks beyond the conventional scope of layered frameworks dominated by silver-based systems. These technologies can assist in the identification of suitable precursors, control of reaction parameters, and optimisation of synthesis pathways, thereby reducing trial-and-error experimentation and enhancing the reproducibility of desired layered materials. Ultimately, this multidisciplinary approach facilitates the rapid and systematic exploration of new layered frameworks, leading to accelerated progress in materials preparation and the discovery of innovative materials with tailored properties.

The exfoliation of materials belonging to this particular class, \textit {albeit} nontrivial, has recently emerged as a viable technique, capitalising on the mechanical exfoliation process. A notable advancement in this regard is the successful mechanical exfoliation of $\rm Ag_2CrO_2$, accomplished through the utilisation of the well-known "scotch-tape" methodology, which presents a highly promising avenue for the fabrication of expansive two-dimensional (2D) monolayer or few-layer thin configurations of such bilayered materials.\cite {taniguchi2018fabrication, taniguchi2020butterfly} Despite the seemingly insurmountable challenges associated with the realisation of exclusively 2D magnetic materials, intriguing reports have surfaced,\cite {radha2021ultrathin} underscoring the feasibility of generating 2D magnetic layers via the exfoliation technique.

A significant majority of layered materials comprising metallophilic bilayers, such as $\rm Ag_2{\it M}O_2$ (where $M$ represents elements such as Co, Ni, Cr, Fe, and Mn), exhibit antiferromagnetic properties,\cite {yoshida2020partially, williams1989neutron, susloparova2022, yoshida2020static, matsuda2012partially, schreyer2002synthesis, nozaki2008neutron, nozaki2013magnetic} making them promising candidates for the design of 2D magnetic materials with potential applications in future spintronics devices. Antiferromagnetic materials have gained considerable attention in the field of spintronics due to their minimal stray magnetic fields, which contribute to robust data storage capabilities. Additionally, antiferromagnetic materials enable high-speed data processing by virtue of their higher magnetic resonance frequency compared to conventional ferromagnetic materials. On the other hand, 2D antiferromagnetic materials exhibit another intriguing characteristic - magnetic frustrations. This phenomenon has emerged as a central topic in contemporary condensed matter physics. Notably, when magnetic atoms are arranged in a triangular (hexagonal) lattice, as observed in $\rm Ag_2{\it M}O_2$ compounds, it is well-established that the magnetic moments experience significant frustration due to geometric effects.

Moreover, when magnetic atoms are arranged in a honeycomb lattice, as observed in compounds like $\rm Ag_6{\it M}_2TeO_6$ (where $M$ represents elements such as Co, Ni, and Cu), an abundance of exotic magnetic phenomena, including the intriguing Kitaev spin liquid, can be anticipated. Furthermore, the presence of magnetic moments, if any, originating from pseudo-spins within metallophilic bilayers, such as Ag bilayers, can be investigated in nearly single crystal 2D layers of materials like $\rm Ag_2F$ or $\rm Ag_6{\it M}_2TeO_6$ (where $M$ corresponds to Mg or Zn), with the latter demonstrating greater stability.\cite{masese2023honeycomb}

\begin{figure*}[!b]
\centering
\includegraphics[width=1.05\columnwidth]{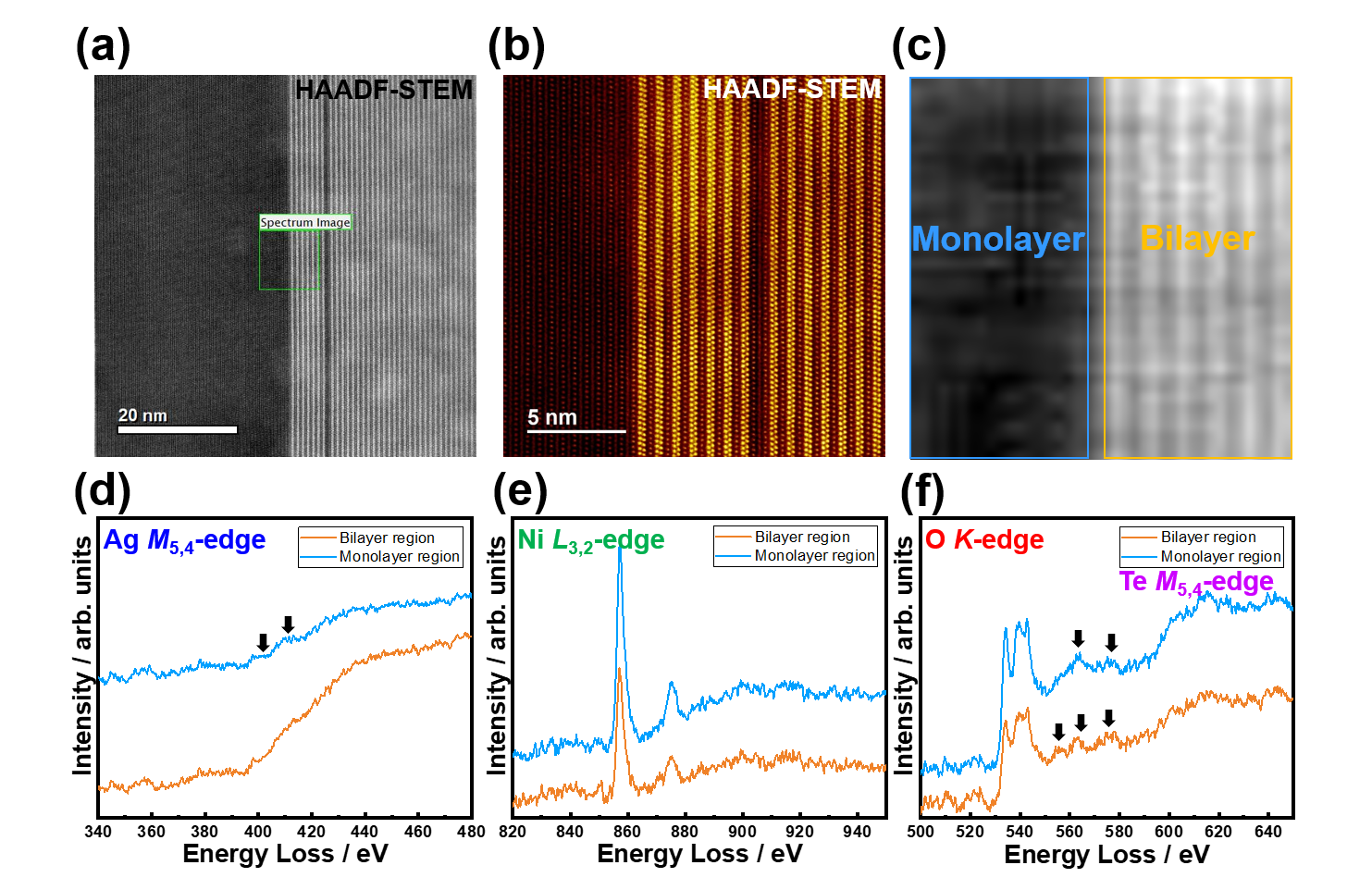}
\caption{Electron energy loss spectroscopy (EELS) of monolayer and bilayer $\rm Ag$ domains in global $\rm Ag_2NiO_2$ composition. (a,b,c) HAADF-STEM images showing $\rm Ag$ bilayer ($\rm Ag_6Ni_2TeO_6$) domains and monolayer domains (${\rm Ag}_{2 - x}\rm Ni_2TeO_6$, $0 < x < 2$). (d,e,f) EELS spectra taken at the $\rm Ag$ {\it M}-, $\rm Te$ {\it M}-, $\rm Ni$ {\it L}- and $\rm O$ {\it K}-edges of monolayer and bilayer $\rm Ag$ domains. Spectral changes are indicated by arrows. Note that, no sophisticated deconvolution procedures were performed on the sample since multiple scattering was found to be substantially suppressed due to the small thickness of the samples studied ($t/\lambda \simeq 0.6$, where $t$ is the thickness value of the samples and $\lambda \simeq 100$ nm is the plasmon mean free path which depends on the acceleration voltage and beam energy (herein, 200 kV and 200 keV respectively)).\cite{egerton2011electron}}
\label{Figure_22}
\end{figure*}

\subsection{Bridging the Glaring Gaps between Theory, Computations and Experiment}

The most straight-forward computations of materials such as $\rm Ag_2Ni_2TeO_6$ do not entirely consider any argentophilic interactions or sub-valency of silver.\cite{tada2022implications} Indeed, typical first principles DFT simulations of honeycomb layered oxides (\textit{e.g.} using VASP) do not incorporate such interactions since the projector-augmented-wavefunction method (PAW pseudopotentials) of the VASP simulation package or equivalent packages associated with bonding energies have to be determined \textit{a priori} via \textit{e.g.} using methods in G. Kresse's extensive body of works\cite{kresse1993ab, kresse1994ab, kresse1996efficient, kresse1996efficiency, kresse1999ultrasoft}, which not only neglect argentophilicity but also yield the valency state of $\rm Ag$ to be $\rm Ag^{1+}$ ($\rm Ag_2Ni_2TeO_6$) hence precluding metallophilic bilayers in typical structural environments exhibited by honeycomb layered oxides, when the Hamiltonian is diagonalised using the Kohn-Sham formalism.\cite{tada2022implications} 
Focusing on VASP, these issues may eventually be overcomed by appropriately setting the NELECT line in the VASP package to consider new physics and chemistry such as silver degenerate states. Given the non-Abelian nature of the proposed $\rm SU(2)\times U(1)$ interactions, this may necessitate a reconstruction of the pseudo-potentials in POTCAR, amongst other non-trivialities that presently prove challenging to computational efforts.

Noteworthy attempts towards successfully simulating metallophilic bilayers involving formalisms such as the electron localisation function (ELF),\cite {becke1990simple} the crystal overlap Hamiltonian population (COHP),\cite {dronskowski1993, deringer2011} or the non-covalent interaction (NCI)\cite {contreras2011} scalar field have proven to be invaluable tools for probing argentophilic interactions and discerning dissimilarities between layered frameworks exhibiting monolayer and bilayer configurations of cations. Nevertheless, it is essential to acknowledge that electronic calculations conducted within such formalisms are underpinned by certain assumptions. 
Analogous to the concept of polyhedral skeletal electron pair theory which applies to polyhedral clusters where the faces are triangular such as in the so-called Wade-Mingos molecular clusters\cite{wade1976, mingos1972nature, mingos1984}, noteworthy treatments have conjectured the existence of excess localised electrons within the $\rm Ag$ tetrahedral structures to account for the subvalency of $\rm Ag$, with certain drawbacks tied to the deployment and subsequent interpretation of ELF in the computations within the context of bond theory.\cite{yin2021reply, lobato2021comment, kovalevskiy2020uncommon} Heuristic considerations assume the excess electrons are placed within the bonding $5s$ bands or local $5s/5p$ skeleton orbitals. Consequently, it is crucial to perceive these computational endeavours as numerical methods, to be distinct from analytic methods such as the $\rm SU(2)\times U(1)$ theoretical framework advocating for $\rm Ag$ degenerate states. In comparison, the $\rm SU(2)\times U(1)$ model predicts $\rm Ag_{16}^{1/2+}B_4^{3+}O_{10}^{2-} = Ag_8^{2+}Ag_8^{1-}B_4^{3+}O_{10}^{2-}$ (instead of the non-equivalent $\rm Ag_{16}^{1/2+}B_4^{3+}O_{10}^{2-} = Ag_{16}^{1+}B_4^{3+}O_{10}^{2-} \times 8e^-$ proposed within polyhedral skeletal electron pair theory) which is already appealing from the onset since bonds between $\rm Ag^{2+}$ and $\rm Ag^{1-}$ (cation-anion bonds) forming stable structures are more intuitive within the context of valence bond theory compared to \textit{e.g.} $\rm Ag^{1+}-Ag^{1+}$ bonds. 

\blue{Given the proposition that a Heisenberg model (eq. (\ref{H_Dirac_eq})) due to the pseudo-spin degree of freedom associated with Ag $4d_{z^2}^1$ orbital ($\rm Ag^{2+}$) ought to exist in all Ag metallophilic bilayers exhibiting the subvalent state $1/2+$, it is extremely plausible that the conventional analyses proposed throughout \textbf{Section \ref{Section: Properties_Functionalities}} and \textbf{Section \ref{Section: Characterisation_Techniques}} such as resistivity, susceptibility and specific heat of Ag-based bilayered materials need to be re-evaluated. Specifically, $\rm Ag^{1/2+}$ in ${\rm Ag_2}M\rm O_2$ ($M = \rm Cr, Mn, Fe, Co, Ni$) is largely assumed non-(pseudo-)magnetic, leading to considerable challenges as to the origin of magnetic order especially in specific situations. Exploring the specific assumptions implicit in such analyses: 
\begin{enumerate}
    \item Similar to $\rm Ag^{1/2+}F^-$\cite{tong2011,wang1991,ikezawa1985,fujita1974,andres1966superconductivity, ido1988,ott1928,kawamura1974,argay1966redetermination}, the $\rm Ag^{1/2+}$ electron band structure is assumed quarter-filled, providing itinerant electrons responsible for the conductivity characteristics in resistivity measurements (resistivity against temperature ($\rho - T$) plots). However, a mechanism for the subvalent state independent of the properties of the electronic state of $M^{3+}$ leading to the quarter-filled $\rm Ag^{1/2+}$ electron band is generally not rigorously incorporated in the analysis of electrical and magnetic properties, with plausibly one or two noteworthy exceptions\cite{schreyer2002synthesis, johannes2007formation};
    
    \item Despite a (potentially distorted) triangular lattice of $M^{3+}$ atoms within the $M\rm O_2$ slabs always ascertained which often leads to geometric frustration, the experimentally observed anti-ferromagnetism associated with a N\'{e}el temperature, $T_{\rm N}$ in majority of ${\rm Ag_2}M\rm O_2$ materials is argued to arise from particular Heisenberg models of $M^{3+}\,\,3d$ electron spins by assuming an additional mechanism - the distortion of the $M\rm O_2$ octahedra/$M^{3+}$ lattice from trigonal/triangular to monoclinic/isosceles due to the Jahn-Teller effect. This has the benefit of aiding anti-ferromagnetic order to out-compete the geometric frustration, aiding to realise \textit{e.g.} A-type anti-ferromagnetism.\cite{yoshida2006spin} The distortion is often associated to discontinuous features in the specific heat and resistivity-versus-temperature plots at a structural transition temperature $T_{\rm s}$, provided the susceptibility-versus-temperature plots are feature-less at $T_{\rm s}$. The orbital ordering is assumed to play a role in determining $e_g$ ($d_{3z^2 - r^2} (d_{z^2})$ and $d_{x^2 - y^2}$) occupancy, which \textit{e.g.} governs the presence or absence of ferromagnetic and/or antiferromagnetic (indirect and/or super-exchange) Heisenberg interactions, spin canting and/or Dzyaloshinskii-Moriya (anti-symmetric exchange).\cite{yoshida2006spin, sugiyama2006Incommensurate, yoshida2020partially, yoshida2020partially, yoshida2008unique, nozaki2013magnetic, nozaki2008neutron, eguchi2010resonant} Alternatively, anti-ferromagnetism at $T_{\rm N}$ is introduced via an XY or 120$^\circ$ (chiral) model of $M^{3+} d$ spins on the triangular lattice whenever experimental orbital ordering can be precluded experimentally.\cite{korshunov1986phase} Other magnetic long range order such as Kosterlitz-Thouless (KT) transition and $Z_2$ vortices are often mentioned for completeness\cite{sugiyama2009static, taniguchi2020butterfly}; 
    
    \item The range of the low-to-high spin value of the $M^{3+}$ cation is determined by Hund's rule, either assuming the presence or absence of a Jahn-Teller distortion of the $M\rm O_2$ octahedra. Crystal field splitting is $3d$ orbitals ($t_{2d}$ and $e_g$) orbitals is always considered, \textit{albeit} Hund's rule sometimes chosen to arrive at high spin in order to explain the observed magnetic order. For instance, in ${\rm Ag}_2^{1/2+}M^{3+}\rm O_2^{2-}$ ($M = \rm Fe$), in place of the low spin $\rm Fe^{3+}$ $t_{2g}^5e_g^0$ ($S = 1/2$) electronic configuration expected from Hund's rule, the high spin configuration $\rm Fe^{3+}$ $t_{2g}^3e_g^2$ ($S = 5/2$) is chosen to explain experimental data.\cite{yoshida2020partially} A counter example is the case for $M = \rm Co$, which at low spin corresponds to $\rm Co^{3+}$ $t_{2g}^6e_g^0$ ($S = 0$). However, in order to explain the observed anti-ferromagnetism, $\rm Co^{2+}$ $t_{2g}^6e_g^1$ is assumed instead, which effectively implies $\rm Ag_2^{1+}Co^{2+}O_2^{2-}$ instead of $\rm Ag_2^{1/2+}Co^{3+}O_2^{2-}$.\cite{yoshida2020static} This would be problematic for the SU(2)$\times$U(1) model which associates Ag bilayers strictly with the subvalent state $\rm Ag^{1/2+}$. Note that, the higher spin configurations $\rm Co^{3+}$ $t_{2g}^5e_g^1$ ($S = 1$) and $\rm Co^{3+}$ $t_{2g}^4e_g^2$ ($S = 2$) were inconsistent with the measured paramagnetic moment $p_{\rm eff} = 1.62 \mu_{\rm B} = g\sqrt{S(S + 1)}\mu_{\rm B}$ with a Land\'{e}-$g$ factor of approximately $1.87$, but consistent with $S = 1/2$;  
    
    \item The interplay between the resistivity and magnetic properties such as the considerably enhanced conductivity below $T_{\rm N}$ are responsible for the heavy fermion (enhanced mass leading to $\rho \sim aT^2$ dependence in eq. (\ref{rho_eq})) characteristic at low temperatures\cite{eguchi2010resonant, ji2010orbital}, and a host of other phenomena such as a significant decline in resistivity.\cite{taniguchi2020butterfly, taniguchi2018fabrication, kida2015transport, yoshida2011novel, matsuda2012partially} Particularly, in ${\rm Ag}_2^{1/2+}M^{3+}\rm O_2^{2-}$ ($M = \rm Cr$), interactions between $\rm Cr$ $3d$ spin and Ag $5s^1$ itinerant electron interactions from $sd$ hybridisation manifest as an indirect exchange (RKKY) term between the thermally-fluctuating spins above $T_{\rm N}$ and anti-ferromagnetically localised spins below $T_{\rm N}$. The anti-ferromagnetism in the triangular lattice is explained by the appearance of a spin-canted partially disordered magnetic state commensurate with (3), \textit{albeit} an unexplained small magnetic moment along the $c$-axis persists below $T_{\rm N}$. Giant magneto-resistivity is attributed to a strong uniaxial anisotropy leading to a spin gap modulated by the magnetic field.\cite{taniguchi2020butterfly}   
\end{enumerate}
It is important to note that the $\rm SU(2)\times U(1)$ model offers a potentially different scenario for (1), (2), (3) and (4). In particular, (1) is accounted for by taking $\rm  Ag^{2+}Ag^{1-} \rightarrow Ag_2^{1/2+}$. In ${\rm Ag}_2M\rm O_2$, this suggests the chemical reaction,
\begin{align}\label{mono_bilayer_chem_eq}
    {\rm Ag^0} + {\rm Ag^{1+}} + M^{3+} \rightarrow {\rm Ag^{1-} + Ag^{2+}} +  M^{3+} \rightarrow 
    {\rm Ag_2^{1/2+}} + M^{3+},
\end{align}
where the silver atom $\rm Ag$ acts as an oxidising agent ($\rm Ag^0 \rightarrow Ag^{1-}$) leading to $\rm Ag^{1+} \rightarrow Ag^{2+}$. Strictly speaking $\rm  Ag^0Ag^{1+} \rightarrow Ag^{2+}Ag^{1-} \rightarrow Ag^{1/2+}Ag^{1/2+} = Ag_2^{1/2+}$ should be treated as a \textit{bona fide} oxidation state due to $\rm SU(2)\times U(1)$ gauge field mixing. `\textit{Bona fide}' here means that the $\rm Ag$ $\rm {\it L}_{III}$-edge XANES spectra of ${\rm Ag}_2M\rm O_2$ such as in \textbf{Figure \ref{Figure_11}b} ought to show no discernable difference between the energy peak positions of $\rm Ag^{1/2+}$ and $\rm Ag^{1+}$, owing to the orbital degeneracy of $4d^95s^2$ and $4d^{10}5s^1$ as depicted in \textbf{Figure \ref{Figure_17}c} and \textbf{Figure \ref{Figure_17}d} for the prismatic case (Strictly speaking, any differences in energy positions would be either as a result of orbital ordering of the $4d_{z^2}$ and $5s$ orbitals, with the difference corresponding to either the energy gap or arising from $\simeq 3.5$ eV because of the additionally plausible electronic configuration in \textbf{Figure \ref{Figure_17}b}. Meanwhile, the intensities $I$ will vary by a ratio of transition probabilities $I_{\rm Ag^{1+}}/I_{\rm Ag^{1/2+}} \propto |\langle 5s|H_{\rm int.}|p \rangle|^2/|\langle 4d_{z^2}|H_{\rm int.}|p \rangle|^2$ with $|\cdots \rangle$ the orbital wave functions and $H_{\rm int.}$ the interaction energy of the Ag $5s, 4d_{z^2}$ and O $p$ orbitals by a relaxation electron). 

Meanwhile, the magnetic order at $T_{\rm N}$ in (2) can potentially be explained by pseudo-spin degrees of freedom arising from the involvement of $4d_{z^2}^1$ electrons.\cite{kanyolo2023pseudo} These pseudo-magnetic degrees of freedom should be of magnetic character for this to be a plausible explanation. In this scenario, the geometry of the $\rm Ag^{1/2+}$ bilayers is considered as a (bifurcated) bipartite honeycomb lattice, thus potentially by-passing significant difficulties with geometric frustration precluding anti-ferromagnetic long-range order. A monolayer-to-bilayer phase transition then occurs at $T_{\rm s}$ and persists at lower temperatures corresponding to eq. (\ref{AgNiO2_Ag_eq}) given by, 
\begin{align}
    {\rm Ag^0} + {\rm Ag^{1+}}M^{3+}{\rm O_2^{2-}} \rightarrow {\rm Ag_2^{1/2+}}M^{3+}{\rm O_2^{2-}},
\end{align}
equivalent to eq. (\ref{mono_bilayer_chem_eq}). This is supported by the significant change in the interlayer distance already observed in $\rm Ag_2NiO_2$,\cite{nozaki2008neutron} \textit{albeit} attributed by the authors to a local/cooperative Jahn-Teller distortion of the $\rm NiO_2$ octahedra in accordance to assumptions in (2). This assumption contradicts $\mu^+SR$ measurement results which are consistent with the $\rm NiO_2$ octahedral structure remaining rhombohedral down to low temperatures.\cite{sugiyama2006Incommensurate} Generally, it can be argued that the bifurcated lattice is (pseudo-spin) ferromagnetic whereas the monolayered honeycomb lattice is (pseudo-spin) anti-ferromagnetic, since the bifurcation is a consequence of a finite pseudo-magnetic field.\cite{kanyolo2023pseudo, masese2023honeycomb} Nonetheless, parity considerations assign antagonistic behaviour between pseudo-spin and actual spin on the bipartite honeycomb lattice resulting in a possible (actual-spin) anti-ferromagnetic state at $T_{\rm N} < T_{\rm s}$, commensurate with experimental results. This also suggests we should always expect a spin $1/2$ Heisenberg model. Moreover, as mentioned in (3), the differing conclusion for the $\rm SU(2)\times U(1)$ model\cite{masese2023honeycomb} about the spin values of $M^{3+}$ mainly concerns $\rm Ag_2CoO_2$, whose oxidation state ought to be $\rm Co^{3+}$, in order to ensure the material has metallophilic Ag bilayers with subvalency $1/2+$. The final assumption in (4) concerns the hybridisation of $M^{3+}$ $d$ and the subvalent Ag $5s$ orbitals responsible for the heavy fermion behaviour, decline of resistivity below $T_{\rm N}$ and magneto-resistivity. Whilst such a hybridisation has not been ruled out in the $\rm SU(2)\times U(1)$ model, it is imperative to note that this explanation relies on anti-ferromagnetic order arising in the $M^{3+}$ $d$ triangular lattice below $T_{\rm N}$. However, difficulty in explaining the $\mu^+$SR data based on $\mu^+$ sites solely at the vicinity of the $\rm MnO_2$ plane in $\rm Ag_2MnO_2$ required the consideration of an addition site at the $\rm Ag_2$ plane, but nonetheless concluded the unlikelihood for both sites simultaneously to contribute to the antiferromagnetic order.\cite{sugiyama2009static} Nonetheless, since A-type anti-ferromagnetism, chirality order and other collinear ordered phases were excluded by the same token, it is very likely that the two relevant $\mu^+$ sites corresponding to two quite varied frequencies differing by a factor of 11.7 might correspond to the varied response of $\rm Ag^{2+}$ and $\rm Ag^{1-}$ within the bilayered structure. This would be consistent with the heavy fermion behaviour, since a bifurcated lattice is equivalent to an additionally generated mass by $\rm SU(2)\times U(1)$ symmetry breaking not only for the Ag cations, but also the itinerant electrons, which should be detectable as a renormalised electron mass term.\cite{kanyolo2023pseudo} The electrons are also considered paired below $T_{\rm s} > T_{\rm N}$, although this may not always lead to superconductivity at low temperatures. Finally, to take into account the magneto-resistivity, the modulation of the bilayered structure by external magnetic fields needs also be fully considered.\cite{kanyolo2023pseudo}

It is imperative to remark that, the above alternative explanations are meant to showcase the somewhat disarrayed state of (1), (2), (3) and (4) relative to experimental findings and the newly proposed $\rm SU(2)\times U(1)$ model. It is more than likely that, the $\rm SU(2)\times U(1)$ model can successfully be incorporated within key aspects of (1)-to-(4) without their complete overhaul, thus eliminating the bleak zero-sum-game scenario showcased in this section. Indeed, there appears to be a possible connection between the residual spin postulated in eq. (\ref{anisotropy_eq}) and the functional form of the pseudo-spin Hamiltonian for the Ag bilayers as postulated in eq. (\ref{H_Dirac_eq}). Finally, given the reports of magnetic and structural phase transitions, these materials are strong candidates for multiferroic properties.\cite{spaldin2005renaissance}}

Proceeding, we shall also focus on the challenges of the $\rm SU(2)\times U(1)$ model towards unequivocal experimental verification. In particular, it is not clear whether existing measurements techniques such as X-ray absorption spectroscopy (XAS) and X-ray photoelectron spectroscopy (XPS) are sensitive enough to distinguish between the degenerate states of silver in reported silver bilayered materials, amongst other novelties. Moreover, depending on the particular honeycomb layered oxide with metallophilic bilayers, the complexity of using XAS and XPS is further compounded by the lack of purity of the sample. For instance, the electronic structure of the newly discovered silver-based honeycomb layered tellurate has a global composition ${\rm Ag_2}M_2\rm TeO_6$ ($M = \rm Ni, Co, Mg, Zn, Cu,  \textit {etc}.$), which is a mixed phase of monolayer and metallophilic bilayers (${\rm Ag_6}M\rm TeO_6$). Nonetheless, spatially resolved spectroscopic characterisation via STEM electron energy loss spectroscopy (EELS) performed on Ag-rich nanocrystallite domains at the Ag {\it M}-, Te {\it M}-, Ni {\it L}- and O {\it K}-edges may prove more viable to distinguish these two phases, enabling the direct visualisation of local changes in valency states and chemical bonding in such intricate materials. Particularly, the utilisation of EELS is distinguished from XAS by inherent benefit of an \aa ngstr\"{o}m-scale electron probe. This technique can facilitate the direct visualisation of minute variations in local valency states and chemical bonding within such intricate materials, thereby yielding unparalleled insights into their structural intricacies. \textbf {Figure \ref {Figure_22}} illustrates the STEM-EELS spectra obtained from bilayer and monolayer $\rm Ag$ domains within $\rm Ag_2Ni_2TeO_6$ global composition, showcasing their distinct electronic structures. Spectra were acquired at the O {\it K}-, Ni {\it L}-, Te {\it M}- and $\rm Ag$ {\it M}-edges to discern variations. Whilst the valency state of Ni ($\rm Ni^{2+}$) remains constant in both the monolayer and bilayer domains, discernible changes in the EELS spectra at the $\rm Ag$ {\it M}-, Te {\it M}-, and O {\it K}-edges are apparent. However, the presence of a high background hampers a definitive assessment of the differences in valency states between the two domains. It should be noted that the imaging contrast of Ag columns in the $\rm Ag$ bilayer ($\rm Ag$-rich) regions exhibits greater brightness compared to that in the $\rm Ag$ monolayer ($\rm Ag$-poor/deficient) regions (\textbf {Figures \ref {Figure_22}a,b,c}). This contrast difference primarily results from the overlapping presence of adjacent $\rm Ag$ columns within the $\rm Ag$ bilayer domains. Additionally, it is important to acknowledge that, given the alignment was centered on the bilayer region, there exists the possibility of imaging deviations from the primary zone axis when examining the monolayer regime, potentially resulting in disparities in electron channeling (contrast). 

\begin{figure*}[!t]
 \centering
 \includegraphics[width=\columnwidth]{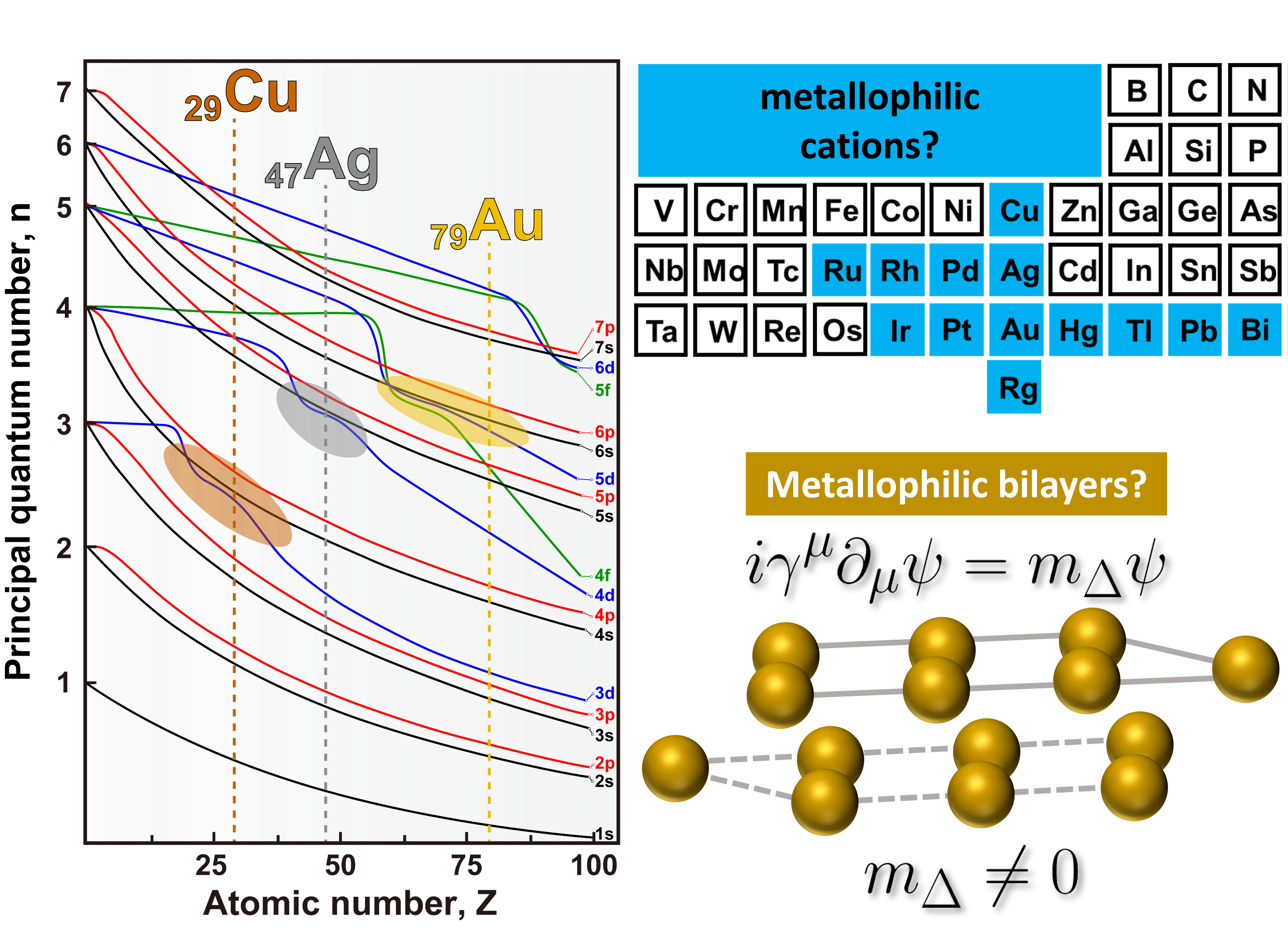}
 \caption{A depiction of the proximity of $s,p,d,f$ orbital energies for various principle quantum numbers ($1 \leq n \leq 7$) versus atomic number ($Z \leq 100$).\cite{Piela2007Electronic, atkins2014atkins} Evidently, the orbitals $nd$ and $(n + 1)s$ orbitals ($n = 3, 4, 5$) are energetically close to each other in coinage metal atoms $\rm Cu, Ag$ and $\rm Au$ corresponding to atomic numbers $Z = 29, 47$ and $79$. A truncated periodic table depicting $A = \rm Cu, Ru, Rh, Pd, Au, Ir, Pt, Au, Hg, Tl, Pb, Bi$ \textit{etc} metals with metallophilicity experimentally reported in literature (\textit{albeit} the report of metallophilicity in $\rm Rg$ is theoretical).\cite{doerrer2010steric} This suggests several candidates for metallophilic bilayers besides $\rm Ag$, which are worthy research prospects for the depicted metallophilic bilayers, such as is the case for the already experimentally reported thallium-based tellurate $\rm Tl_2MnTeO_6$,\cite{nalbandyan2019Tl2MnTeO6}. The bifurcated honeycomb lattice is depicted with $A$ atoms corresponding to gold coloured spheres, $i\gamma^{\mu}\partial_{\mu}\psi = m_{\Delta}\psi$ the Dirac equation, $\psi^{\alpha} = (A^{\alpha'}, A^{\alpha''})^{\rm T}$ the Dirac spinor, $m_{\Delta}$ the Dirac mass, and $\alpha, \alpha'$ the oxidation states of $A$ after spontaneous symmetry breaking (\textit{e.g.} $\alpha = 1/2+$, $\alpha' = 2+$ and $\alpha'' = 1-$ in the case of $A = \rm Cu, Ag, Au$).}
 \label{Figure_23}
\end{figure*}

On the theoretical front, the $\rm SU(2)\times U(1)$ model predicts there is an energy cost in particular $\rm SU(2)\times U(1)$ interactions which should be existent in measurement data (XPS or XAS), for instance whenever $\rm Ag^{1+}$ is available and transmutes into $\rm Ag^{1-}$ or {\it vice-versa}. One such cost is associated to $sd$-hybridisation, estimated to be of order 3.5 eV,\cite{blades2017evolution}  which leads to degeneracy of the valence band ($5s^2$-orbitals) and conduction bands ($4d_{z^2}^1$ orbitals) on the (bifurcated) honeycomb lattice. Lifting this degeneracy corresponds to a metal-semiconductor/metal-insulator (monolayer-bilayer) phase transition, in the spirit of peierls instability\cite{garcia1992dimerization}, analogous to Cooper-pair instability, which is related to paired electrons resulting in an energy gap.\cite{anderson1973remarks, little1964possibility} Rigorously, the $\rm SU(2)\times U(1)$ model introduces the Gellmann-Nishijima formula $Q = 2I + Y$, where $Q$ and $Y$ are the generators of the relevant charges $q_{\rm eff}, q_{\rm e}$ of the interacting particles/fields and $I = \sigma_3/2$ the iso-spin generator with $\sigma_3$ is the $z$ component of the Pauli vector, $\Vec{\sigma} = (\sigma_1, \sigma_2, \sigma_3)$. Particularly, $Q$ corresponds to the generator of the oxidation states of $\rm Ag$ after symmetry breaking (eq. (\ref{SU2U1_eq2})), which yields $+1, +2$ and $-1$ respectively for appropriate $I$ and $Y$ values,\cite{masese2023honeycomb} which in principle should be directly measured in actual experiments with effective elementary charge, $q'_{\rm eff}$ given in eq. (\ref{q_eff_dash_eq}). Recall that, $q_{\rm w}$ is the coupling constant of the $\rm SU(2)$ gauge field $W_{\mu}$ in eq. (\ref{SU2U1_eq}) which is \textit{a priori} unknown. Meanwhile, $q_{\rm e} \simeq 1.6\times 10^{-19}$ C is the familiar elementary charge of the electron (the coupling constant of the $\rm U(1)$ gauge field, $A_{\mu}$). Linking the $sd$-hybridisation to 3.5 eV (the energy gap between the orbitals) implies we can calculate $q_{\rm w}$ from,
\begin{align}\label{mass_WZ_eq}
    3.5 \,\,{\rm eV} \sim m_{\rm W}v_{\rm F}^2 = \frac{q_{\rm w}}{\sqrt{q_{\rm e}^2 + q_{\rm w}^2}}m_Zv_{\rm F}^2,
\end{align}
provided the screening length $\ell_{\rm s} = h/2\pi m_Wv_{\rm F}$ of electromagnetic fields arising form interaction of the form, $\rm Ag^0 = Ag^{1+}Ag^{1-}$ (assumed to give rise to argentophilicity in elemental silver) can be determined, with $h$ Planck's constant, $m_Z$ the mass of the $Z$ boson and $v_{\rm F}$ the Fermi speed. In the $\rm SU(2)\times U(1)$ model, we set $q_{\rm w} = q_{\rm e}$, which means the effective charge in eq. (\ref{q_eff_dash_eq}) becomes $q'_{\rm eff} = q_{\rm e}/2$, which translates into the existence of the half oxidation state, $\psi^{1/2+} = ({\rm Ag^{2+}, Ag^{1-}})^{\rm T}$. Such a monolayer-bilayer phase transition is yet to be experimentally observed. Nonetheless, reports in related metallophilic bilayers such as $\rm Ag_{16}^{1/2+}B_4^{3+}O_{10}^{2-}$ with unexpected semi-conductive properties and paired electrons\cite{kovalevskiy2020uncommon} encourage this prediction, suggesting there is cause for further investigation. 

On the other hand, there is need to predict the appropriate charge distribution expected in such experiments, as well as incorporating these new ideas into DFT calculations, particularly the Kohn-Sham formalism. The simplest treatment may eventually involve incorporating a Heisenberg pseudo-spin Hamiltonian term in pseudo-potentials to be diagonalised in the Kohn-Sham equations. Within this treatment, the bifurcation of the honeycomb lattice into a pair of hexagonal lattices constituting the metallophilic bilayer can be computationally treated using methods typically employed in Heisenberg spin Peierls distortion.\cite{said1984nonempirical, garcia1992dimerization} However, since the expectation is that the continuous symmetry breaking of $\rm SU(2)\times U(1)$ symmetry and such pseudo-spin models would arrive at complementary results, there is need to unequivocally show the equivalence of these two models. This endeavour ought to reveal unexpected results that should lead to further testable predictions. Moreover, there is a question of the nature of pseudo-spin interactions with (pseudo-)magnetic field terms such as strain, stress and/or external magnetic fields. Such magnetic terms may lead to the tuning of the argentophilic bond of silver fermionic pairs, analogous to the tuning of Feshbach resonance of paired fermions in condensates by external magnetic fields.\cite{chin2010feshbach} Finally, it is known that $nd$ and $(n + 1)s$ orbitals ($n = 3, 4, 5$) are energetically close to each other in coinage metal atoms.\cite{blades2017evolution} Since this proximity encourages $sd$ hybridisation, one should expect metallophilicity and hence non-Abelian gauge theories/pseudo-spin models on the bipartite honeycomb lattice for atomic numbers at or close to $29, 47$ and $79$ corresponding to $\rm Cu, Ag$ and $\rm Au$ respectively.\cite{blades2017evolution} Thus, the $\rm SU(2)\times U(1)$ model predicts coinage metal atoms in honeycomb layered materials can also form metallophilic bilayers. Other candidates for such metallophilic bilayers, such as the already experimentally found thallium-based tellurate $\rm Tl_2MnTeO_6$,\cite {nalbandyan2019Tl2MnTeO6} ought to involve metals at close proximity to the group 11 atoms in the periodic table, as displayed in \textbf{Figure \ref{Figure_23}}. 

Finally, one must acknowledge that a diversity of computational/numerical, analytical and experimental methods are expected to \textit{all} play crucial roles in the eventual understanding of the unexplained physics and chemistry of metallophilic bilayered structures. Ultimately, the validation of any proposed theoretical and/or computational framework should primarily rest upon experimental results.

\subsection{\label{Section: Conclusion} Concluding Remarks}
In the relentless pursuit of comprehending not only the intricate atomistic mechanisms but also venturing into the realms of material functionalities, this \textit {Review} has endeavoured to illuminate the recent strides achieved in the development of layered frameworks manifesting metallophilic bilayers from the viewpoint of synthesis, characterisation methodologies and ultimately simulation and theory. With an eye for meticulous scrutiny, we have sought to unravel the vast expanse of crystal chemistry, predominantly governed by compounds exhibiting Ag bilayers, whilst delineating the specific techniques employed to elucidate the diverse functionalities harboured within this remarkable class of materials. Through an extensive retrospective examination spanning an excess of six decades, we appraise the unparalleled progress witnessed in the development of these layered frameworks. However, amidst this expansive journey, a lingering inquiry emerges, beckoning us to contemplate the horizon of possibilities that lie ahead in this enigmatic universe of metallophilic bilayered materials within the vast multiverse of possible honeycomb layered frameworks and quantum matter in general.\cite{geim2016rise} Quantum matter refers to materials whose defining properties cannot be adequately understood via low-level quantum mechanics\cite{geim2016rise}, and hence may involve high-level quantum mechanical models and techniques \textit{e.g.} ubiquitous in high-energy physics or quantum gravity research. Our \textit{Review} has elucidated on various connections of the defining condensed matter physics processes in layered materials with metallophilic bilayers to concepts in high-energy physics and quantum gravity research such as electro-weak theory and black hole thermodynamics respectively, heralding their possible deployment as analogue systems especially in cases where the high-energy physics and quantum gravity counterparts are inaccessible in the lab. Will further advancements grace this domain, propelling it beyond a transient fascination and transforming it into an enduring and transcendent subject of exploration? It is our view that this \textit{Review} has convincingly answered this question to the affirmative.

\newpage
\section*{Acknowledgements}
The authors would like to acknowledge the financial support of AIST Edge Runners Funding, TEPCO Memorial Foundation, Japan Society for the Promotion of Science (JSPS KAKENHI Grant Number 23K04922) and Iketani Science and Technology Foundation. 
\newpage

\section*{Conflicts of interest}
There are no conflicts to declare.

\newpage

\bibliography{rsc} 
\bibliographystyle{rsc} 

\end{document}